# Fluctuations and Long-Term Stability: From Coherence to Chaos

## Maria K. Koleva

Institute of Catalysis, BAS, 1113 Sofia, Bulgaria
e-mail: mkoleva@bas.bg

## **C O N T E N T**

# Preface: What is This Book About?

Fluctuations are part of our daily life: traffic noise, heartbeat, public opinion, currency exchange rate, electrical current, chemical reactions - they all permanently irregularly vary in space and in time. Though this intuitive description has an ample understanding, we face the question whether there is a conceptual definition of a fluctuation independent of the enormous diversity of the "driving "forces": human emotions, economics, and physical interactions. The existence of such definition opens the door to a powerful general study whose outcomes are relevant to each system regardless to its origin and particularities. This makes the subject an indispensable part of the fundamental science. Indeed, an abstract definition of the fluctuations does exists for a long time - irregular deviations from the average that leaves the system stable; the average stays steady while the amplitude and the frequency of the deviations are nor precisely predictable. The established in the literature definition reads that the behavior of a fluctuating system is matched trough the properties of an irregular time series. A lot of efforts and ingenuity has been involved to study their characteristics such as the frequency and size of the fluctuations, first exit time etc. The delineation of the enormous number of papers is that they are specific to the statistics of the fluctuations. That is why the mean stream of interest has been concentrated on the onset of this or that statistics and the associated classification according to appropriate parameters.

However, no attention has been paid to the issue about the long-term stability of the systems permanently exerting fluctuations. The question is how the system regulates the characteristics of the fluctuations in order to stay stable. Again our daily experience helps to adjust our definition to the new constraint - the stability is provided if the fluctuations do not exceed the thresholds of stability of the system. Otherwise the system undergoes either qualitative changes or collapses. Evidently, the thresholds of stability are specific to the system. Furthermore, the "distance" to them can be subjective as in the case of our emotions. Yet, we assert that any "distance" to the thresholds of stability is always finite. This confinement is grounded on the general physical conjecture that all the objects in our Universe are created by involving finite amount of energy and/or matter and correspondingly each of them can be destroyed by involving finite amount of energy and/or matter.

An immediate outcome of above conjecture is that the functional relation between exerting fluctuations and long-term stability of a system implies two-fold comprehension. To begin with, the fluctuations should be bounded in one sense or another. For example, only limited by the thresholds of stability amount of energy and/or matter is involved in each fluctuation of the natural systems. In addition, the existence of thresholds of stability in the extended systems calls for a coherency of the behavior of the local fluctuations. Indeed, if the local fluctuations were unbounded and developed independently from one another, they would give rise to local defects. Since the spatio-temporal configurations of the fluctuations would be permanently unstable and would vary in uncontrolled way, they would yield the collapse of the system. Therefore, the coherence of the local fluctuations is necessary for providing long-term stability. Another important aspect is that it makes possible the observation of any fluctuation. Otherwise, if the local fluctuations were uncorrelated, their spatio-temporal average would yield permanently zero.

I shall prove that, to the most surprise, the bounded irregular time series exhibit strong chaotic properties that are remarkably insensitive to the particularities of the fluctuations. Let just recall that the properties of the unbounded time series are specified by the statistics of the fluctuations. On the contrary, I shall make evident that the bounded ones exhibit properties that are shared at arbitrary statistics if only the correlation size of the fluctuations is limited. Furthermore, it turns out that the chaotic properties of the bounded irregular series are insensitive to their mathematical origin. Thus, the chaoticity is not anymore hallmark of simple non-linear deterministic systems only. The boundedness demonstrates an extraordinary universality - it is not only universal in its conceptual necessity for providing long-term stability but it gives rise to property that is shared on the same basis. This justifies the title of the book: the coherence and boundedness are necessary conditions for the long-term stability of any object/subject whilst the chaoticity is its hallmark.

The universality of the boundeness opens up a lot of questions. Among the most important ones is the one that concerns its asymptotic: what happens when a fluctuation reaches the thresholds of stability - does it make a "U-turn" or bumps into them?! Is the "U-turn" related to specific for the system process? If so, how to incorporate it among the universal properties? If not, is there a functional operation that provides the "U-turns" insensitively to the involved physical processes? Besides, how to conceptualize the coherence? What makes the local fluctuations at distant points to behave coherently when the interactions among the entities are only short-ranged? If the coherence does exist, is there any operational equivalence of its functioning in different systems?

The present book is a systematic study that aims to consider the above topics. On the way to build a comprehensive framework I shall discuss controversies whose solution requires the introduction of radically new viewpoints on a number of widely accepted so far issues.

I shall start with considering the operational equivalence of the coherence in the social and natural systems. Now I shall present its implementation in the social systems. Thus I can make my point more universally understandable.

Social behavior considered as a realization of numerous strategies by different agents provides one of the greatest mysteries in the world. Some local events bound to instability by the current religious, historical, cultural, etc. milieus suddenly become definitive in the rearrangements of the social priorities that take place in a given community. Rationality and various reductionist social models based on automatic interplay of social groups, modes of production, and functions of social structures do not provide sufficient answers that can explain these sudden social changes.

I shall try to outline the paths leading to such 'non-linear' behavior of sudden changes of the public opinion. Dominant systems of community values shape individual opinions and behavior. Opinion polls, however, show diversified responses and their results remain away from a simple sum of individual answers. Let me start with some social preferences that modify personal opinions. We are very positive on certain topics, beliefs, and predispositions, but remain hesitant in some others. For example most of people believe that there exist alien forms of life in the Universe. But to the question: 'Are aliens friendly to us?' public opinion splits approximately in a half**.** For the purpose of this book it is enough to know that I have definite opinion in some issues, while others make me feel uncertain. My aim is to describe the general path that leads from destabilization of positive personal opinions into social instability that rearranges some priorities in a given community. Such factors may have different nature, but their essential quality is that they operate locally or globally. The local behavior is formed by the closest circle of people we love, family, friends that constitute our 'neighborhood'. Normally, we share same opinions. Yet, the neighborhood may be split on important issues and sends disorienting signals to its members. Opinions can change from positive to negative and vice versa. Moreover, the uncertainty is "contagious". Our "neighbors "infect" their other "neighbors" and so the process of destabilization spreads throughout larger parts of the society. Now the question whether the destabilization is limited to scattered non-overlapping areas or it is spread over the entire society comes. The importance of this question becomes evident from the following: in the case of isolated "islands" of uncertainty embedded in a "sea" of certainty, the dominant opinion comes from the "sea". Thus, it is insensitive to the current events and does not change significantly with the course of time. If, however, the uncertainty prevails, i.e. it is above the percolation threshold, the vote exhibits minor random variations around the "neutral" line - 50:50. This makes the social support "blind" - the winner is elected by a "close margin' or sometimes is unpredictable as "a tossing of a coin". Out next task is to illustrate that there is a way to global stabilization and that it goes through global coherence. The latter provides uniformity of every "neighborhood" and thus ensures the stabilization of each individual. Further, it is important to consider what are the factors and implements that help stabilization.

Our experience teaches us that the psychological and emotional arguments make us more confident in ourselves than any logic. However, isolated and non-correlated episodes of individual stabilization do not help the global one. The second major source of influence on the individual opinion, the global one, comes from the viewpoint of our "idols", movie stars, journalists, favorite shows, delivered instantly to every individual by the wireless technologies (radio, TV, Internet). Thus, if some local event of a great psychological and/or social impact is presented as emblematic and is strongly supported by the celebrities, its effect stabilizes simultaneously the public opinion through the stabilization of each individual. The hallmark of the stabilization is the coherence - all the individuals vote for same alternative, i.e. alternative "chosen" by the effect of the emblematic "hot case". Thus, the society supports that decision-making strategy whose top priority is situated at best to the current emblematic event(s).

The role of the psychological aspect of the local events is well known and is widely exploited nowadays. But it has been exploited before the TV time as well. Let us go back to the medieval times, to the discovery of America by C. Columbus. The new continent was named not after him but after the name of Amerigo Vespucci, the man who has become popular because he proclaimed the discovered by C. Columbus land "The New World". The magic of the collocation "New World" gave a strong hope for a new life for the old world though the people knew a little or noting about it.

The general framework of the coherence drawn from the above considerations is as follows:

(I) the coherence operates in those many-body systems that are subjects to both local and global destabilization. The understanding of the local one, derived from its "social" context, implies that the interaction of each entity with its neighborhood exerts multi-valued response so that one selection, arbitrarily chosen among all available. At the next instant, the entity responses via another selection, again arbitrarily chosen. As a result, the spatio-temporal configuration of the local fluctuations permanently varies which gives rise to the global destabilization.

(ii) the coherence is achieved through a very fast process that imposes the characteristic of a single selection throughout the entire system. As a result, all the entities share the characteristic of the same selection.

Since the succession of the selections is arbitrary, it is to be anticipated that the global characteristics exhibit permanent variations with time. It should be stressed on the point that the process of coherence eliminates the destabilization through making all the entities to share the properties of a single one.

Yet, the application of the above protocol to the natural many-body systems inevitably challenges our notions about "interaction" and "stability". Since our undergraduate times, we are used to believe that every complex dynamical interaction is additively decomposable to irreducible simpler contributions. This idea has been widely exploited to account for the effect of the media on every single entity. However, the *additivity* of the dynamical contributions always gives rise to a single-valued effect. That is why, despite its ingenuity and elaboration, every of the developed so far approaches to the many-body interactions brings about a single-valued result. Then, how it is possible at all, to have a multi-valued response? Actually, it has already emerged in the diagrammatic approach though it has not been recognized as a source of destabilization. Indeed, each $n-$body interaction contributes by several different diagrams. However, that additivity is strongly grounded on the assumption that the local fluctuations due to the establishing of different "diagrams" at distant points are automatically damped. Our next task is to illustrate that the development of the local fluctuations produces a self-sustained destabilization whose elimination demands global coherence.

To illustrate the problem I present it for the case of surface reactions. They are an important and wide class of chemical reactions that proceed on interfaces gas/solid. The generic properties of the surface reactions have been successfully modeled by the use of the lattice-gas approach. The surface is approximated by a lattice at whose vertices identical active sites are displaced. The prerequisite of any reaction is the adsorption of the reactants on the active sites. The reaction proceeds only between already adsorbed entities of required types brought about at the same site by their mobility. A very important property of the adentities is that they exert hard-core repulsion - only one entity can be adsorbed on a single active site. The modeling of the hard-core repulsion sets the relevance of the lattice-gas approach to a large variety of other many-body systems.

Suppose now that the lattice is exposed to a steady flow of reactants. It means that free entities hit the surface and some of them become trapped in one of highly excited states of the active sites. Let us consider an entity trapped in a vacant site. Its further relaxation to the ground state can be interrupted by an already adsorbed entity that arrives at the same site and most probably occupies it. Thus the adentity violates the further trapped entity relaxation at that site since no more than one entity can be adsorbed at a single site. The trapped entity can complete the adsorption if and only if after migration it finds another vacant site. The impact of the adentity intervention to the trapped entity probability for adsorption is twofold: first, it cannot be considered as a perturbation, since it changes the adsorption potential qualitatively, namely from attractive it becomes repulsive. That is why, that type of interaction has been called diffusion-induced non-perturbative interaction. Second, the lack of coherence between the trapping moment and the moment of adentity arrival makes the probability for adsorption multi-valued function: each selection corresponds to a certain level of relaxation at which a diffusion-induced non-perturbative interaction happens. Therefore, this interaction brings about fundamental duality of the probability for adsorption (and of the adsorption rate correspondingly): though each selection can be computed by an appropriate quantum-mechanical approach, the establishing of a given selection is a stochastic process since it is a random choice of a single selection among all available.

Since the diffusion-induced non-perturbative interactions are local events, the non-correlated mobility of the adentities produces a lack of correlation among the established selections at any distance and at any instant. As a result, the produced adlayer is unstable - the lack of correlation among the local adsorption rates produces non-correlated variations of the local concentrations of the adentities. Furthermore, the enduring mobility of the adentites permanently sustains the development of unbounded both in size and in amplitude fluctuations of the concentrations. Thus, the local fluctuations certainly give rise to local defects such as overheating, sintering, reconstruction of the surface etc. If the fluctuations remain unlimited and non-correlated and the spatio-temporal configurations succeed in arbitrary order, the destruction rapidly confines larger and larger areas and eventually the system collapses. So, the prevention of the breakdown ultimately demands long-range correlation among the local fluctuations so that their development to "respect" the thresholds of stability.

Coupling of the fluctuations is an essentially non-local event and thus requires the involvement of spatially extended excitations. Since the interactions among reacting entities are short-ranged, the only available non-local excitations are the cooperative excitations of the lattice (surface, interface). A successful coupling needs a feedback that acts toward evening of the initially non-identical rates making the coupled entities "response" to further perturbations coherent. It should be grounded on a strong coupling adlayer-lattice, namely: the energy of colliding entities dissipates to local cooperative excitations of the lattice. In turn, the impact of these local modes on the colliding entity is supposed large enough to induce a new transition that dissipates through the excitation of other local cooperative modes and so on. The feedback ceases its action whenever the colliding entities response becomes coherent.

However, there are two seemingly highly contradictive requirements to the feedback. On the one hand, it should operate at every reaction, i.e. regardless to the chemical identity of the reactants. On the other hand, it should preserve the chemical identity and the specific stoicheometry of any particular reaction. The reconciliation is non-trivial and needs radically novel viewpoint on a number of notions and approaches.

The major attention of the book is focused on the mechanisms that guarantee boundedness and coherence in the physical systems. I put forward crucial arguments that they cannot be integrated coherently in any of the existing so far approaches. My leading priority is the development of a self-consistent theory that is grounded on the conjecture of boundedness and coherence.

# Acknowledgments

Usual practice in writing monographs is the cooperation with a number of colleagues, sharing ideas and discussion on hot issues. The other side of that practice is reading a huge amount of literature. Therefore every book is result of collective efforts that should be acknowledged by the authors. Its official expression goes via explicit listing of the names of the people with whom the main stream of discussion that have been going on. The acknowledgement that comes from the literature is expressed trough appropriate citation.

I am going to share that practice and I am starting with my gratitude to my good friend Prof. Donald L. Bennett from Copenhagen who did his best to provide financial support for my participation to the $8-th$ Workshop "*What Comes beyond the Standard Model*" that took place in Bled, Slovenia in July 2005. At that workshop my book was intensively discussed by the participants which resulted in significant progress on a number of issues. I would like to thank all the participants and in particular to Prof. Holger B. Nielsen from Niels Bohr Institute in Copenhagen for his effective involvement and comments on various topics that are subject of the book.

I am also indebted to Prof. N. Peskov from Lomonosov University in Moscow who granted to me his computer codes by the use of which I have worked out the embedding dimension in Chapter 8.

The reader might be surprised of the very few citations in the book. The purpose is that I put forth an original concept that has not been discussed elsewhere. That is why I have focused the attention rather on its development in a self-consistent theory than on its comparison with other ideas and theories.

# Chapter 1: Bounded Irregular Sequences

## 1.1. Why to Read Chapter 1

Let us start the journey into the central conjecture about the boundedness and coherence by my most general claim: the universality of chaotic properties of bounded irregular sequences (BIS). More precisely, I assert that all BIS of infinite length share same asymptotic properties regardless to when the fluctuations commence and the details of their development. Alone the assumption of boundedness allows us to understand common properties of systems as diverse as quasar pulsations, financial time series, heartbeat and fluctuations in electric current.

The first step on the road is to specify the notion of a BIS. A BIS is any stochastic sequence the terms of which succeed in arbitrary order but the size of the terms is limited to be within given margins. The terms correspond to the fluctuations and the margins are dictated by the thresholds of stability. An example of a BIS is the record of daily variations of the currency exchange rates.

The length of a BIS is determined by the time interval during which the system stays stable. In this chapter we consider only BIS of infinite length. Applied to the real world this means that system in question stays stable permanently. Here human intuition runs into a puzzle - what happens when a large enough fluctuation reaches the thresholds of stability? On the one hand, if the fluctuation "bumps" into the threshold transversely, it destabilizes the system. So, depending on the "bumping" angle, the further evolution is either terminated due to instability or requires a specific response in order to stabilize the system. However, the stabilization is always temporary because one of the next "bumps" will most likely cause the collapse of the system. Hence, in order to provide an arbitrarily long-term stability, the fluctuations must make a "U-turn" at their approach to the thresholds of stability. But this seems rather odd - it means that the fluctuations do not "feel" the thresholds because there is something that drives the U-turn. This is a key question in the present book. The mechanism behind the U-turn brings about coherence and the stretching and folding, the mechanism that brings about the deterministic chaos in a simple dynamical system, in an intriguing interplay. Still, I provide rigorous study on the matter only for the natural many-body systems. This discussion is postponed to Chapter 3.

Since the general mechanism for the "U-turns" requires only very weak conditions about the properties of the fluctuations and their origin, it is reasonable to postulate the relevance of the U-turn mechanism for a broad spectrum of systems. Although the specification of these conditions is postponed until Chapter 3, the idea associated that the "U-turn" is that the development of any given fluctuation does not initiate a process behind any specific response of the system. Hence, the sequence is scale-free, i.e. all the time scales contribute uniformly to the properties of the system.

Now I can formulate the most fundamental property of the BIS of infinite length: each BIS remains bounded and scale-free upon coarse-graining. The latter implies replacing of the detailed structure of an object, domain etc. by a "smoothed" one according to an appropriate protocol. Coarse-graining is inevitable in any recording of experimental results. Despite the sensitivity and elaborativity of the latest experimental equipment, its complexity makes the results obtained "coarse-grained" in one sense or another. Here we use the notion "coarse-graining" in a broader sense - it encompasses processes ranging from local averaging, to local amplification and/or local damping, filtering, resolution effects etc., i.e. the coarse-graining involves non-linear operations that acts non-homogeneously on the original BIS. The "non-linearity" introduced in the recording of data has a crucial effect on every stochastic sequence since it distorts the values of its members. Hence, does the "coarse-graining" effects a "transformation" of one stochastic sequence into another?! This suggestion is of primary importance both from the conceptual and the practical point of view. Can we ever hope to separate the specific characteristics of any system from the "noise" if we are not able to "record" any signal precisely?! This problem outlines the immense importance of our statement that BIS have certain common properties and makes the goal of our book two-fold - along with establishing the general properties of the fluctuations, we must outline the route to the specific ones.

### 1.2. Transformational and Scaling Invariance

Now I shall illustrate how the coarse-graining operates by means of one of its "tools", namely local averaging. Given a BIS that may be either discrete, (i.e. its members are $x_i$ placed at time points $t_i$) or continuous, (i.e. to be a bounded stochastic function $x(t)$). For the sake of convenience I will develop what follows in terms of functions.

The local average is defined in the window $[T_1, T_2]$:

$$\bar{x}_{T_1 T_2} = \frac{1}{(T_2 - T_1)} \int_{T_1}^{T_2} x(t) dt \tag{1.1}$$

Recalling that the expectation value of an arbitrary sequence is defined as:

$$\langle x \rangle = \lim_{T \to \infty} \frac{1}{T} \int_0^T x(t) dt \tag{1.2}$$

the question of how "close" the local average is to the expectation value arises naturally. Obviously, when the window $[T_1, T_2]$ goes to infinity, it is to be expected that the local average converges to the expectation value. Suppose that for the BIS the expectation value does exist and is finite. This will be proved later. Now, along with this question I address the issue what happens when the window is finite. The latter problem is explicitly related to data recording. In general, a successful modeling of a wide variety of effects like inertia, finite resolution, amplification/damping that give rise to "non-linearities" in the recording, entails into partitioning of a BIS into windows of variable length. As a result, the "recording" constitutes a new stochastic sequence the members of which are the local averages with the length of the BIS rescaled by an arbitrary parameter $n$. But is it indeed another BIS?! Yes, it is. To prove it, let us have a closer look at the parameters that govern the value of a local average. It reads:

$$\bar{x}(\tilde{t}) = \frac{1}{(T_{i+1} - T_i)} \int_{T_i}^{T_{i+1}} x(t) dt \tag{1.3}$$

where $T_i = T_{i+1} - T_i$ is the length of the $i-th$ window and $\tilde{t}$ is the rescaled time. Our first task is to show that new stochastic sequence remains bounded. Let $r_{corr}$ be the correlation radius of the fluctuations. The largest possible value of $\int_{T_i}^{T_{i+1}} x(t) dt$ cannot exceed $r_{corr} x_{thres}$ at $T_i > r_{corr}$. Then, all the deviations of the local averages from the expectation value remain bounded within the following margins:

$$0 < \bar{x}(\tilde{t}) - \langle x \rangle \le \frac{r_{corr} x_{thres}}{T_{min}}. \tag{1.4}$$

So indeed the new function is bounded if only $T_{min}$ is non-zero.

Furthermore, the value of each member of the offspring BIS can be made equal to any *apriori* given one by an appropriate choice of the corresponding window length. Hence, the possibility for "designing" an offspring BIS is due to the variability of the window length. However there is another possibility for "designing". The local averaging maps an entire interval (window) onto a single point as to associate the value of the local averaging with this point. However, the mapping is not diffeomorfic since each point of the window is possible target. This is an immediate outcome of our supposition that all the scales and all the points uniformly contribute to the properties of the BIS. Thus, I am free to associate the averaged value with any point in the window. However, each particular choice constitutes a different offspring BIS. As a result the "digitizing" of a continuous signal (function) is mapped into discrete sequence the detail of which is not uniquely specified.

In summary, the "coarse-graining" can indeed modify every parameter that characterizes a given BIS, but leaves intact the property of the BIS that is to be bounded. I call this property transformational invariance. In more abstract terms it reads: the boundedness of a BIS is invariant under coarse-graining. Furthermore, the transformational invariance implies and justifies the transitivity of a set of BIS, i.e. each BIS can be reached by any other by the application of a sequence of appropriate coarse-grainings.

It should be stressed that the transformation of one BIS into another does not interfere with their self-similarity. Let me recall that self-similarity is a property that has nothing to do with the coarse-graining! Indeed, the self-similarity involves a simultaneous rescaling of the offspring function and the time by the same parameter in linear and homogeneous way, i.e. it acts on each member in the same *linear* manner. On the contrary, coarse-graining is essentially "non-linear" and its action is non-homogeneous.

My amazing story about the coarse-graining is not yet finished. The first part was about how it transforms one "beast" into another. The next one will be about how it makes all the "beasts" into the same kind. In other words, I shall show that the BIS are scaling invariant.

Let us have closer look at the upper bound in eq.(1.4). Since both $r_{corr}$ and $x_{thres}$ have fixed values determined by the particularities of the BIS, the only parameter that governs the value of the local average is $T_{\min}$. Evidently, when it monotonically increases, the term $\frac{r_{corr} x_{thres}}{T_{\min}}$ becomes smaller and smaller. In turn, the margins within local averages deviates from the expectation value become more and more narrow. So, the first conclusion is that the expectation value does exist and it is finite. The second one is that local average is scale invariant. The next task is to prove this. According to the widespread interpretation, the scaling invariance is validated when the following relation holds:

$$\frac{1}{T}\int_0^{T/\delta} dt \int_t^{t+\delta} \left( x(\xi) - \langle x \rangle \right) d\xi = \varepsilon(T) = \varepsilon(\delta)\varepsilon\left(\frac{T}{\delta}\right). \tag{1.5}$$

Decoded into simple words, this reads: given a window of length $T$ equipartitioned into subwindows of length $\delta$, the local averaging can be done in two ways. The first one is by averaging over the original BIS. The second one involves two steps. The first one is the averaging over the subwindows due to coarse-graining that results in an averaging over the subwindows and time rescaling by parameter $\delta$ so that a new BIS is produced. The second step is the averaging over the offspring BIS. The claim of eq.(1.5) is that the result should not depend on the choice of the parameter $\delta$. Formally, it is achieved if and only if all the $\varepsilon$-functions are power functions of the same power and the sequence is scale-free, i.e. there is no scale (length) that has a specific contribution to the average.

But, the relation:

$$\varepsilon(\delta)\varepsilon\left(\frac{T}{\delta}\right) = \varepsilon(T) \tag{1.6}$$

seems unclear because the association of the $\varepsilon$-functions with the local average renders them ill-defined. Indeed, the local average varies from one window to another. Yet, the very meaning of the $\varepsilon$-functions as defined by the *l.h.s* of (1.5) is that they are to be associated with the local average! The only way out goes via the identification of the $\varepsilon$-functions with the upper limit of the local average determined by eq.(1.4). Indeed, since $\varepsilon(\delta)$ is local average, it cannot exceed $\frac{r'_{corr} x'_{tresh}}{\delta}$. By the same reasoning, $\varepsilon\left(\frac{T}{\delta}\right)$ cannot exceed $r''_{corr} x''_{tresh} \frac{\delta}{T}$ and $\varepsilon(T)$ cannot exceed $\frac{r_{corr} x_{tresh}}{T}$. Now, the substitution of the corresponding upper limits into (1.6) yields:

$$\frac{r'_{corr} x'_{tresh}}{\delta} r''_{corr} x''_{tresh} \frac{\delta}{T} = \frac{r_{corr} x_{tresh}}{T} \tag{1.7}$$

Evidently, the relation (1.7) confirms the scaling invariance of the original BIS, i.e. the independence from partitioning parameter $\delta$. However, (1.7) says more than that: after the reduction of the scale parameters $\delta$ and $T$, it becomes an explicit expression of transformational invariance! Indeed, it relates the specific parameters $r_{corr}$ and $x_{tresh}$ of the original BIS and the specific parameters $r''_{corr}$ and $x''_{tresh}$ of the offspring BIS obtained by the coarse-graining.

So, the coarse-graining applied to a BIS entangles the transformational with the scaling invariance! Remarkable!

Let us now consider another aspect of the scaling invariance. I assert that (1.5) is a necessary condition for having the uniform convergence of the local average to the expectation value at $T \to \infty$. Intuitively, it seems that this is redundant because compliance with (1.4) is enough for this purpose. However, (1.4) is enough for the original BIS because its parameters $r_{corr}$ and $x_{tresh}$ are well-defined by our choice. But, how about its offsprings? Do they also have well-defined correlation size and thresholds of stability? The relation (1.7) gives an affirmative answer because it explicitly relates the parameters of the original and the offspring BIS. Since (1.7) allows a range of $(r_{corr}, x_{rresh})$ combinations, it gives rise to an infinite number of well-defined BIS that can transform into each other. So, the scaling invariance validates the uniform convergence of the local average to the expectation value for the entire set of the BIS.

It is to be anticipated that the entanglement between the scaling and transformational invariance is what gives rise to those properties that are insensitive to the particularities of the BIS. Indeed, if one such property is established for a BIS subject to a scaling invariance, the relation (1.7) insures that this property is automatically shared by the entire set of its offsprings.

We now start the study of those properties of the BIS that are insensitive to the details of the fluctuations. The pivotal point is that the desired properties emerge from interplay of the boundedness and the scaling invariance but remain independent of the statistics of the fluctuation succession. As the scaling invariance is to play an important role we need first to investigate the properties of a BIS in a window of arbitrary but finite length $T$ and then to take the limit $T \to \infty$.

### 1.3 Autocorrelation function

I start with a very important characteristic of every irregular sequence, namely its autocorrelation function whose definition reads:

$$G(\eta, T) = \lim_{T \to \infty} \frac{1}{T} \int_0^{T-\eta} \left( X(t+\eta) - \langle x \rangle \right)\left( X(t) - \langle x \rangle \right) dt . \tag{1.8}$$

The autocorrelation function is a measure of the average correlation between any two points in the sequence separated by a time interval $\eta \leq T$. An intriguing interplay of the boundedness and the scaling invariance will be developed. On the one hand, the uniform contribution of all time scales restrains persistence extent without signaling out any specific one. On the other hand, the boundedness limits the persistence because every deviation inevitably "turns back". Note that the uniform contribution of all time scales is very different from random contributions. The latter is characterized by the lack of any systematic correlations whilst the former is characterized by persistent but parameter-free correlations. But how far away is the persistence spread? And how the boundedness affects its extent?

The uniform contribution of all time scales gives rise to the following estimation of the autocorrelation function. An immediate outcome of (1.4) and (1.7) is that the local average in a window of length $T$ is of the order of $\frac{1}{T}$. This means that the "distance" between the successive "zeroes" of the BIS, i.e. the points where the fluctuations cross the average, is bounded and is independent of $T$. This result applied to the autocorrelation function $G(\eta, T)$ yields the following estimation of $G(\eta, T)$:

$$H\left(\frac{\eta}{T}\right) \propto \sigma \lim_{T\to\infty} \frac{T-\eta}{T} = \sigma \lim_{T\to\infty}\left(1-\frac{\eta}{T}\right) \tag{1.9}$$

where $\sigma$ is the variance of the fluctuations. But how well does $H\left(\frac{\eta}{T}\right)$ approximate $G(\eta,T)$? The advantage is that $H\left(\frac{\eta}{T}\right)$ depends on $\eta$ and $T$ only through their ratio. This immediately makes the autocorrelation function independent from the particular choice of $T$. Note that $\frac{\eta}{T}$ is always confined to be in the interval $[0,1]$! In turn, the independence of the autocorrelation function $G(\eta,T)$ from $T$ meets the requirement to be scale-free. However, the evasive point is that the permanent boundedness does not participate to $H\left(\frac{\eta}{T}\right)$?! This statement needs a clarification since so far we have been considering only *bounded* sequences!

The fact is that the limitation over the size of the distance between the zeroes is not unambiguously related to the permanent boundedness. Indeed, a bounded distance between the zeros can be sustained by the unbounded sequences as well - through making the rate of developing of the infinite fluctuations appropriately large. However, for the unbounded sequences, the only factor that governs the distance between the zeroes is the rate of the fluctuations development. On the other hand, the boundedness straightforwardly introduces an additional factor - it "enforces" the fluctuations to make a "U-turn" at the thresholds of stability regardless to their current rate. So, the "U-turns" increase the frequency of the zeroes. In turn, the growth of that frequency results in an additional reduction of the long-range correlations. Note that the intensity of the long-range correlations is inversely proportional to the frequency of the zeroes.

To outline, the action of the boundedness on the autocorrelation function results in weakening of the long-range correlations. Note, however, that this weakening is not result of any physical process!

Hence, the additional reduction of the long-range correlations renders the autocorrelation function to decrease faster than $H\left(\frac{\eta}{T}\right)$. But how faster? In order to find out it we suggest that provided the limit $T\to\infty$ is taken, the shape of $G(\eta,T)$ reads:

$$G(x) = \sigma\left(1-(x)^{\nu(x)}\right) \tag{1.10}$$

where $x=\frac{\eta}{T}$ and $x\in[0,1]$. The suggested shape of $G(x)$ meets the requirement that the autocorrelation function is parameter-free. Our next task is the specification of $\nu(x)$. To begin with, let us get use of $H(x)$. It is good approximation for both $G(x)$ and $\frac{dG(x)}{dx}$ at $x\approx 0$. So, by means of $G(0)=H(0)$ and $\left.\frac{dG(x)}{dx}\right|_{x=0} = \left.\frac{dH(x)}{dx}\right|_{x=0}$ its contribution to the specification of $\nu(x)$ reads $\nu(x) = 1 \pm px^n$. Further, the selection of $p$ and $n$ is made on the grounds of the requirement for uniform contribution of all time scales. The latter demands non-transversal approach of $G(x)$ at $x$ going to $1$. Then $\frac{dG(x)}{dx}$ goes to zero when $x$ goes to $1$. So, the non-transversality sets $\nu(x) = 1-x^n$, but leaves $n$ arbitrary. The latter is determined by the circumstance that the boundedness is not associated with any specific physical process. Indeed, note that while $\frac{dG(x)}{dx}$ monotonically decreases on $x$ going to unity for every $n$, already $\left|\frac{d^2G(x)}{dx^2}\right|$ monotonically increases for every $n$ apart from $n=1$. Obviously, any increase of $\left|\frac{d^2G(x)}{dx^2}\right|$ interferes with our suggestion about the lack of any physical process associated with the course of the long-range correlations. So, our evaluations select $v(x)=1-x$ as the only exponent that fits all the requirements. Summarizing, the obtained exponent is the only one that ensures a non-transversal approach of $G(x)$ to zero so that all time scales perform uniformly and no special physical process is associated with the boundedness.

Note that the derivation of the shape of the autocorrelation function of a BIS does not require any information about its statistics. This makes the obtained property universal, i.e. insensitive to the nature of the BIS and to the particularities of the fluctuation succession. Thus, the time series that comes from systems as diverse as quasar pulsations, DNA sequences, financial time series etc. share the same autocorrelation function.

### 1.4 Power Spectrum

Another important characteristic of every irregular sequence is its power spectrum $S(f)$. Though it is the Fourier transform of the autocorrelation function $G(\eta,T)$, it cannot be evaluated by the straightforward application of the Fourier transformation. To elucidate the difficulties, let us have a look at the definition:

$$S(f) = \int_0^1 G(x)\exp(-ifx)dx = \int_0^1 \left(1 - x^{1-x}\right)\exp(-ifx)dx \qquad (1.11)$$

The usual trick to estimate such integrals is to rescale the variable $x$ to $x' = fx$. The purpose is to make the integral free from $f$ and to collect the dependence on $f$ into term that multiplies the integral. However, now this trick the does not work because the non-constant exponent $v(x)=1-x$ prevents the "elimination" of the $f$-dependence from the integral. Indeed:

$$S(f) = \frac{1}{f}\int_0^f \left(1 - \left(\frac{x'}{f}\right)^{1-\frac{x'}{f}}\right)\exp(-ix')dx' \qquad (1.12)$$

Intuitively, the problem seems only technical. However, it has a fundamental aspect as well because the Fourier transformation is an essentially non-linear operation. So it is likely to expect local "amplifications", "stretching" etc. Yet, the very definition of the operation makes these non-linearities to appear only when certain time scale has specific contribution. Let us just remember that the Fourier spectrum of every periodic function is discrete and its components are concentrated around its periods and their harmonics. Indeed, any long-range correlation between two time scales appears as single line whose amplitude is proportional to their correlation. Recalling that on the one hand, the uniform contribution of all time scales restrains persistence extent without signaling out any specific one and on the other hand, the boundedness limits the persistence because every deviation inevitably "turns back", the question becomes how the interplay of the long-range correlations introduced by the boundedness and the scaling invariance of the time scales shapes the power spectrum. Our first expectation is that the power spectrum must be a monotonic function so that not to signal out any specific component. Indeed, a closer look at (1.12) indicates that the power spectrum is a strictly decreasing power function. But is the exponent again non-constant? And if so, is their some property of that exponent that remains invariant under the Fourier transformation? Further, is the exponent again insensitive to the statistics of the BIS?

Eq(1.12) indicates that $S(f)$ is a decreasing power function with non-constant exponent:

$$S(f) \propto \frac{1}{f^{\alpha(f)}} \tag{1.13}$$

where the shape and the properties of $\alpha(f)$ are to be worked out. The accomplishment of this task is made on the basis of the permanent boundedness of $X(t)$ by the use of the Wiener-Khinchin theorem. It relates the power spectrum $S(f)$ and $y_T(f)$, i.e. the Fourier transform of $(X(t)-\langle x \rangle)$ over a window of length $T$. More precisely, the relation reads:

$$S(f) = \lim_{T \to \infty} \frac{1}{T} \langle |y_T(f)|^2 \rangle \tag{1.14a}$$

Actually, since the power spectrum is discrete for any finite $T$, (1.14) states that $\frac{1}{T}|y_T(f)|^2$ uniformly fits the shape of $S(f)$ as $T$ approaches infinity. So, $X_T(t)$ is majorized by:

$$I_T(t) = \sqrt{T} \int_{1/T}^{\infty} \frac{\cos ft}{f^{\alpha(f)/2}} df + \sqrt{T} \int_{1/T}^{\infty} \frac{\sin ft}{f^{\alpha(f)/2}} df \tag{1.14b}$$

where the Fourier coefficients are constructed on the basis of (1.13) and (1.14). In order for $I_T(t)$ to serve as an estimate of $X_T(t)$ it should be finite for every $t$ and $T$. For this purpose it is enough to find out at what values of $\alpha(f)$

$$I_T(0) = \max|I_T(t)| = \sqrt{T} \int_{1/T}^{\infty} \frac{1}{f^{\alpha(f)/2}} df \tag{1.15}$$

is finite and its value does not depend on $T$.

The integration of a power function of non-constant exponent is highly contraintuitive and involves a non-trivial step. I account for the details in the Appendix at the end of this chapter because their presentation is rather lengthy and stays a little apart from the mean stream of the section. Though, I urgently recommend its studying.

Further, I utilize the obtained result, namely:

$$\int_a^b \frac{df}{f^{\alpha(f)}} = \frac{b^{-\alpha(b)+1}}{1-\alpha(b)} - \frac{a^{-\alpha(a)+1}}{1-\alpha(a)} \tag{1.16}$$

Actually, the application of (1.16) to the *r.h.s.* of (1.15) selects that $\alpha(f)$ which renders $I_T(0)$ to be bounded in margins that are independent of $T$. The formal expression of this property is through:

$$I_T(0) = \lim_{T \to \infty} \sqrt{T} \int_{1/T}^{T} \frac{df}{f^{\alpha(f)/2}} = \lim_{T \to \infty} \sqrt{T} \left. \frac{f^{1-\frac{\alpha(f)}{2}}}{1-\frac{\alpha(f)}{2}} \right|_{1/T}^{T} =$$

$$= \lim_{T \to \infty} \frac{T^{\frac{3-\alpha(T)}{2}}}{1-\frac{\alpha(T)}{2}} - \lim_{T \to \infty} \left(\frac{1}{T}\right)^{\frac{1-\alpha\left(\frac{1}{T}\right)}{2}} \frac{1}{1-\frac{\alpha\left(\frac{1}{T}\right)}{2}} = T^0 \tag{1.17}$$

Evidently, the second line of (1.17) holds when $\alpha(f)$ is a continuous strictly increasing function between the following limits:

$$\alpha\left(\frac{1}{T}\right) = 1 \text{ and } \alpha(\infty) = p \tag{1.18}$$

where $p$ is arbitrary but $p \geq 3$.

But we can go further in the specification of $\alpha(f)$! The scaling invariance applied to the power spectrum implies that neither of its frequencies has specific contribution. This is another way to say that all time scales contribute uniformly in the power spectrum. Therefore, neither $\alpha(f)$ nor any of its derivatives should select any specific time point. Simple calculations show that it is possible if and only if $\alpha(f)$ is a linear function. Any non-linearity in $\alpha(f)$ makes its derivatives to change their sign at certain frequencies. However, the change of the sign makes the corresponding time scale specific and requires additional physical process that "makes" the sign to change.

Above we found out that both $v(x)$ and $\alpha(f)$ are strictly monotonic linear functions of their arguments. In both cases the strict monotonicity along with the linearity of the exponents comes out from the uniform contribution of all time scales. Therefore the linearity of the exponents is property invariant under the Fourier transformation. Indeed, non-trivial and unexpected!

Note that the derivation of the non-constant exponent $\alpha(f)$ does not involve any reference to the statistics and the length of the BIS. This allows concluding that the shape of the power spectrum of the BIS is also universal, i.e. it is insensitive to the details of the fluctuations succession and is invariant under the length of the time series.

So, the power spectrum of each BIS has the following properties:

(i) the spectrum has an infrared cut-off at the frequency $f_{min} = \dfrac{1}{T}$ where $T$ is the length of the time series;

(ii) $\alpha(f)$ is linear function so that $\alpha(f) = 1 + \kappa(f - f_{min})$ where $\kappa$ is a specific to the system parameter that is strongly related to the properties of the "U-turns".

Actually, $\kappa$ is proportional to the frequency of "U-turns". The major contribution to that frequency comes from the probability that the current fluctuation reaches the thresholds of stability. According to the Lindeberg theorem [1.1] every BIS of infinite length has finite expectation value and finite variance. As a result, the fluctuation statistics approaches normal distribution truncated by the thresholds of stability. Thus:

$$\kappa \propto \exp\left(-\dfrac{X_{thres}^2}{\sigma^2}\right) \qquad (1.19)$$

It seems that the persistent shape of the power spectrum give an easy way to determine the value of the thresholds of stability. The way goes through the extraction of $\kappa$ from the power spectrum. However, nobody has ever proceeded this way! The answer is that for a lot of practical cases $\kappa$ is extremely small. To illustrate the smallness of $\kappa$, let us have a look at the following table:

| $\dfrac{X_{thres}}{\sigma}$ | $\left(\dfrac{X_{thres}}{\sigma}\right)^2$ | $\kappa$ |
|---|---|---|
| 2,25 | 5 | 0.006 |
| 3 | 10 | $10^{-5}$ |
| 4 | 16 | $10^{-7}$ |
| 4.5 | 20 | $10^{-9}$ |
| 7 | 50 | $10^{-22}$ |

Evidently, the smallness of some $\kappa$ goes beyond the most elaborate ways for its extraction from the power spectra. Then what we have in hand from the power spectra that come from the experimental records. Do they support our considerations? Above we have established that the power spectrum of every BIS fits the same shape regardless to the details of the fluctuation succession and the length of the window. Does the physical world support our considerations?!

Indeed, the power spectra of all fluctuating systems persistently fit the same shape established to be $1/f$ in large frequency interval. The persistency of the shape even named the phenomenon - $1/f$ noise. This happens to be one of the most ubiquitous and widely studied phenomena in the world. Though it has been systematically studied for more that a century, two of its major properties remain mystery. The first one is the insensitivity of the power spectrum shape to the details of the fluctuations statistics. One of the most striking examples is that the power spectra of both traffic noise and a Beethoven symphony fit the same shape! The second one is the insensitivity of the shape to the length of the time series. It means that if the power spectrum of some time series of length $T$ fits $1/f$ shape, the power spectrum of the same time series but of different length $T_1$ fits the same shape. Moreover, $T_1$ can be made dozens of orders greater than $T$.

Evidently, these enigmas are successfully revealed by the considerations in the present section. However, the problem whether the exponent $\alpha$ is constant or function is crucial for the conjecture of boundedness! Indeed, let us come back to (1.17). Simple calculations show that if $\alpha$ is constant, neither its value can provide permanent boundedness. Indeed, an ultraviolet divergence emerges for $\alpha \leq 1$, while for $\alpha > 1$ there is an infrared one. So, only an exponent that changes with the frequency can provide permanent boundedness! I shall come to this topic again in Chapter 8 where I shall present experimental evidences that $\alpha$ is a linear function. Besides, if the fluctuations were unbounded, they would go beyond the thresholds of stability that would result in collapse of the system! On the other hand, that $1/f$ noise has been spanned over several dozens of orders. Hence, it is rather to be related to the stability of the system. This is an important point in our score! But the major one, the determination of $\kappa$ is still to come! Now I shall present an elegant way for its determination.

### 1.5 Variance. Setting of $\kappa$

A crucial test for all our considerations is the evaluation of the power spectrum amplitude. Indeed, we can make it into two ways - once through its definition:

$$\sigma = \lim_{T \to \infty} \frac{1}{T} \int_0^T |X(t)|^2 dt \qquad (1.20a)$$

and on the grounds of the spectral density (power spectrum):

$$\sigma_s = \lim_{T \to \infty} \frac{1}{T} c \int_{1/T}^{\infty} \frac{df}{f^{\alpha(f)}} \qquad (1.20b)$$

where $c$ is the amplitude of the power spectrum that is to be determined. The permanent boundedness of $X(t)$ automatically provides the existence of variance calculated by means of (1.20a). However, $\sigma_s$ is problematic since:

$$J = \lim_{T \to \infty} \frac{1}{T} \int_{1/T}^{\infty} \frac{df}{f^{\alpha(f)}} \qquad (1.21)$$

seems to have singularity in its lower limit, namely:

$$J = \lim_{T \to \infty} \frac{1}{T} \frac{\left(\frac{1}{T}\right)^{1-\alpha\left(\frac{1}{T}\right)}}{1 - \alpha\left(\frac{1}{T}\right)} = \lim_{T \to \infty} \frac{1}{T} \frac{T^0}{0} \qquad (1.22)$$

To reveal the singularity we should determine the rate by which $\alpha(f)$ goes to $1$. For this purpose let me recall that the power spectrum of every BIS of arbitrary but finite length is discrete. So, the distance between any of its two components cannot be made smaller than $\frac{1}{T}$. So, the rate by which $\alpha(f)$ approaches unity is proportional to $\frac{1}{T}$. Let us now evaluate $J_1$:

$$J_1 = \lim_{T \to \infty} \frac{1}{T} \int_{2/T}^{\infty} \frac{df}{f^{\alpha(f)}} \tag{1.23}$$

The purpose is twofold: on the one hand, when $T$ goes to infinity, $J_1$ converges to $J$. On the other hand, the "shift" of the lower limit in $J_1$ allows involving the rate by which $\alpha(f)$ goes to $1$. Indeed:

$$J_1 = \lim_{T \to \infty} \frac{1}{T} \frac{\left(\frac{1}{T}\right)^{1-\alpha\left(\frac{2}{T}\right)}}{1-\alpha\left(\frac{2}{T}\right)} = \lim_{T \to \infty} \frac{1}{T} \frac{1}{\kappa \frac{1}{T}} \left(\frac{1}{T}\right)^{-\frac{k}{T}} = \lim_{T \to \infty} \frac{1}{\kappa} \left(\frac{1}{T}\right)^{-\frac{\kappa}{T}} \tag{1.24}$$

How annoying! The removal of one singularity brings about another one!? What the behavior of $x^{-x}$ is at $x$ going to zero? To find out, we use the exponential presentation, namely:

$$x^{\pm x} = \exp(\pm x \ln x) \tag{1.25}$$

Further, since $x$ goes faster to zero than $\ln x$ diverges:

$$\lim_{x \to 0} x^{\pm x} = \exp(\pm \lim_{x \to 0} x \ln x) = 1 \tag{1.26}$$

Finally, the variance calculated on the grounds of the spectral density reads:

$$\sigma_s = \frac{c}{\kappa} \tag{1.27}$$

Since the variance can be determined independently by means of (1.20a), (1.27) sets the value of $c$, i.e. the amplitude of the power spectrum. As a result we have established all the characteristics of the power spectrum and its final specification reads:

$$S(f) = \sigma\kappa \frac{1}{f^{\alpha(f)}} \tag{1.28}$$

where $\alpha(f) = 1 + \kappa\left(f - \frac{1}{T}\right)$ and $f \geq \frac{1}{T}$.

Now we can utilize the last property, namely the existence of infrared cutoff of the power spectrum. A closer look at (1.28) shows that the amplitude of the cutoff component is $A_c = \sigma\kappa T$. This gives any easy way to extract $\kappa$: it is proportional to the slope of the plot of $A_c$ on the length of the time series $T$. The linearity of that plot makes the task particularly easy.

The proportionality of the power spectrum amplitude on $\kappa$ reconciles another mystery of the $1/f$ noise. It has been established that $1/f$ spectrum is well discerned only for extremely long time series! A brief look at the amplitude of the cutoff component $A_c \approx \kappa T$ tells that it becomes of the order of unity at $T \propto \frac{1}{\kappa}$. And the extreme smallness of $\kappa$ explains why the mysterious and ubiquitous $1/f$ noise emerges only if one is patient enough to record very long time series! And this is another point in our score! To make the things more clear let us suppose that the amplitude of the power spectrum does not comprise $\kappa$ and is of the order of unity. Then $A_c \propto T$. Further, suppose that the amplitude of a specific for the system process is also unity. Consider now a record that comprises $10^3$ points. Keeping in mind that the variance is unity, the amplitude of the noise band would be thousand times greater than that of any specific to the system process!? And how about a record of billion points? Thus, thanks to the extreme smallness of $\kappa$, the "noise" band in the power spectra can be separated from the contributions that comes from the specific for a system processes. However, the noise band hides extremely valuable specific information - that about the thresholds of stability. The practical importance of that information is enormous: there is no need to wait for a "bump" into the thresholds of stability to find out whether a collapse is to be expected.

## 1.6. Dynamical Boundedness. Embedding dimension

So far we have established that the power spectrum and the autocorrelation function of the BIS exhibit remarkable universality - their shape and characteristics are free from any specification of the fluctuation dynamics. Though this ubiquitous insensitivity has been rigorously proven, our intuition remains unsatisfied and I am strongly tempted to suggest that the power spectrum and the autocorrelation function are too "coarsen" characteristics to take into account the particularities of the fluctuations dynamics. Indeed, the dynamics of the fluctuations comes out from the specific for every system processes associated with their development. Yet, now I shall show that the fluctuation dynamics is also subject to certain universality.

The usual way to reveal the dynamics of the correlations in a stochastic sequence is to study the properties of its phase space. The procedure involves dividing of the phase space volume into small size cells and counting the points at which the trajectory intersects each of them. It is certain that this elaborate operation helps much in the examination of the correlation dynamics. Alongside, it is to be expected that the obtained information is highly specific. However, the general aim of this section is to study the universal properties of the BIS and their relation to the chaoticity. Our major goal is the systematic derivation of all chaotic properties from the boundedness. That is why now I focus our attention on the question how the boundedness is incorporated in the dynamics of the correlations.

I start with substantiating the general constraint that the boundedness imposes on the fluctuation dynamics. The notion about boundedness is associated with the conjecture that the each object/subject comprises finite amount of energy and matter. The boundedness of the fluctuation size is one of its manifestations. However, it is not enough to keep a system stable. Likewise it is necessary that the amount of energy/matter exchanged with the environment is also permanently limited. This requirement imposes permanent boundedness of the rate of the fluctuations development. Further I call the boundedness of the fluctuation rate dynamical boundedness while the boundedness of the fluctuation size is named static boundedness.

Our expectation is that the interplay of the dynamical and static boundedness renders every BIS confined in a finite phase space volume of specific dimension. It is easy to understand how the limitation over the volume of the attractor appears. Indeed, since every BIS is comprised by bounded variations around the expectation value, the attractor is a compact set that has a fixed point matched by the expectation value. The upper bound in each direction is set on the thresholds of stability. So, the static boundedness indeed provides finite volume of the attractor but it is not enough to prompt its dimension.

The specification of the phase space dimension looks hard and confusing. The usual practice is to make guess on the grounds of the temporal behavior of a single or a few variables. But, does the guess involve specific arguments in each particular case or there is a general rule?! My next task is to show how the interplay of the static and dynamical boundedness provides finite dimension of the attractor whose value is specific to the dynamics of the BIS but always finite.

To begin with, let us mention that the dynamical boundedness makes every fluctuation extended in the time course. More precisely, it means that every fluctuation is approximated by a trajectory that starts at the expectation value at a given moment, pass trough maximum and turns back to it for the first time after certain time interval called hereafter duration. Further, the duration of that interval is related to the size of the corresponding fluctuation so that the rate of development is permanently bounded. Hence, both the duration and the size of all fluctuations are finite and limited by the size and the duration of the largest one.

Now I am ready to start the study of the phase space properties. My first task is to show that its dimension is always finite. Given a one-dimensional rescaled BIS $X(t) = \frac{\tilde{X}(t) - \langle \tilde{X} \rangle}{\sqrt{\sigma}}$ where $\tilde{X}(t)$ is the original BIS, $\langle \tilde{X} \rangle$ is its expectation value and $\sigma$ is the variance. By means of this operation the only specific to the BIS parameter is its threshold of stability. Let us now apply the widely used in the time series analysis procedure of time delay embedding. The purpose is to make the comparison to the well established results straightforward. The time-delay embedding implies that the vectors $\vec{R}_n(t) = (X(t), \ldots X(t - i\tau), \ldots X(t - n\tau))$ are embedded in a $n$ dimensional Euclidean space. $X(t)$ are the successive points that come from a given time series; $\tau$ is a small delay. Further, the phase space is divided into small size cells and the vectors whose ends are inside each cell are counted. The population in lg-lg scale is plotted against the cell size. Applied to a BIS, the time delay embedding is a noting more than a particular way of coarse-graining. So, the ends of the vectors $\vec{R}_n(t)$ also construct a BIS. Then I am able to associate the value $\overline{X}(t - i\tau)$ of the coarse-grained BIS with the $i-th$ axis, where $i \in [1, n]$.

Note that the association of the delay $i\tau$ with the topological dimension is equivalent to the parameterization of the size of the fluctuation through the phase space angle. But the size of a fluctuation is already parameterized trough its relation with its duration. The evident parity of both parameterizations selects a topological dimension so that a fluctuation forms a closed continuous curve by a single revolt. The topological dimension that renders the largest fluctuation to appear as a loop is called embedding dimension. The hallmark of this dimension is that all the smaller fluctuations are closed continuous loops so that each loop has its own embedding dimension smaller than the embedding dimension of the entire attractor. And the boundedness of both the size and duration makes the embedding dimension always finite.

To make the above considerations more evident let us point out that when the topological dimension $n$ is smaller than the embedding dimension, the larger fluctuations appear as complicate helices not as closed curves. Depending on the ratio between the fluctuation and the topological revolt angle, the helix is first unwinding then turns to winding. The unwinding happens whenever the current revolts are not enough for the fluctuation to reach its full size. The winding is associated with the development of the fluctuation from its full size to zero.

But how to illustrate definitively the established structure of the attractor?! Can we exhibit its property to stretch and fold the distance between successive points of the BIS? Indeed, when the helix unwinds, the distance stretches while its winding results in folding. That is why it is to be expected that the stretching and folding strongly affects the density of the attractor. But how? I start the task by establishing of the factors that govern the "stretching and folding" mechanism. The first one is the topological dimension $n$. It selects a size of the fluctuation $l_n$ so that its embedding dimension coincides with $n$. So, the fluctuation whose size is $l_n$ forms a loop. Thus $l_n$ acts as the demarcation where stretching turns to folding. Put it in geometrical terms, the topological dimension drives the "angle" of the helix. The factor that drives the step of the helix is the finite rate of the fluctuations development. Indeed, though the helicoidal behavior is purely geometrical effect, it should integrate the limited rate of the fluctuations development. The incorporation of the limited rate into the helicoidal behavior is through the following non-linear relation of power type between the angle $u$ and the step $l$:

$$l = l_n u^{\gamma_n(u)} \tag{1.29}$$

where $u$ is rescaled so that $u \in [0,1]$, $\gamma_n(u)$ are chosen so that to fit two conditions, the first of which is that $l = l_n$ at $u = 1$. Actually, this condition means that the helix turns to loop of size $l_n$ by a single revolt. The second one is that it should provide specific for the system dynamical rate. Further, it should provide the same dynamical rate at each topological dimension. Simple geometrical considerations show that it is achieved by setting $\gamma(u)$ dependent on the topological dimension $n$. It is worth mentioning also that the meeting of the above conditions makes $\gamma(u)$ to increase on the increase of $n$. Note that the power type relation between the step and the angle of the helix is parameter-free! In other words, it does not select any size that has specific properties. Thus, neither scale has specific contribution to the stretching and folding mechanism!

Now we come to the matter about the density of the points that are subject to stretching and folding. Evidently, the major factor that sets their participation is the correlation between the points. I have already found out that the average probability that two points separated by a time interval $\eta$ have the same value is the autocorrelation function $G(x)$. Since both points have the same value, they can be considered as belonging to the same effective fluctuation whose duration is $x = \dfrac{\eta}{T}$. The size of the fluctuation $\tilde{l}$ is limited by $l_n$ and the relation between $x$ and $\tilde{l}$ is set on the dynamical boundedness. Further, the requirement that both the size and the rate of development of a fluctuation remain bounded without involving any specific process (scale) renders the power type relation between $\tilde{l}$ and $x$:

$$\tilde{l} = l_n x^{\tilde{\gamma}(x)} \qquad (1.30)$$

The apparent similarity between (1.29) and (1.30) is actually equivalence! Indeed, since both $x$ and $u$ are confined in the same range $[0,1]$, the same dynamical rate imposes $\gamma(u) = \tilde{\gamma}(x)$ for every $x = u$. The equivalence of (1.29) and (1.30) proves explicitly our heuristic arguments about the parity between the parameterization of a fluctuation through its duration and through the phase space angle.

Therefore the probability that two points separated by distance $l$ are subject to stretching and folding is $G(u)$. On the other hand, $(1 - G(u))$ is the probability that the distance remains $l$. Therefore, the probability that the distance between a pair of point is less than $l$ reads:

$$C(x) = 1 - \int_o^u G(x)dx = 1 - P(u) \qquad (1.31)$$

where

$$P(u) = \int_0^u G(x)dx = \int_0^u \left(1 - x^{1-x}\right)dx = x - \frac{x^{2-x}}{2-x} \qquad (1.32)$$

The plot of the function $P(u)$ given in Fig.1.1 reveals peculiar behavior: it goes through maximum! Does it mean selection of a specific scale?! To uncover the mystery, let us start step by step: the increasing part of the plot is obvious - on the increase of $u$ the correlation $P(u)$ increases until current size of the unwinding helix reaches its "loop" size $l_n$. Further the helix turns to winding. However, the winding shrinks the distance and so brings about an effective decrease of the correlations that participate stretching. Thus, the plot of $P(u)$ strongly supports the idea about the boundedness: the stretching is limited by certain size where it turns to folding. Note, that while expressed through $u$, the boundary between the stretching and folding is universal and independent of the topological dimension , its size $l_n$ is specific to the attractor and depends on the current topological dimension $n$ as well.

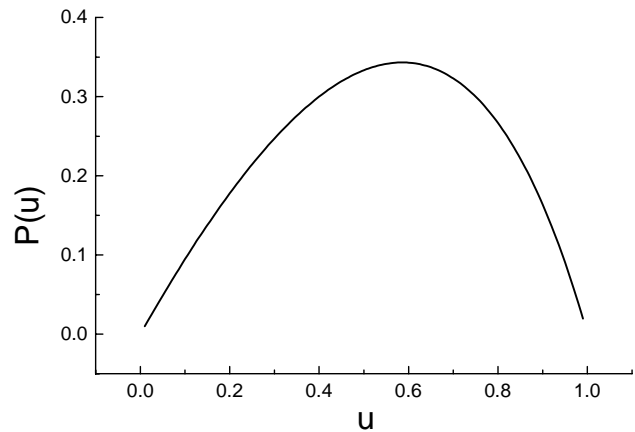

Fig.1.1 The probability $P(u)$ for a pair to participate the stretching and folding

It is to be expected that the fine structure of the attractor is highly non-trivial and very different from the coarse-grained one. This statement is very well illustrated by the population structure. Indeed, let us divide the phase space into cells of size $l$ and count all points inside each of them. The measure of the corresponding number is the plot of $C(l)$ vs. the size $l$ where $l$ is related to $u$ trough (1.29). To reveal better the difference between fine and coarse-grained partitioning, the plot is presented in lg-lg scale. Note that since $C(l) \equiv C(u)$ where $C(u)$ is set on (1.31), the plot is actually parameterized by $u$ trough the relation (1.29).

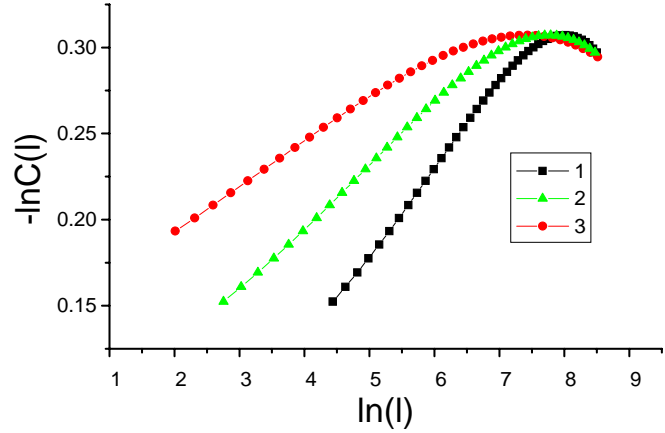

Fig.1.2. The plot of the phase space population on the degree of coarse-graining.
*curve* 1: $l = 10.3u^{3.2}$; *curve* 2: $l = 11u^{4.5}$; *curve* 3: $l = 12u^{6.6}$.

The values of $l_n$ and $\gamma_n$ in Fig.1.2 are set so that to correspond to different topological dimensions. Though we do not specify the actual topological dimension $n$, it is easy to find out that both $l_n$ and $\gamma_n$ grow on $n$ increasing. Note that the growth is not arbitrary but $l_n$ and $\gamma_n$ are related so that to leave the dynamical rate intact. The plot exhibits the following characteristic properties of the population behavior:
(I) at every topological dimension the plot manifests nearly linear part. Accordingly the population behaves as $l^{-D_n}$ where $D_n$ decreases on $n$ increasing.
(ii) at every topological dimension the population reaches saturation;
(iii) the saturation value is insensitive to the value of the topological dimension.
The linear part of the plot is apparently associated with the prevailing role of the stretching. Further, since the folding shrinks the distance between the points, it effectively weakens the correlations necessary for stretching. In other words it acts towards stochastisation of the points. Thus the population tends to remain constant and insensitive to any further growth of the cell size. Furthermore, remember that $C(u)$ is independent of the topological dimension. This immediately renders the saturation value insensitive to the topological dimension as well. However, the plot of $C(u)$ vs. cell size $l$ is good approximation for the population only up to the saturation. It should be remembered that $C(l)$ is the measure for the correlations that give rise to the stretching. However, since the folding contributes to the stochastization, $C(l)$ ceases to be the measure of the population for the cell sizes larger than the cell size at which the saturation emerges.

Let us now estimate the value of the cell size at which the saturation occurs. Let me recall that $C(l) \equiv C(u)$. An immediate look at $C(u)$ from (1.31) shows that the saturation is achieved at $u$ very close to $0.6$. Simple calculations show that $r(u) = \dfrac{l(u = 0.6)}{l_n} = 0.03$ on the curve $3$. So, the population changes its behavior at cell size that is only 3% of $l_n$! Hence the fine structure of the attractor becomes evident only at very fine partition of the phase space. Yet, note that the value of $r(u)$ strongly depends on the rate of development of the fluctuations.

So far so good. But do the time series that come for the real events match such behavior? Yes, they are. There is plenty of time series coming from a broad spectrum of systems whose population is studied by means of the time-delay embedding and exhibits exactly the same behavior as that depicted in Fig.1.2. Moreover, this behavior is the one recently taken as decisive for the chaos!

Though the chaos has been intensively studied, a rigorous and non-ambiguous definition is still missing. My ambition is to illustrate that alone the static and dynamical boundedness give rise to the major chaotic properties established so far. The greatest advantage is that the physical conjecture about the boundedness is available for a wide variety of systems. This corresponds to the enormous ubiquity of the chaotic behavior. It is typical for phenomena of very different nature including engineered, natural and social ones. So, if there is a general idea behind the chaos it must give rise to all its aspects established so far. Our first task in this direction is to prove that the boundedness alone indeed brings about its major characteristics: autocorrelation function, power spectrum and embedding dimension from the viewpoint of the boundedness.

## 1.7 Chaos and Boundedness

Historically the deterministic chaos has been observed for the first time by Lorentz at numerical simulation of a simple system of 3 deterministic non-linear ordinary differential equations (ODE) that fits the Lipschitz conditions. He has established that at certain parameters its solution behaves in highly non-trivial way: it widely varies and exhibits strong sensitivity to initial conditions. This sensitivity implies that every infinitesimal deviation from the initial conditions diverges to finite size in finite time interval. Some authors choose this property as definitive for the chaos. Still, it lacks enough universality since it is straightforwardly related only to the deterministic dynamical systems. Moreover, the sensitivity towards the initial conditions suffers a great disadvantage - though nowadays it has clear explanation, the irregularities of the solution remain mystery. The greatest ambition of my book is to explicate that the chaotic properties appear as a result of the cooperation between the coherence and the boundedness. Since the sensitivity towards the initial conditions and the irregularities of the solution are characteristics of the temporal evolution, their relation to the matter is postponed to the Chapter 3 where the integration of the coherence and the boundedness in the evolution of the many-body systems is studied. Further in this section I shall consider only those aspects of the chaotic behavior that are not readily related to their mathematical origin.

Let me start with the idea of the homoclinic orbits. It implies that the phase space is a dense set of orbits that has a fixed point and whose period is infinite. In turn, the homoclinic orbits give rise to a continuous power spectrum. Sounds familiar? Indeed, let us remember that the phase space of a BIS is a set of "loops" each of which is fixed to the origin. Remember that since the fixed point is the expectation value of the BIS, each "loop" is fixed to it. Further, the size of the loops continuously fills the range from zero to the thresholds of stability. The period of the loops is infinite - though each loop is repeated in finite time intervals, its *period* is infinite. Let me recall that the notion of a period implies a regular repetition in a given period while the repetitions of the loops succeed in finite time intervals but following in an irregular manner. Thus, the association of the loops with the homoclinic orbits is straightforward. Moreover, the power spectrum of a BIS is also continuous band. Note, that the inifinity of the "loops" periods substantiates the property of the power spectrum to be a continuous band! Yet, the infinity of all loop periods cannot set the monotonic decrease of the power spectrum. But the boundedness can! It specifies not only the monotonic decrease but settles the precise shape as well!

Another very important aspect of the chaoticity was considered in the previous section. It is believed that the phase space of each chaotic system has specific but finite dimension. Further, this dimension can be extracted from the specific behavior of the phase space population depicted in Fig.1.2. The missing point, however, is the interrelations among those three aspects of chaoticity. The lack of such interrelations gives rise to the following puzzle – one the one hand, the infinite periods of the homoclinic orbits render an infinite density of the attractor. Indeed, the folding of an arbitrary long time series in the finite volume of its attractor makes the density infinite. On the other hand, the behavior of the population depicted in Fig.1.2 implies inifinite repetitions of the trajectory so that the number of the discernible points remains permanently finite and insensitive to the length of the time series. The major advantage of the idea about the static and dynamical boundedness is that it manages not only to unify the idea of the homoclinic orbits, continuous power spectrum and the embedding dimension but to reveal the above discrepancy. Let me first recall that though the "loops" have infinite periods, they repeat in finite but irregular time intervals. Further, since the distances between the zeroes of a BIS are finite and independent of the BIS length, the frequency of the repetitions is insensitive to the length of the BIS. Evidently, this consideration is enough to conclude that the population of the discernible points remains finite and insensitive to the length of the BIS.

Another very important and decisive characteristic of the chaotic behavior, not yet considered by me, is the so called Kolmogorov entropy or K-entropy for short. By definition it reads:

$$K = -\lim_{\tau \to 0} \lim_{l \to 0} \lim_{N \to 0} \frac{1}{N\tau} \sum_{i_0 \ldots i_N} P_{i_0 \ldots i_n} \ln P_{i_o \ldots i_N} \qquad (1.33)$$

where $P_{i_0 \ldots i_N}$ is the joint probability that the $\vec{R}(t=0)$ is in the $i_0$ cell, $\vec{R}(t=\tau)$ in the $i_1$ cell,..., $\vec{R}(t+n\tau)$ - in the $i_N$ cell. The sum is taken over all possible partitionings of the phase space. This makes $K$ measure of the average information necessary for precise setting of the motion in the phase space. The value of K-entropy is definitive for the chaos since it is zero for the deterministic motion infinite for the stochastic one and finite for the deterministic chaos. So, what is its value for the BIS?! Let us consider the function:

$$K(u) = -(1 - P(u))\ln(1 - P(u)) \qquad (1.34)$$

Since $(1 - P(u))$ is the probability that any two points separated by time interval $u$ are not stretched, $K(u)$ is the information necessary to fix that distance between the initial and the finial point to $u$. Note that this probability does not depend on the details of the trajectory. Since it is impossible to predict better the motion on the attractor, $K(u)$ covers the meaning of K-entropy. But is it finite?! Evidently, it is. To make it apparent, let us have a look on its plot demonstrated in Fig.1.3.

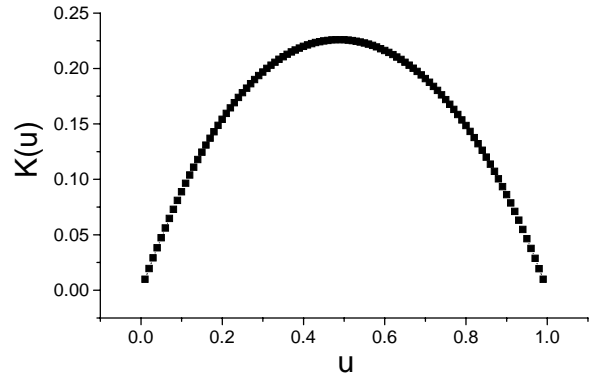

Fig.1.3. K-entropy for a BIS

$K(u)$ has two very important properties. The first one is that its behavior is straightforwardly related to the stretching and folding. The explication of that relation becomes evident by setting $K(u)$ through $(1 - P(u))$, where $P(u)$ is the probability for participation to the stretching and folding. Then the plot of $K(u)$ vs. $u$ has the following meaning: the increase of the probability for stretching at small $u$ requires more information for setting the distance between the initial and the final point of any trajectory. Further, the stretching reaches its maximum and turns to folding which, however, shrinks the distance between the initial and the final point of the trajectory. Thus the folding contributes to better precision of the position of the motion. In turn, better precision needs less information for specifying the characteristics of the trajectory. As a result, the K-entropy turns to decrease.

The second property is that we do *not* relate $K(u)$ to the so called Lyapunov exponent. Why? Let us first explain what Lyapunov exponent is. It is the measure of the stretching and folding presented in an exponential form. However, it should be stressed that this presentation is not parameter-free while the stretching and folding associated with the dynamical and static boundedness is presented by the parameter-free power form (1.29) and (1.30). Let us outline that the power type form is the only possible one that meets the interplay of the requirements about the lack of a specific time scale and the dynamical boundedness. On the contrary, though the exponential stretching does not select specific time scale, it is not able to meet the dynamical boundedness. So, the correct definition of the Laypunov exponent must take into account the dynamical boundedness. I postpone the consideration until &3.5 since then I will be able to associate explicitly the Lyapunov exponent with the automatic execution of the U-turns.

### 1.8 What Comes Next

So far I have established that the static and the dynamical boundedness indeed make every BIS to manifest the major chaotic properties. Moreover, the greatest advantage of the idea of the boundedness is that it manages to unify self-consistently all those properties. However, some very important topics still remain assumptions. To begin with, this is the matter about the uniform contribution of all time scales. It is plausible if and only if the U-turns happen without involving any specific physical process and without destabilization of the system. The obvious relation between stretching and folding mechanism and the stability of the system prompts the way to solve the problem - to associate the issue about the U-turns with the stretching and folding. But is it sufficient? We will see that the successful explication needs some additional information about the mechanism that drives the fluctuations. Further in the line stands the assumption about the irregularity of the fluctuations succession, i.e. what drives the fluctuations to appear irregularly. How this matter is related to the boundeness and the coherence? When the fluctuations persist infinitely long time and when they are damped? How to incorporate the boundedness and the coherence in the mathematical description of the fluctuations? These are some of the major topics that are milestones in the entire concept of the book. Since my further attention is focused on the many-body systems, several controversial topics of the equilibrium and non-equilibrium statistical mechanics are revisited in the next chapter. The purpose is to present certain aspects from the viewpoint of our concept.

### Appendix

The task of this appendix is to find out how to integrate a power function with a non-constant exponent. So, the next item is the calculation of:

$$J(a,b) = \int_a^b x^{\pm \nu(x)} dx . \tag{A.1}$$

where $\nu(x)$ is a continuous function with finite everywhere derivatives. First I consider the case when $\nu(x) \neq 1$ for any $x \in [a,b]$. The interval $[a,b]$ is divided into subintervals of equal length $\varepsilon$ and the value of $\nu(x)$ is set constant equal to its value at the lower limit of any subinterval. Then:

$$J(a,b) = \frac{b^{1 \pm \nu(b)}}{1 \pm \nu(b)} - \frac{a^{1 \pm \nu(a)}}{1 \pm \nu(a)} + \lim_{\varepsilon \to 0} \sum_{k=1}^{[(b-a)/\varepsilon]} R_k , \tag{A.2}$$

where:

$$R_k = (a+k\varepsilon)^{\pm \nu(a+(k-1)\varepsilon)+1} - (a+k\varepsilon)^{\pm \nu(a+k\varepsilon)+1} \approx$$
$$\approx (a+k\varepsilon)^{\pm \nu(a+k\varepsilon)+1}\left((a+k\varepsilon)^{-\delta_k \varepsilon} - 1\right),$$

where:
$$\delta_k = \nu'(a+k\varepsilon).$$

Since it is supposed that $\nu(x)$ has finite everywhere derivatives, $\delta_k$ is finite for every $k$. In turn, this provides that $\left((a+k\varepsilon)^{-\delta_k \varepsilon} - 1\right) \to 0$ at $\varepsilon \to 0$. So each term $R_k = 0$ at $\varepsilon = 0$. Thus:

$$J(a,b) = \frac{b^{1\pm \nu(b)}}{1 \pm \nu(b)} - \frac{a^{1\pm \nu(a)}}{1 \pm \nu(a)}. \quad (A.3)$$

Let us now consider the case when $\nu(x)$ is a continuous function with finite everywhere derivatives such that it crosses 1 at the point $x=c$ and $c \in [a,b]$. Then:

$$J(a,b) = \lim_{\varepsilon \to 0}\left[\int_a^{c-\varepsilon} x^{\pm \nu(x)}dx + \int_{c+\varepsilon}^b x^{\pm \nu(x)}dx\right] =$$
$$= \frac{b^{1\pm \nu(b)}}{1 \pm \nu(b)} - \frac{a^{1\pm \nu(a)}}{1 \pm \nu(a)} + \lim_{\varepsilon \to 0}\left((c-\varepsilon)^{\pm \nu(c-\varepsilon)+1} - (c+\varepsilon)^{\pm \nu(c+\varepsilon)+1}\right). \quad (A.4)$$

Taking into account that:
$$|\nu(c-\varepsilon)| = 1 - \nu'(c)\varepsilon$$
$$|\nu(c+\varepsilon)| = 1 + \nu'(c)\varepsilon,$$
$$\lim_{\varepsilon \to 0}\left((c-\varepsilon)^{\pm \nu(c-\varepsilon)+1} - (c+\varepsilon)^{\pm \nu(c+\varepsilon)+1}\right) \approx \lim_{\varepsilon \to 0}\left(c^{-\nu'(c)\varepsilon} - c^{\nu'(c)\varepsilon}\right) \to c^0 - c^0 = 0.$$

Therefore again:
$$J(a,b) = \int_a^b x^{\pm \nu(x)}dx = \frac{b^{1\pm \nu(b)}}{1 \pm \nu(b)} - \frac{a^{1\pm \nu(a)}}{1 \pm \nu(a)}. \quad (A.5)$$

However, the result seems controversial because the derivative of $J(a,b)$ deviates from the *l.h.s.* of (A.5). Next a proof that this argument is alias and (A.5) holds whenever $\nu(x)$ is a continuous function with finite everywhere derivatives is presented.

To simplify the further calculations let us present (A.5) in a slightly modified form, namely:

$$J(a,b) = \int_a^b (\pm \nu(x))x^{\pm \nu(x)-1}dx = b^{\pm \nu(b)} - a^{\pm \nu(a)} \quad (A.6)$$

Indeed, the application of the formal differentiating rules to the function $x^{\pm \nu(x)}$ in its exponential presentation:

$$x^{\pm \nu(x)} = \exp(\pm \nu(x)\ln(x)) \quad (A.7)$$

yields:

$$\frac{dx^{\pm \nu(x)}}{dx} = \pm \nu(x)x^{\pm \nu(x)-1} + \nu'(x)\ln(x)x^{\pm \nu(x)} \quad (A.8)$$

Obviously (A.8) deviates from the *l.h.s.* of (A.6) by the factor $\nu'(x)\ln(x)x^{\pm \nu(x)}$.

Let us now apply the rigorous definition of a derivative. In our case it reads:

$$\frac{dx^{\pm \nu(x)}}{dx} = \lim_{\varepsilon \to 0}\frac{(x+\varepsilon)^{\pm \nu(x+\varepsilon)} - x^{\pm \nu(x)}}{\varepsilon} =$$
$$\lim_{\varepsilon \to 0}\frac{\exp(\pm \nu(x+\varepsilon)\ln(x+\varepsilon)) - \exp(\pm \nu(x)\ln(x))}{\varepsilon} \quad (A.9)$$

For each $x$ fixed and non-zero one can define the small parameter $\dfrac{\varepsilon}{x}$. Hence, the Taylor expansion of $v(x+\varepsilon)\ln(x+\varepsilon)$ to the first order reads:

$$v(x+\varepsilon)\ln(x+\varepsilon) = (v(x)+\varepsilon v'(x))\left[\ln\left(1+\dfrac{\varepsilon}{x}\right)+\ln(x)\right] =$$
$$= (v(x)+\varepsilon v'(x))\left(\dfrac{\varepsilon}{x}+\ln x\right) = v(x)\ln(x) + v(x)\dfrac{\varepsilon}{x} + \varepsilon v'(x)\ln(x) \tag{A.10}$$

Then:

$$\exp(\pm v(x+\varepsilon)\ln(x+\varepsilon)) = x^{\pm v(x)} \exp\left(\pm v(x)\dfrac{\varepsilon}{x}\right)\exp(\varepsilon v'(x)\ln(x)) \tag{A.11}$$

The expansion of $\exp\left(\pm v(x)\dfrac{\varepsilon}{x}\right)$ to the first order of $\dfrac{\varepsilon}{x}$ is always available since by the limit $\varepsilon \to 0$ in (A.9) is taken at fixed value of $x$. It yields $\left(1 \pm v(x)\dfrac{\varepsilon}{x}\right)$. However, the expansion of the term $\exp(\varepsilon v'(x)\ln(x))$ into series is incorrect because of the divergence of $\ln(x)$ at small $x$. Furthermore, $x$ can always be made arbitrarily small by appropriate choice of the unites. That is why the meeting of the scaling invariance requires taking the limit $\varepsilon \to 0$ at fixed but still finite $x$. In result this makes the term $\exp(\varepsilon v'(x)\ln(x))$ equal to 1 at each value of $x$. So, (A.11) becomes:

$$\exp(\pm v(x+\varepsilon)\ln(x+\varepsilon)) = x^{\pm v(x)} \exp\left(\pm v(x)\dfrac{\varepsilon}{x}\right) \tag{A.12}$$

Note that this operation provides scaling-invariant form of the small parameter $\dfrac{\varepsilon}{x}$ in (A.11). Note also that the value of $\dfrac{\varepsilon}{x}$ does not depend on the choice of units! So it turns out that the derivation preserves the scaling-invariance!

Thus, (A.8) yields:

$$\dfrac{dx^{\pm v(x)}}{dx} = \pm v(x) x^{\pm v(x)-1} \tag{A.13}$$

So, (A.13) matches our result (A.6).

# Chapter 2: Statistical Mechanics Revisited

### 2.1. Why to Read Chapter 2

Nowadays the statistical mechanics is viewed as a bridge between the microscopic dynamics and the evolution of the many-body systems. It is widely accepted that the local dynamics does not contribute straightforwardly to the macroscopic behavior. It is supposed that because of its complexity, the evolution is governed by parameters that have no dynamical analog - entropy, concentration etc. Their introduction requires probabilistic manner of describing.

Now I shall view the problem from different prospective. The statistical mechanics is supposed to explain the following 3 phenomena:

(I) the dissolution of the salt into water (soup). Our everyday experience tells that regardless to the way one puts the salt into the soup, it is always homogeneously dissolved. Moreover, neither the ingredients nor the amount of the soup affects the homogeneity. Even eating does not have an effect on it - every other sip gives the same sense of saltiness. Why the saltiness never exhibits fluctuations?!

(ii) the dissolution of pollens into water (Brownian motion). It is well known that each pollen moves randomly in all directions. It is suggested that this motion is provoked by the collision with the water molecules. However, since a pollen is immensely larger than a water molecule, every its displacement requires coherent "efforts" of the water molecules. What makes them to exhibit cooperative behavior? Moreover, the coherence is temporary since each pollen executes random walk independently of the random walks of the other pollens. So the "coherence" of the solvent appears as formation of well-separated from one another finite-size fluctuations that have finite lifetime. This gives rise to the question what makes the water to behave differently under the replacement of salt by pollens?

(iii) the third phenomenon is the antithermodynamic behavior of certain type open systems. Now I shall convince you that this claim is not absurd. I start with recalling the zeroth law of the thermodynamics that asserts: equilibrium always exists. The scientific community has taken for granted the validity of this statement and never try to escape it. However, now I pose the question about the exceptions from the proclaimed ubiquity of that assertion. Consider now open systems, e.g. a chemical reaction exposed to steady external constraints (control parameters). Usually, the reaction products are immediately removed from the system. In turn, this keeps the system permanently out of equilibrium. To remind that chemical equilibrium requires detailed balance, i.e. each step to be in equilibrium with its reverse one. However, the removal of the reaction products makes certain steps irreversible. Thus, in general, the system reaches steady state that, however, is not equilibrium. Moreover, one of the major results of I. Prigogine is that out-of-equilibrium there is no Lyaponov functional that guarantees automatic damping of the fluctuations. Besides, it is well established that the open systems evolves non-monotonically on varying the control parameters: they can be in a steady state, to manifest self-sustained oscillations or exhibit wild irregularities in their temporal behavior. This results in opening the door for transitions between states of different entropy that are executed by means of the fluctuations. For example, let a fluctuation size covers the "distance" between a steady state and the state of self-sustained oscillations. Obviously, both states have different entropy. As a result, considering the fluctuation as a trajectory that starts and ends at the same state, the deviation in the direction from the state of larger entropy to the state of the less one is antithermodynamic.

I select the above 3 phenomena because they have much in common: they consist of huge amount of entities and their macroscopic states are characterized by such parameters as temperature, pressure, volume, concentration, etc, i.e. the parameters that have no dynamical analog but are natural for the thermodynamics. Another common property is that their dynamics involves only short-range interactions. That is why I am tempted by the prospective about unified description of all these phenomena. But is it possible? Here we encounter one of the greatest paradoxes of the thermodynamics: on the one hand, the thermodynamics asserts that once a system reaches equilibrium, it stays there forever. However, another fundamental assumption of the thermodynamics that is: the dynamics of the interactions is supposed invariant under the reverse of time, implies that the dynamics is not a constraint for exerting significant deviations from the equilibrium. Besides, this is not only a serious conceptual puzzle but the reality splits between both assumptions – the homogeneity of the soup supports the thermodynamics while the Brownian motion and the antithermodynamical behavior of the open systems call for new viewpoint on the dynamics and its integration in the macroscopic evolution. The need of a new viewpoint is supported also by the notion of a fluctuation - the very idea of macroscopic fluctuation requires coherent behavior. Otherwise, if the local fluctuations in extended systems are uncorrelated in space and time, their average always turns to zero and no fluctuation can ever be observed. However, the short-range dynamics is not able to give rise to the target coherent behavior. Now it is clear that a unified frame for all three phenomena should be built on fundamentally new grounds. The first step in this direction is reconsidering of the idea of interaction. Later, in &2.5 I shall introduce a new broader idea of interaction and shall elucidate how its helps to solve partially the above paradox. However, further the question how it is related to the desired coherence at the Brownian motion and at the formation of macroscopic fluctuations in the open systems comes. The answer to this question goes via demonstration that unified frame for all 3 phenomena is possible only on the grounds of new approach that integrates boundedness as an indispensable part of its fundament. The first step is to demonstrate that the traditional statistical mechanics suffers serious flaws that do not allow self-consistent incorporation of the boundedness concept. Let me start with some crucial arguments that explicitly reveal the necessity of new ideas in the thermodynamics. I shall continue this task in Chapter 4.

### 2.2 Equilibrium State

Equilibrium state is a fundamental notion in the thermodynamics and the statistical mechanics. A many-body system, exposed to certain constraints such as fixed volume, temperature, pressure, is supposed to be in an equilibrium state if after some time it arrives in it and stays there arbitrarily long time. The equilibrium state is supposed to be unique asymptotic state and global attractor, i.e. whatever the initial conditions are, the system finally arrives in it. Further, the behavior of the system in the equilibrium is governed by a few parameters, called state variables. They are, pressure, volume, temperature etc. Moreover, the relations between these variables are functionally insensitive to the chemical identity of the entities that constitute the system. The insensitivity of the equilibrium state to the details of the dynamics strongly contributes to its super-universality.

The major assumption of the thermodynamics is its zeroth law: the equilibrium state always exists and can be reached at whatever the external constraints imposed on the system are. Then the equivalence of the thermodynamical description in terms of microcanonical, canonical and grand-canonical ensemble requires establishing of the so called thermodynamical limit. This implies that the limit:

$$n = \lim_{\substack{N \to \infty \\ V \to \infty}} \frac{N}{V} = const \tag{2.1}$$

exists and does not depend on the order of taking the approaches, i.e. it is independent whether first the number of entities $N$ goes to infinity or first goes to infinity the volume $V$.

It is established that the thermodynamical limit and the equivalence of the 3 ensembles impose certain limitations on the type of the dynamical interactions. However, these restrictions do not provide uniformity of the limit (2.1). Actually, the uniformity is related straightforwardly to the issue about the fluctuations and the stability. It naturally demands finite spatial size of the fluctuations as well as the boundedness of the amplitude of each of them. Note that these requirements ensure existence of finite $N_0$ and $V_0$ so that for every $N > N_0$ and $V > V_0$ the uniformity of the limit (2.1) is guaranteed. I address the problem from the following prospective: is there any aspect of the dynamics that provides the finite size of spatially extended fluctuations? Note that I put the focus on the relation between the dynamics and the boundedness and stability, relation that has not been considered so far.

Though the above arguments calls for revision of the thermodynamics, they still do not convince that its fundamental change is unavoidable. That is why let me now present explicitly the first of my crucial arguments. Let us follow the conventional derivation of the necessary conditions that a chemical reaction at fixed pressure and temperature should obey. I promise to persuade you that it leads to an obvious absurd. In equilibrium the amount of each of the reactants and products is constant and corresponds to the stoicheometric relation:

$$\sum_{i=1}^{S} \nu_i A_i = 0 \tag{2.2}$$

where $\nu_i$ is the stoicheometric coefficient of the $i-th$ sort, $A_i$ is its amount and $S$ is the number of the $i-th$ sort entities.

Eq. (2.2) is equivalent to the following relation:

$$\frac{\partial \Phi}{\partial N_i} + \frac{\partial \Phi}{\partial N_2}\frac{\partial N_2}{\partial N_1} + \ldots + \frac{\partial \Phi}{\partial N_i}\frac{\partial N_i}{\partial N_1} + \ldots + \frac{\partial \Phi}{\partial N_S}\frac{\partial N_S}{\partial N_1} = 0 \tag{2.3}$$

where $\Phi$ is the Gibbs energy, $N_i$ is the number of entities of the $i-th$ sort.

Let us now rewrite (2.3) in slightly different form:

$$\sum_{i=1}^{S} \frac{\partial \Phi}{\partial N_i} \frac{\nu_i}{\nu} = 0 \; . \tag{2.4}$$

By definition:

$$\frac{\partial \Phi}{\partial N_i} = \mu_i \tag{2.5}$$

where $\mu_i$ is the chemical potential.

Thus, at equilibrium the following relation holds:

$$\sum_{i=1}^{S} \nu_i \mu_i = 0 \tag{2.6}$$

Further, since the Gibbs energy is thermodynamical potential in the so-called linear domain around the equilibrium state, i.e. in its the nearest neighborhood, the condition for its smoothness:

$$\frac{\partial^2 \Phi}{\partial N_i \partial N_j} = \frac{\partial^2 \Phi}{\partial N_j \partial N_i} \tag{2.7}$$

leads to the following relation:

$$\frac{\partial \mu_i}{\partial N_j} = \frac{\partial \mu_j}{\partial N_i} \tag{2.8}$$

where $\mu_i$, $\mu_j$ are the chemical potentials of the $i-th$ and $j-th$ sort entities correspondingly. Let us now have a closer look on (2.8) - it turns out that (2.8) makes the chemical identity dependent on the number of the entities! Note that according to its meaning (2.5), the chemical potential is predetermined by the chemical identity of the entities. That is why, it is hardly to be expected that it would depend on the current number of the entities of the other type. Otherwise, the medieval alchemist dream would come true - the chemical identity changes under the change of the number of entities!

So, obviously the definition of the chemical potential needs serious revision. In the following chapter I propose new definition associated with the properties of the state space. The advantage is that it is straightforwardly related to the issue of the stability. The fundamental reason to associate the notion of the chemical potential with the stability is complexity of the interaction of a system and the environment that can entangle them so that sometimes their individuality is blurred. That is why it is important that the notion of the chemical potential involves the functional relations introduced by the interaction among the system and the environment.

### 2.3 Brownian Motion and the Chemical Potential

Let us now consider the most widespread approach to the Brownian motion. It is accepted that the random walk of each pollen is a result of a random "kick" that comes from the solvent. It is assumed that the "kicks" are result of the action of random "sources" that originate from the discrete atomic structure of the solvent. The intensity of the random sources is supposed to vary in an infinite range. Then the random walk of each pollen is successfully described by adding stochastic force to the deterministic friction in the equation of motion so that the stochastic force to match the behavior of the random "kicks". This equation is called Langvin equation. Further, the transition from the Langvin to Einstein-Smoluhowsky equation for the diffusion on coarse-grained time scale demands well-defined concentration of the pollen species. However, this requirement immediately demands finite spatial size and finite amplitude of the fluctuations, random "kicks" included. So, it poses the question that there must be mechanism that guarantees the destruction of arbitrarily large fluctuations. On the other hand, since the size of the pollen is much larger than the size of the solvent molecules, it is obvious that the "sources" must be spatially extended. Thus, the mechanism that destroys the large fluctuations should be "flexible"-it should select "good" small fluctuations and destroy the "bad" large ones.

The current status-quo cannot help because the chemical potential in the thermodynamics accounts only for the chemical identity of a single entity and allows associating of arbitrarily large number of entities which immediately allows growing of a fluctuation to arbitrarily large size. In &3.7 I shall advance a new definition of the chemical potential that involves functional relations among entities and relates the number of associated entities with the stability of the system. One of the major outcomes of that definition is that it always provides finite size and finite lifetime of extended fluctuations.

Yet, one might say: "OK, let us put forward a new definition of the chemical potential. But there is still a good chance for the thermodynamics to survive." Let me now present my heaviest weapon: a fundamental controversy that blows the very core of that theory.

### 2.4 Small Fluctuations

The fundamental property of the many-body systems is that their macroscopic description involves variables that have no dynamical analog. Such basic variable is the entropy. It participate each thermodynamically potential, i.e. it governs the behavior of a system at any external constraint (temperature, pressure etc.). Since the equivalence of the description by different ensembles requires holding of the thermodynamical limit, the notion of entropy must also be consistent with the idea of the thermodynamical limit. The latter implies that the entropy must be insensitive to the partitioning of the system into subsystems. Moreover, in order to meet the thermodynamical limit it should be additive with respect to the partitioning: the total entropy of a system partited into subsystems is sum of the entropies of the subsystems. In addition, the total entropy must be independent of the way the partitioning is made.

The importance of the additivity is that it makes the entropy well-defined single-valued characteristic of any state. In turn, this renders any thermodynamical potential to approach monotonically the equilibrium state. This opens the door for the universal behavior of the small fluctuations around the equilibrium.

Indeed, any deviation of a state variable from its equilibrium value is called fluctuation. We are able now to define their distribution. Indeed, since in the equilibrium the thremodynamical potentials have extremum (maximum of the entropy for the isolated systems and minimum for the Gibbs and Helmholtz energy) the first non-zero term in their Taylor series expansion is the second one:

$$\Phi = \Phi_{eq} + a(x - x_{eq})^2 \quad (2.9)$$

It is well known that the Boltzmann-Gibbs distribution is exponent of the corresponding thermodynamical potential. Then the distribution of the small fluctuation around equilibrium reads:

$$f_{B-G} \propto \exp\left(-\frac{(x - x_{eq})^2}{\sigma^2}\right) \quad (2.10)$$

A remarkable property of eq. (2.10) is that its derivation does not involve any information about the dynamics. For a long time it has been another hallmark of the super-universality and power of the thermodynamics.

Let us come back and have a closer look on the issue about the additivity of the entropy. Can it tell more about the fluctuations? Yes, it can. We start with a system partitioned into subsystems. Let us consider the following two cases:

(I) each subsystem has dynamical invariant(s) such as energy. Applied to the partitioning, it means that the there exists time span so that the subsystems does not interact. This helps to define the notion of isolated system. Indeed, a system is considered isolated when it does not exchange matter and energy with the environment. Then, its total energy is invariant.

(ii) the second case is when the boundary effects are negligible compared to the volume effects of the subsystems.

In both cases the fluctuations in the subsystems are independent from one another. Then the distribution of the fluctuations is Poissonian:

$$f_p = \frac{\lambda^x \exp(-\lambda)}{x!} \quad (2.11)$$

where $\lambda$ is the expectation value set by the thermodynamical limit. So, by means of the same assumptions, namely the additivity of the entropy and the thermodynamical limit, we derive two different distributions for the small fluctuations: the Gaussian one ((2.10)) and the Poissonian one ((2.11))! The problem is that they have different asymptotic when $x$ goes to zero. Indeed, $f_{B-G}$ asymptotic reads:

$$f_{B-G} \propto 1 - \frac{(x - x_{eq})^2}{\sigma^2} \tag{2.12}$$

while the $f_p$ asymptotic is:

$$f_p \propto 1 - x \tag{2.13}$$

It is evident that the Gaussian distribution goes to zero quadraticaly while the Poissonian one linearly. But how this ambiguity emerges?

The above considerations clearly indicate that the core of the problem lies in the requirement about the additivity of the entropy and the ignorance of the dynamics. Indeed, meeting of the thermodynamical limit demands insensitivity to the partitioning. Justification of that conjecture from microscopic viewpoint is grounded on the existence of local invariants and ignorance of the boundary effects. But is it possible at all to have dynamical invariants for subsystems? I will discuss this problem in &2.5. Now I would like to discuss the plausibility of the requirement about the insensitivity to partitioning.

Let us first stress that the insensitivity to the partitioning implies lack of specific for the system spatial scale. Otherwise, the value of the entropy would strongly depend on the way the partitioning is made. In turn, this immediately yields non-monotonicity of the entropy. This happens because the entropy is defined up to additive factor that is determined by the partitioning. However, any non-monotonicity of the entropy violates the core of the thermodynamics - the uniqueness of the equilibrium state! The requirement of lack of any specific spatial scale implies perfect homogeneity and lack of any boundary effects. So, in order to meet those constraints the system needs perfect mixing. Many authors put forward diffusion as the implement that provides it. However, in the reality the diffusion has finite rate which makes it ineffective in a number of cases. For example, the chemical reactions are certainly out of such frame because the reaction-diffusion coupling always creates specific scale so that the boundary effects due to the diffusion become compatible with the volume effects of the reaction. Hence what left are ideal gases! And the real systems that work according to the thermodynamics are only steam engines! It is worth to stress that the solvent and the Brownian motion cannot be explained by the thermodynamics because they require a dynamical mechanism that suppresses fluctuations in the first case and restricts them to a finite size in the second one!

Thus, the thermodynamics fails in the description of the behavior of more complicated systems such chemical reactions, living organisms and social systems, i.e. with systems that have specific temporal and spatial scales.

Let us now tempt your curiosity and focus the attention to what happens under the supposition that the entropy is non-monotonic. It is obvious that it becomes a multi-valued function - more than one state corresponds to a single value of the entropy. Another reading of this statement is that the rates of the changes of the state variables are multi-valued functions so that each selection corresponds to one of the equientropical channels. Put it formally, it reads:

$$\frac{d\vec{x}}{dt} = \hat{A}_i(\vec{x}) - R_j(\vec{x}) \tag{2.14}$$

where $\vec{x}$ is the vector of the state variables, $\hat{A}_i(\vec{x})$ is the rate of the springs and $R_j(\vec{x})$ are sinks. The springs are the volume effects while the diffusion through the boundary is to be considered as a sink. The indices $i$ and $j$ indicate the current choice of the equientropical channel(s). Since all the channels have the same entropy, it is most likely to expect random choice of a single channel among all available. This introduces an indispensable duality stochasticity-determinism of the system response to any change of the state variables - though all the selections can be computed, the system randomly "chooses" one of them. Sounds surprising and unexpected, is not it? But let us make just one small step further. Let us suppose that all the selections are bounded, so that only finite energy and/or matter are involved in the system. Then, $\hat{A}_i(\vec{x})$ is a BIS and always has expectation value. So, (2.14) can be presented in the form:

$$\frac{d\vec{x}}{dt} = \hat{A}_{av}(\vec{x}) + \hat{\eta}_{ai}(\vec{x}) - \hat{R}_{av}(\vec{x}) - \hat{\eta}_{rj}(\vec{x}) \tag{2.15}$$

where $\hat{A}_{av}(\vec{x})$ and $\hat{R}_{av}(\vec{x})$ are the rate averages; $\hat{\eta}_{ai}(\vec{x})$ and $\hat{\eta}_{rj}(\vec{x})$ are zero-mean BIS; the index $i$ and $j$ are put to stress the stochastic nature of that terms. But, let have a closer look to eq.(2.15). It is a stochastic equation! Moreover, it describes a system that permanently fluctuates! So, indeed, we come to completely new result: the system in question permanently fluctuates! So, we open the door to permanent fluctuations but there are a lot of questions that come. Among the most important ones is that about the dynamics that ensures such behavior. My goal is to establish the dynamical foundation of the evolution governed by eq.(2.15). It is supposed to give satisfactory explanation of both homogeneous saltiness of the soup and restricted size and amplitude of the "sources" in the Brownian motion. Let us now start with advancing of a radically novel viewpoint on the dynamics of the interaction.

### 2.5 Stochastising Interactions

So far the dynamics of the many-body systems considers predominantly the two-body interactions. The reason for that is that they are much more frequent than the $3-, 4-$ and other $n-$body interactions. Another strong reason for the ignorance of $n-$body interactions $(n>2)$ is the postulate that they do not contribute anything fundamentally new to the two-body dynamics. I do not argue the first argument but do indeed the dynamics of the $n-$body interactions is the same as that of the two-body interactions?!

Let me start with the two-body interactions. Both in the classical and quantum case the energy and the momentum conservation unambiguously defines the outgoing velocities (scattering waves) and energies. Furthermore, this unambiguouty gives rise to the time-reversible invariance of the two-body dynamics. It implies that under the change of the time arrow the velocities "turn back"- the outgoing ones become ingoing ones. This fact constitutes the major contradiction between the dynamics and the thermodynamics: on the one hand, the thermodynamics asserts that a system monotonically approaches the equilibrium state and once reaching it stays there permanently. On the other hand, the dynamical time-reversibility makes possible significant departures from the equilibrium state.

Let us now consider $3-$body interactions. A simple check shows that in this case the laws of energy and momentum conservation do not unambiguously determine the outgoing energies and velocities. Indeed, the classical energy and momentum conservation for the $3-$body interaction read:

$$\frac{m_1 \vec{V}_1^2}{2} + \frac{m_2 \vec{V}_2^2}{2} + \frac{m_3 \vec{V}_3^2}{2} = \frac{m_1 \tilde{\vec{V}}_1^2}{2} + \frac{m_2 \tilde{\vec{V}}_2^2}{2} + \frac{m_3 \tilde{\vec{V}}_3^2}{2} \qquad (2.16)$$

$$m_1 \vec{V}_1 + m_2 \vec{V}_2 + m_3 \vec{V}_3 = m_1 \tilde{\vec{V}}_1 + m_2 \tilde{\vec{V}}_2 + m_3 \tilde{\vec{V}}_3 \qquad (2.17)$$

where $m_i$, $i=1,2,3$, are the masses of the entities; $\vec{V}_i$ are their initial velocities and $\tilde{\vec{V}}_i$ are the final velocities. However, we have 4 equations for the 9 components of the final velocities. Hence indeed the energy and momentum conservation are not enough to determine non-ambiguously the final velocities even in the simplest case of elastic $3-$body interaction. Thus, the theory of interaction calls for further fundamental assumptions about the details of the process of interaction.

It is natural to suppose that any $3-$body interaction is a two-stage process: at first stage two of the entities collide and start interaction; the second stage starts after the arrival of third entity. Note that the simultaneous collision of all three entities implies existence of spatial correlations among the entities! Further, I suppose that the first two entities create a compound temporary Hamiltonian. When the third entity arrives, it changes that Hamiltonian non-smoothly. The key point is that since moment of arrival of the third entity is arbitrary, the final outcome of the interaction becomes a multi-valued function whose selections correspond to the level of the compound Hamiltonian at which the third entity has arrived. Hence the total Hilbert space of the interaction is the following direct sum:

$$\hat{H}_{tot} = \hat{H}_{2i} \oplus \hat{H}_3 \qquad (2.18)$$

where $\hat{H}_3$ is the Hilbert space after the third entity arrival; $\hat{H}_{2i}$ is a projection of the entire Hilbert space of the initial $2-$body interaction onto one of its subspaces. The properties and the dimension of that subspace are determined under the following rule: the arrival of third entity before completing the $2-$body interaction makes only part of the states of the entire $2-$body Hamiltonian available for this interaction. Then, the $2-$body interaction participates to the total Hilbert space by one of its subspaces. Furthermore, since the moment of arrival of the third entity is arbitrary, $\hat{H}_{tot}$ becomes a multi-valued function whose selections are determined by the moment of arrival of the third entity; it should be stressed that the arrival moment sets the properties and the dimension of $\hat{H}_{2i}$. Therefore each selection can be computed by choosing a particular subspace of $\hat{H}_2$; the number of selections corresponds to the number of the subspaces. Note that the multi-valuedness immediately introduces dualty determinism-stochastisity: though each selection can be computed by an appropriate quantum-mechanical approach, their realization in a single event is random choice of one selection among all available.

I will present the physical idea why the $2-$body interaction participate to the total Hamiltonian trough its subspaces in Chapter 4. The reason for the postponing is that there this problem is related with the notion of chemical identity.

The multi-valuedness of the total Hilbert space (Hamiltonian) $\hat{H}_{tot}$ is very important novel property because it gives rise to the violation of the dynamical time reversal invariance. Indeed, since the moment of arrival of the third entity stays random both in forward and backward directions of time, a number of outgoing trajectories corresponds to each ingoing scattering trajectory. Multi-valuedness of the outgoing trajectories emerges straightforwardly from (2.18). It is to be expected that the violation of the dynamical time-reversal invariance has crucial impact on the fundments of the statistical mechanics and the thermodynamics. The major goal of the present book is use of the multi-valuedness of the dynamical interactions and the boundedness concept as grounds to built a successful approach to the considered in the &2.1 three phenomena .

The random uncorrelated motion of the entities makes the arrival moment of the third entity to vary both in space and time. In turn, it brings about strong spatial non-homogeneity because different selections are established even at closest points. On the other hand, the spatial non-homogeneity breaks any long-range interactions among positions and velocities. In sequel, it violates the formation of local invariants. Now we can explain the lack of fluctuations in the soup. The $n-$body interactions behave like "propellers" that permanently homogenize the solution and permanently destroy the long-range correlations. That is why we call these interactions stochastising ones.

But do the considerations about the $3-$body interactions sound familiar, do not they? Yes, indeed they do - in the Preface we have already considered something very similar, the diffusion-induced non-perturbative interactions. However, then we considered them in order to settle the necessity of coherence. Though there is a great difference between that case and the "soup": why we do not even mention any coherence in the latter. The reason is simple but it demarcates two very important "roles" of stochastising interactions. The first one is their role as homogenizers. This happens when their concentration is much below the percolation threshold. Then they are rapidly "dispersed" in the sea of the $2-$body interactions and thus they do not destabilize the system. Let us remind that the amount of the salt is very few - usually we put about a tea spoon in a liter or more of water. I will be back to this issue in Chapter 4. On the contrary, in the case of the surface reactions considered in the Preface, diffusion-induced non-perturbative interactions happen in the immediate neighborhood of each chemisorbed entity. Thus, they certainly destabilize the system since their concentration is kept high enough due to the permanent bombardment of the surface by the gas entities. Then, the system can survive breakdown if and only if there is mechanism that makes all the entities to behave coherently. I will consider the details and conditions of its operating in Chapters 5 and 6. It is obvious that the mechanism that gives rise to the coherence also needs radically novel ideas because the coherence implies establishing of long-range correlations among distant points. On the other hand, the physical interactions in most many-body systems far from the points of phase transitions are supposed short-range ones. So, the very idea of coherence encounters us to the following puzzle: how short-range interactions give rise to long range correlations?!

### 2.6 What Comes Next

The task of the present chapter has been to show explicitly that the thermodynamics needs revision on both dynamical and macrolevel. The new viewpoint on the $n-$body $(n>2)$ interactions as stochastising ones essentially changes the dynamical background of any macroscopic description. The special attention must be focused on the case when they appear as destabilizing factor. Then, the stabilization is possible only if there are long-range correlations among distant points. But do these long-range correlations give rise to a single state that is analog to the equilibrium one or it has entirely new properties? Moreover, if equilibrium is established, can a system fluctuate at all? Further in my book I shall show that the mechanism of coherence leads to establishing of behavior that has very new macroscopic properties such that certain macroscopic characteristics exhibit permanent fluctuations in the time course.

In the course of the book the primary importance of the boundedness shall appear evidently. In the Chapter 1 the statical and dynamical boundedness are suggested. In the Chapter 5 and 6 they will appear naturally from the concept of coherence. That is why now it is important to establish the universal properties of the solution of eq.(2.15) that are due to the static and dynamical boundedness. Among the very important universal properties is the distribution of the macroscopic fluctuations. In &2.4 I have shown that the classical thermodynamics is controversial on the issue about the distribution of the fluctuations. The Gaussian distribution that comes from Boltzmann-Gibbs ensemble interferes with the Poissonian one that comes from the additivity of the entropy. That is why it is important to prove that the boundedness itself guarantees existence of asymptotic distribution that is insensitive to the details of the fluctuations. This is made in the next chapter.

The major aim of the present book is to study both microscopic and macroscopic evolution of the system whose behavior is governed by the boundedness. The next major aim is to derive these equations starting from the dynamical point of view, namely to built the foundations of the theory of the coherence and to prove rigorously that it gives rise to eq.(2.15). This will be made in Chapter 6 and 7. The goal of the next chapter is to prove that the state space of a system whose macroscopic behavior is governed by eq.(2.15) is strongly chaotic. Thus, the rest of the book justifies the second part of our title: From Coherence to Chaos. Further, the premises and the details of the coherence are considered in Chapters 4, 5 and 6. The dynamical basis of the coherence and its link to eq.(2.15) is systematically derived.

My considerations open more questions than give answers but I focus your attention on those topics that helps creating a self-consistent theory that successfully explains all 3 phenomena listed in &2.1.

### References

I present of list of selected papers that has helped me to constitute my arguments for the necessity of revision the thermodynamics. I did not cite any of them in the chapter because its content is not straightforwardly related to them.

# Chapter 3: State Space and Long-Term Stability

### 3.1. Why to Read Chapter 3

Destabilization induced by stochastising interactions requires explicit involvement of the factors that ensures long-term stability. Our goal is to elucidate how far they change the fundament of the statistical mechanics.  The task of the present chapter is to consider the motion in the state space under the most general constraint that provides the long-term stability:  the boundedness. This conjecture has two-fold comprehension: the first one is boundedness of the fluctuation size. This idea suggests that largest fluctuations make "U-turns" without involving any special physical process. In Chapter 1 we took for granted this assumption but now I aim to work out the general condition for automatic execution of the "U-turns". The second comprehension of the boundedness conjecture is that the boundedness of the fluctuation amplitude is not enough to keep the system stable. Likewise it is necessary that the amount of energy/matter exchanged with the environment is also permanently limited. This requirement imposes permanent boundedness of the rate of the fluctuations development. To remind, we call the boundedness of the fluctuation rate dynamical boundedness while the boundedness of the fluctuation amplitude is named static boundedness.

In addition to the dynamical and static boundedness, the stability of extended systems needs also spatial coherence among local fluctuations at distant points. Indeed, suppose that a system develops under local rules and short-range interactions. Then, there is noting to suppress the unlimited expansion both in size and in amplitude of the local fluctuations. Indeed, the lack of long-range interactions renders local transitions non-correlated both in space and in time. This however, immediately introduces local strain, local overheating, sintering etc. The relaxation of these defects is many orders of magnitude slower than the thermalization which in turn sustains their further formation. Due time course the interaction among them produces local reconstruction, creates mechanical defects etc. Eventually the process yields the system breakdown. Therefore, the long-term stability calls for a mechanism that makes distant fluctuations to respond coherently.

There is a wide spectrum of works aimed to explore an effect of correlations of the fluctuations in extended systems as interplay among noise correlations, non-linearity and spatial coupling. However, all the developed so far approaches model the stochastic variables and noise sources as Wiener process whose increments are independent and unbounded. Thus, though cooperation of the fluctuations is available, the sequence of spatio-temporal configurations through which the system arrives to global coupling varies in uncontrolled way that it is incompatible with the idea of boundedness.

The question now is whether there is a general rule that selects the transitions that do not violate the stability of the system. Obviously, the transition between any two states does not violate the stability of a system if and only if the local rates of energy and/or matter exchange permanently do not exceed the thresholds of stability. The above considerations illustrate that this is possible only if the local fluctuations are coupled so that to behave coherently. In Chapter 6 I shall prove that the execution of the coherence mechanism predetermines a range of admissible states to every given one. The most important properties of that range is that all admissible states are at finite distance to the given one; the range is determined by the evolutionary equations of the system. In Chapter 6 and 7 I shall make evident that they are stochastic equations of the type (2.15). The evolutionary equations are balance equations for the state variables whose behavior is governed by the rates of the elementary processes. I shall demonstrate that the coherence mechanism gives rise to duality determinism-stochasticity of the rates of the elementary processes: they become multi-values functions whose selections are deterministic and always bounded but their realization at any instant is random choice among all available. In sequel, the motion in the state space is governed by the boundedness and multi-valuedness of the rates. These properties of the rates bridge the spatial coherence and the dynamical and static boundedness as joint aspects of the same concept. Note that the boundedness is radically novel constraint to the motion in the state space. It selects a range of admissible states to each given one making the velocity of the motion finite as it is established in &3.4.2. This is in sharp contrast with the hypothesis, taken for granted in the statistical mechanics, that all the states are accessible from any given one which, however, makes the velocity of the motion in the state space to be infinite [3.1-3.2]. The dilemma whether the velocity is finite or arbitrary is straightforwardly related with the issue of the dynamical boundedness. Indeed, an arbitrary velocity implies involving unlimited amount of energy/matter in a transition and its spreading through space and time with arbitrary velocity. On the contrary, bounded velocity implies involvement of a limited amount of energy/matter in each transition and its spreading through space and time with finite velocity.

It is to be expected that the boundedness introduces radically novel understanding of the major characteristics that are related to the properties of the state space. The turning point is the definition of the chemical potential so that to involve not only the chemical identity as in the thermodynamics, but also the functional relations created in the interaction. In reality the interacting systems are entangled by the processes that proceed among them and create functional relations with the environment. Hence, sharp separation between a system and environment is impossible which makes important to define the chemical potential so that to involve the functional relations created in process of the interaction. Since they are not constant during the interaction, their modification results in variations of the chemical potential. Among all, they can even turn it to zero which brings about an immediate collapse of the system. The latter apparently calls for explicit relation between the chemical potential and the stability of a system. In &3.7 I shall present definition of the chemical potential that straightforwardly relates the association/dissociation of entities with the stability of the system. It turns out that this relation makes possible association/dissociation of only limited amount of entities so that the system stays stable. Note that this result strongly interferes with the traditional definition of the chemical potential which makes possible associating/dissociating of arbitrary number of entities without any effect on the stability of the system.

The major assertion about the joint action of all 3 aspects of the boundedness, namely the static and the dynamical boundedness and the spatial coherence, taken for granted in this Chapter, is that the corresponding system permanently exerts macroscopic fluctuations. Gradually in the book I shall present decisive arguments in favor of this assertion. Furthermore, as verified in Chapter 7, this behavior is spread on every time scale starting from the fundamental one over which the spatial coherence proceeds. In sequel, for systems subject to spatial coherence, there is no "equilibrium" state: they permanently exert motion in the state space. My present goal is to find out the properties of the motion in the state space subject to the joint influence of all 3 aspects of the boundedness conjecture: the static and the dynamical boundedness and the spatial coherence.

### 3.2. Evolutionary Equations

A very important consequence of the spatial coherence is that it makes expected to preserve the notion of state variables. Indeed, if the entities are correlated, it is always possible to define their "concentration" in the sense that the limit (2.1) holds. Hence, we keep on the idea of the state variables. It lets the macroscopic evolution of a fluctuating system to be described in terms of state variables. Now I put the stress on the central conjecture: any system subject to joint influence of the static and dynamical boundedness and the spatial coherence permanently fluctuates and has no equilibrium state.

I shall demonstrate in Chapter 6 that the evolution of these systems is governed by the most general frame of balance equations for all springs and sinks. The most important property of their transition rates is that they are multi-valued functions all selections of which are bounded at every instant in a specific to the system range. Further, it implies that at any given instant a single selection, randomly chosen among all available, is established. At next instant another selection again randomly chosen among all available is established. Since all the selections are bounded, we come to eqs. of the type (2.15). Since I need them further in this section, I present them again:

$$\frac{d\vec{x}}{dt} = \hat{A}_{av}(\vec{x}) + \hat{\eta}_{ai}(\vec{x}) - \hat{R}_{av}(\vec{x}) - \hat{\eta}_{rj}(\vec{x}) \tag{2.15}$$

where $\hat{A}_{av}(\vec{x})$ and $\hat{R}_{av}(\vec{x})$ are the spring and sink rate averages; $\hat{\eta}_{ai}(\vec{x}) = \hat{A}(\vec{x}) - \hat{A}_{av}(\vec{x})$ and $\hat{\eta}_{rj}(\vec{x}) = \hat{R}(\vec{x}) - \hat{R}_{av}(\vec{x})$ come from the current selections and are zero-mean BIS; the index $i$ and $j$ are put to stress the stochastic nature of that terms.

It should be stressed that eqs.(2.15) radically differs from the mean-field approach because, eqs.(2.15) are brought about by a long-range coupling process while the mean field approach is founded on the premise that any system can be divided into subsystems whose fluctuations are independent one from another.

Eqs.(2.15) have a very important property with surprising and far going consequences that will be discussed a bit latter. It is that the power spectrum of its solution comprises additively two parts. The first one comes from the solution of the following system of differential equations:

$$\frac{d\vec{x}_{\det}}{dt} = \hat{A}_{av}(\vec{x}_{\det}) - \hat{R}_{av}(\vec{x}_{\det}) \tag{3.1}$$

Being a system of non-linear ordinary or partial differential equations, eqs. (3.1) give rise to different dynamical regimes and coherent structures depending on the values of the control parameters. Hence, this result covers the behavior that still puzzles the modern science and is supposed to be complicated effect of the non-linearity integrated in the traditional thermodynamics. Note that we come to the same result without any reference to it!

It should be stressed that eqs.(3.1) do not yield thermodynamical equilibrium even in the case when they have a single stable solution. This is because the thermodynamics requires not only the existence of a single solution but it should have a global attractor as well. However, the solution of (3.1) has not any global attractor because its phase space is separated into basins of different solutions whose margins are governed by corresponding domains in the control parameter space. Furthermore, it is impossible to construct a general thermodynamical potential that has the property to damp fluctuations because the Laypunov functional does not exist for the majority of the dynamical systems as it is well known from the mathematical literature. Moreover, even if the Lyapunov functional exists for certain class of equations, the execution of the stochastic terms makes the system to deviate permanently from the "equilibrium".

The stochasticity of the eq.(2.15) poses the question whether its solution has properties that are reproducible. This is very important question since the only reliable information about a system is the reproducible one. Evidently, it is not the solution itself since it varies with the realizations. On the other hand, the solution of eq.(3.1) is certainly reproducible. Its power spectrum is also. The boundedness of the stochastic terms in eqs.(2.15) gives us hope that the power spectrum of the stochastic part is insensitive to the fluctuation statistics which makes its reproducibility highly expected. Then, along with the reproducibility of the power spectrum of the solution of (3.1) it ensures reproducibility of the entire power spectrum. The key point is the additivity of both power spectra because it provides unambiguous separation of the specific to the system information concentrated in the solution of (3.1) from the "noise" focused in the stochastic part.

This makes our present task to be establishing of the additivity of the both parts in the power spectra. It becomes obvious by the following consideration:

$$\frac{d\bar{x}}{dt} = \frac{d}{dt}\lim_{T\to\infty}\frac{1}{T}\int_0^T x(t)dt = \lim_{T\to\infty}\frac{1}{T}\int_0^T \frac{dx(t)}{dt} =$$
$$= \lim_{T\to\infty}\frac{1}{T}\int_0^T \left(A_{av}(x) - R_{av}(x)\right)dt + \lim_{T\to\infty}\frac{1}{T}\int_0^T \left(\eta_{ai}(x) - \eta_{ri}(x)\right)dt \quad (3.2)$$

The term
$$\lim_{T\to\infty}\frac{1}{T}\int_0^T \left(\eta_{ai}(x) - \eta_{ri}(x)\right)dt \to 0 \quad (3.3)$$

because of the random appearance of the selections. Put it differently, the random choice of the selections and boundedeness makes correlation size of the stochastic terms always finite. Hence, when being averaged over an infinite time interval, the average of the stochastic terms always turns to zero. To remind that the sequences $\eta_{ai}(x)$ and $\eta_{ri}(x)$ are defined as zero-mean BIS.

Therefore the power spectrum of the solution of eq.(2.15) comprises the power spectrum of the solution of (3.1) that at appropriate values of the control parameters is a discrete band. On the other hand, the power spectrum of the solution of eqs.(2.15) always comprises the continuous band that comes from the BIS constituted by the stochastic part $(\vec{x} - \vec{x}_{\det})$. We already know from &1.4 that its power spectrum fits the shape $1/f^{\alpha(f)}$ where $\alpha(f) \to 1$ at the infrared edge and linearly increases as $f$ approaches infinity. The additivity of the both parts and the robustness of the shape $1/f^{\alpha(f)}$ makes possible extraction of the specific to the system information that is concentrated in $\vec{x}_{\det}$ from the "noise" part associated with both natural noises and human intervention during recording and/or monitoring. Note that the robustness of the shape $1/f^{\alpha(f)}$ to the statistics of the BIS renders reproducibility of the power spectrum that in turn makes the separation of the signal from the noise unambiguous and reproducible. Furthermore, note that the robustness of the shape $1/f^{\alpha(f)}$ and the additivity of the discrete (specific to the system) band and the "noise" (continuous) one is an immediate result of the boundedness. It should be stressed that its lack makes the shape of the power spectrum strongly dependent on the statistics of the time series. In turn, not only the time series would not be reproducible, but the power spectrum would entangle the genuine signal with the unavoidable human intervention during recording and/or monitoring.

My major task is to distinguish the novelty of our description from the thermodynamical one. That is why we start our journey in the properties of the state space subject to boundedness with a comparison to the properties of a "traditional" state space.

### 3.3. State Space Subject to Boundedness. General Properties

A superficial look at (2.15) says that one should not expect anything new because the stochastic terms have Markovian property and then the successive steps in the state space also share that property. Hence, the Chapmen-Kolmogorov equation is fulfilled. It gives rise to the Fokker-Plack equation which results in going back to the traditional statistical physics. But is it indeed so? Now I shall present a crucial argument that the boundedness violates the Markovianity. The question is whether the state space is indeed connected by a Markovian chain under the imposed by the boundedness constraints. If so, we are back to the traditional statistical mechanics. If no, one should anticipate radically new properties.

Let us suppose that the transition from state $j$ to neighbor state $i$ depends only on whether the transition to $j$ has happened. However, the transition to $j$ depends on whether it comes from the range of the admissible states, i.e. those ones that does not violate the boundedness – let us denote them by $k$. Hence the transition to $i$ is set on the chain of the previous transitions …$l…kj$. Therefore, is the process non-Markovian though the Chapman-Kolmogorov relation holds?! It seems Markovian because the transition from $j$ to $i$ depends only on $j$. However, it is non-Markovian, because any admissible transition depends on the succession of the previous ones. (Examples of non-Markovian chains that fulfill Chapman-Kolmogorov relation are presented in [3.3]). As a result, we should look for a new approach to establish the properties of a state space subject to boundedness.

At first, let me specify some features of the state space trajectories. The multi-valuedness of the transition rates renders existence of more than one admissible state to any given one. This is matched in eqs.(2.15) through the random choice of one selection from all available at every instant. Indeed, though only one transition takes place, the set of all available ones constitute the current range of the admissible transitions. The outcome is that the trajectory appears as a kind of a fractal Brownian "walk". In addition, the stochastisity induced by the random choice of one selection among all available breaks any possible long-range periodicity. So, the trajectories in the state space are BIS subjects to incremental boundedness. The latter implies that the transition rates are not only bounded, but their correlation size along any trajectory is finite. The incremental boundedness supplements the notion of the dynamical boundedness at the point about the finite memory size of the transition rates along the state space trajectories. Let me remind that the derivation of the chaotic properties in Chapter 1 is made under the *apriori* assumed static and dynamical boundedness along with the lack of long-range physical correlations among the time scales. Keeping in mind that the combination of these assumptions is equivalent to the incremental boundedness, it becomes clear that the assumptions on which the chaotic properties of the BIS were established naturally agree with eqs.(2.15).

At this point I must outline a very important difference between our approach to motion in the state space and those applied in the statistical mechanics, namely: the latter assumes that the dynamical degrees of freedom enter the macroscopic description being already somehow averaged, already stochastized while our approach involves multi-valued functions whose selections are entirely deterministic; the stochasticity in our approach is brought about by the permanent random choice of a single selection among all available.

The incremental boundedness distinguishes a scale $a$ above which all scales up to the thresholds of stability contribute uniformly to the creation of the fluctuations. Then, it is to be expected that the coarse-grained structure of the attractor exhibits universal properties insensitive to the details of the transition rate statistics. Furthermore, it is most likely that these properties are similar in every direction of the attractor. So, I can study the one-dimensional sequence produced by the projection of a trajectory onto any direction. I must recall that every such sequence belongs to the class of BIS subject to incremental boundedness defined above.

In the next section I shall study those universal properties shared by every trajectory that define the coarse-grained structure of the attractor. Bearing in mind that the state space trajectories are BIS, I shall demonstrate that the coarse-grained over $a$ structure of each trajectory is a succession of well separated one from other successive excursions. To remind, an excursion is a trajectory of walk originating at the expectation value of a given sequence at moment $t$ and returning to it for the first time at the moment $t + \Delta$. The characteristics of each excursion are amplitude, duration and embedding interval. The latter is a property introduced by the boundedness and has no analog for the unbounded sequences. It implies that each excursion is loaded in a larger interval whose duration is interrelated with the duration of the excursion itself. The major role of the embedding is that it does not allow overlapping of the successive excursions and thus prevents growing of the excursion amplitude to arbitrary size. It results in permanent preserving of both static and the dynamical boundedness.

The incremental boundedness renders that each excursion has certain duration interrelated with its amplitude. By the use of this relation, I shall prove in &3.4.2 that the velocity of the motion in the state space is always finite. The dilemma whether this velocity is finite or arbitrary is straightforwardly related with the issue of the dynamical boundedness. Indeed, an arbitrary velocity implies involving unlimited amount of energy/matter in a transition and its spreading through space and time with arbitrary velocity. On the contrary, bounded velocity implies involvement of a limited amount of energy/matter in each transition and its spreading through space and time with finite velocity.

### 3.4. Properties of the State Space Trajectories

The major task of the present section is to establish those properties of the state space subject to boundedness that built the grounds of the universality of its structure and its insensitivity to the details of the transition rate statistics.

Let me start with the reminding that the state space trajectories are BIS subject to incremental boundedness. Recalling that the latter distinguishes a scale above which all scales up to the thresholds of stability contribute uniformly to the creation of the fluctuations, it is to be expected that the properties of the coarse-grained trajectories are insensitive to the particularities of the transition rate statistics. In turn they constitute the universality the coarse-grained structure of the attractor.

In the next subsection I shall prove that every coarse-grained trajectory appears as a sequence of separated by non-zero intervals successive excursions. It should be stressed that this structure is result of the boundedness alone and does not depend on the statistics of the original trajectory.

Each excursion is characterized by its amplitude, duration and embedding interval as sketched in Fig.3.1.

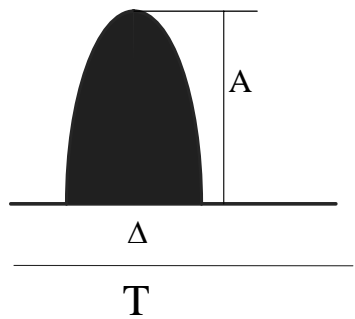

Fig.3.1 Characteristics of the excursions

The separation of the successive excursions means embedding of each of them in a larger interval so that no other excursions can be found in that interval. Below I shall demonstrate that the duration of the "embedding" interval is a multi-valued function whose properties are strongly related to the duration of the embedded excursion $\Delta$ itself: the range and the values of the selections are set on $\Delta$; the realization of any embedding interval is always associated with the realization of the corresponding excursion. Since the duration of the each embedding interval is a multi-valued function, its successive performances permanently introduce stochasticity through the random choice of one selection among all available. Thus, the multi-valuedness of the embedding induces stochasitity that breaks any possible long-range periodicity (i.e. long-range memory) and helps the excursion sequence to preserve the chaotic properties established in Chapter 1 on coarse-grained scale.

The relation $\Delta \leftrightarrow A$ is set on "blob" structure of the walk that produces the coarse-grained trajectory, namely: because of the finite memory size, on coarse-grained scale any fractal Brownian walk can be considered as symmetric random walk of "blobs" created by subwalks whose size counterparts the memory size.

Our first task is to work out explicitly the relations $T \leftrightarrow \Delta \leftrightarrow A$, i.e. the relations between duration of the embedding time interval $T$, respectively the duration $\Delta$ and the amplitude of the corresponding excursion $A$.

### 3.4.1 Embedding Time Interval. Relation $T \leftrightarrow \Delta$

The major role of the "embedding" is that it does not allow overlapping of the successive excursions and therefore prevents growing of the excursion amplitude to an arbitrary size. So, the "embedding" permanently "holds" the trajectories confined in a finite attractor.

The present task is to work out explicitly the relation between the duration of the embedding intervals and the duration of the corresponding excursions. That relation is based on the notion of excursion: trajectory of walk that originates at the expectation value of a given sequence at time $t$ and returns to it for the first time at time $t + \Delta$. Therefore, the probability for excursion of duration $\Delta$ is determined by the integral probability for all pair of points to be separated by distance smaller than $\Delta$. Recalling that the probability that any two points separated by time interval $\eta$ have the same value is given by the autocorrelation function $G(\eta, T)$, the probability that an excursion of duration $\Delta$ happens in an interval $T$ reads:

$$P(\Delta, T) = \frac{1}{T} \int_0^\Delta G(\eta, T) d\eta = P(\Delta, T) = \int_0^{\Delta/T} \left(1 - x^{\nu(x)}\right) dx \tag{3.4}$$

We have already met the function $P(\Delta, T)$! Remember that in Chapter 1 it was the function $P(u)$ that gives the probability for a pair of points to participate to the stretching and folding. The different role in which it appears now comes from the difference in the definition of a BIS. In &1.3 we have considered the original BIS, while now we consider its coarse-grained counterpart. Therefore, the meaning of $P(\Delta, T)$ becomes different, namely both $\Delta$ and $T$ are associated with a single excursion, while for the original BIS they are associated with the long-range correlations. Yet, the parity between the short-range statistics of a single effective excursion in the coarse-grained counterpart and the universality of the long-range correlations in the original BIS renders the universality of the coarse-grained properties.

The $P(\Delta, T)$ dependence only on the ratio $\Delta/T$ in (3.4) verifies the assumption that every excursion of duration $\Delta$ is "embedded" in an interval of duration $T$ so that no other excursion happens in that interval.

The next step is to work out the shape of $P(\Delta, T)$. Its role is crucial for the behavior of the excursion sequences. To elucidate this point let us consider the following extreme cases:

(I) $P(\Delta, T)$ is a sharp single-peaked function. Then it ensures single value of the most probable ratio $\frac{\Delta}{T}$. So, when the trajectory involves identical excursions, their appearance manifests rather periodic behavior which however, inevitably introduces long-range correlations.

(ii) $P(\Delta, T)$ has gently sloping maximum. Then, the relation between $\Delta$ and $T$ behaves as multi-value function: range of nearly equiprobable but different values of $T$ corresponds to the same $\Delta$. Consequently, the identical excursions are embedded in the time intervals whose durations are randomly chosen among all equiprobable that correspond to their duration. In turn, the variability of the embedding time intervals induces stochasticity that breaks any long-range correlations along the trajectory.

The establishing of the shape of $P(\Delta, T)$ requires explicit knowledge about the shape of $\nu\left(\frac{\Delta}{T}\right)$. It is worked out on the grounds of the assumption that all time scales contribute uniformly to the properties of the BIS. Let me recall that we already have established in &1.3 that this requirement sets $\nu\left(\frac{\Delta}{T}\right)$ to be a linear function of its argument:

$$\nu\left(\frac{\Delta}{T}\right) = \left(1 - \frac{\Delta}{T}\right). \tag{3.5}$$

The plot of $P(\Delta, T)$ with the above shape of $\nu\left(\frac{\Delta}{T}\right)$ is presented in Fig.1.1. It has a gently sloping maximum: indeed, the values of $P(\Delta, T)$ in the range $\frac{\Delta}{T} \in [0.25, 0.4]$ vary by less than 7%. Outside this range $P(\Delta, T)$ decays sharply. Thus, though $P(\Delta, T)$ is single-valued function, it provides nearly multi-valued relation between the most probable values of $\Delta$ and $T$, namely: certain range of nearly equiprobable values of $T$ is associated with each $\Delta$. In the course of the time the multi-valued relation is exerted as random choice of the duration of the "embedding" intervals. Thus, the execution of the multi-valuedness prevents the formation of any long-range correlations even when the sequence comprises identical excursions.

### 3.4.2. Relation $A \leftrightarrow \Delta$. Symmetric Random Walk as Global Attractor for the Fractal Brownian Motion. Finite Velocity

The incremental boundedness gives rise to the anticipation that there is a general relation between the amplitude of each excursion and its duration. The relation amplitude $\leftrightarrow$ duration determines not only a property of an excursion but the "velocity" of the "motion" on that excursion as well. In turn, the latter sets the velocity of the motion in the state space.

It has already been established that each state space trajectory can be considered as a fractal Brownian walk. The latter provides the following general relation between the amplitude $A$ and the duration $\Delta$ of an excursion, namely: $\sqrt{\langle A^2 \rangle} \propto \Delta^{\beta(\Delta)}$, where $\beta(\Delta)$ is set on the particularity of the transition rate statistics; the averaging is over the sample realizations. The dependence of $\beta$ on $\Delta$ comes from the interplay of the finite radius of the correlations $a$ and the amplitude of the excursion itself that is limited only by the thresholds of stability.

Because of the finite memory size, on coarse-grained scale any fractal Brownian walk can be considered as a symmetric random walk of "blobs" created by subwalks whose size is counterpart of the memory size. Indeed, the finite size of the memory renders the blob creating subwalks to have finite length $m$; the particularities of the transition rate statistics determines the exponent $\rho$ so that that the $\sqrt{m.s.d.}$ (where $m.s.d.$ stands for mean square deviation) of the blob creating subwalks equals $m^\rho$. Then, the large excursions are approximated by symmetric random walk with constant step equal to the blob size. Thus, the dependence of any large scale excursions on its duration reads:

$$\sqrt{\langle A^2 \rangle} \propto N^{0.5} m^\rho \qquad (3.6)$$

where $N$ is the number of the blobs.
It is obvious that when $N \gg m$ the dependence tends to:

$$\sqrt{\langle A^2 \rangle} \propto N^{0.5} a \qquad (3.7)$$

where $a$ is considered constant independent of $N$. So, the symmetric random walk with constant step appears as global attractor for any fractal Brownian motion regardless to whether it is super- or subdiffusional. Note, however, that this is valid only when the fractal Brownian walk is subject to the boundedness and does not hold for arbitrary walks!

Let me now focus your attention on the limitations that the dynamical boundedness imposes on the relation between the size and the duration of each excursion - the finite rate of development of every excursion requires diffeomorfism between them, i.e. each excursion of a finite size must have finite duration so that the rate of development to stay bounded in the prescribed range. This is automatically provided only for non-zero but finite values of the exponent $\rho$. To make it clear, suppose $\rho = 0$. It makes the blob size insensitive to the short-range statistics which contradicts the major assumption that the blobs are fractal Brownian walks set on the short range statistics. Let us now suppose the other extreme: $\rho \approx \infty$; it immediately turns the blob size to infinity regardless to the details of the transition rate statistics. Note that non-zero but finite values of $\rho$ ensure non-zero but finite value of $\beta(\Delta)$. The velocity of the motion in the state space reads:

$$\frac{dA}{dt} = \frac{d(\Delta^{\beta(\Delta)})}{d\Delta} \frac{d\Delta}{dt} \propto \beta(\Delta) \Delta^{\beta(\Delta)-1} \rho m^{\rho-1} \qquad (3.8)$$

Obviously, every finite and non-zero combination $(\beta(\Delta), \rho)$ provides not only finite velocity but sets it non-zero as well. Thus, the motion in the state space is permanent but is executed with finite velocity. Let me recall that this result is in sharp contrast with the traditional statistical mechanics where the velocity is arbitrary. The dilemma whether the velocity is finite or arbitrary is straightforwardly related with the issue of the dynamical boundedness. Indeed, an arbitrary velocity implies involving unlimited amount of energy/matter in a transition and its spreading through space and time with arbitrary velocity. On the contrary, bounded velocity implies involvement of a limited amount of energy/matter in each transition and its spreading through space and time with finite velocity.

An important outcome of the above considerations is that the finite size of the "blobs" ensures uniform convergence of the average to the expectation value of the original trajectory. Indeed, the distinctive property of any fractal Brownian walk is that any exponent $\rho \neq 0.5$ arises from arbitrary correlation between the current increment $\varsigma_i$ and the corresponding step $\tau_i$. Then the average $\overline{A}$ reads:

$$\overline{A} = \sum_{i=1}^{N} \varsigma_i(\tau_i)\tau_i = \sum_{i=1}^{N} (-1)^{\gamma_i i} \tau_i^{\rho_i} \tag{3.9}$$

and correspondingly the *m.s.d.*:

$$\langle A^2 \rangle \propto \left\langle \sum_{i=1}^{N} (\varsigma_i(\tau_i)\tau_i)^2 \right\rangle = \left\langle \sum_{i=1}^{N} \tau_i^{2\rho_i} \right\rangle \tag{3.10}$$

where the averaging is over the different samples of the trajectory. The property of the above relations is that whenever the probabilities for $\gamma_i$ to be odd and even are not permanently equal; there is correlation between the increment and the corresponding step. So $\overline{A}$ is certainly non-zero that immediately makes that the deviation from the expectation value non-zero. Moreover, eq.(3.9) yields that $\overline{A}$ can become arbitrarily large on increasing $N$. On the contrary, permanent equal probability for $\gamma_i$ odd and even means independence from one another of the increments and the steps. It yields $\overline{A} = 0$ which guarantees the uniform convergence of the average to the expectation value. Hence, on a coarse-grained scale and when (3.7) holds, $\overline{A} = 0$ always holds. In turn, it confirms once again that the coarse-grained state space trajectories are BIS – they share the property of BIS already established in &1.2. Moreover, the uniform convergence of the average to the expectation value makes the excursion sequence a homogeneous process. I shall utilize this property in the next subsection.

### *3.4.3. Distribution of the Excursions. Invariant Measure*

A major property of every BIS is the existence of expectation value and variance regardless to the details of the transition rate statistics. Then, every BIS is subject to the Central Limit Theorem which implies that the amplitudes of the excursions are normally distributed. This is very important result because it provides existence of a unique invariant measure of the state space. It should be stressed that the universality is available for the coarse-grained structure only. The short-range one remains subject to the transition rate statistics. Thus the state space manifest remarkable duality individualism-universality: the specific properties are set on the transition rate statistics and are revealed on short-range scale while the large-scaled ones are universal and insensitive to it. It is to the point to stress that the dualism individualism-universality is brought about by the multi-valuedness of the transition rate execution which breaks any long-range correlations and bounds the memory size. Note that the uniqueness of the invariant measure explicates the ubiquity of the normal distribution that is generic for a broad spectrum of systems of different nature. In addition, it comes as sequel of the boundedness alone and thus avoids the controversy, discussed in &2.4, between the normal and Poisonian distribution that takes place in the thermodynamics.

Another major consequence is that the excursion sequence is a homogeneous process. Indeed, the boundedness and the finite-size memory render uniform convergence of the average to the expectation value of every BIS as it has been established in the previous subsection. In turn, it provides the homogeneity of the excursion sequence. Then, the frequency of occurrence of an excursion of size $A$ is time-independent and reads:

$$P(A) = cA^{1/\beta(A)} \frac{\exp(-A^2/\sigma^2)}{\sigma} \tag{3.11}$$

The required probability $P(A)$ is given by the duration $\Delta = A^{1/\beta(A)}$ of an excursion of amplitude $A$ weighted by the probability for appearance of excursion of that size (normal distribution). $\sigma$ is the variance of the BIS; $c = \dfrac{1}{\sigma^{1/\beta(\sigma)}}$ is the normalizing term. The homogeneity of the excursion appearance ensures that $P(A)$ has the same value at every point of the sequence.

The stationarity of the excursion occurrence renders the set of excursion sequence dense set of periodic orbits. Moreover, it provides the transitivity of that set as well: starting anywhere in the attractor every sequence of excursions reaches every other point in it. Some authors [3.4-3.5] list these properties as definition of the chaos. Here they appear as a result of the incremental boundedness. It is worth noting that the chaotic properties of the BIS established in Chapter 1 are derived under the *apriori* assumed static and dynamical boundedness along with the lack of long-range physical correlations among time scales. Let me recall that the combination of these assumptions is equivalent to the incremental boundedness. It supports the suggestion about the paramount role of the boundedness at defining the chaoticity.

The behavior of any state space trajectory is inherently related to the transition rate statistics trough the explicit dependence of $P(A)$ on $\beta(A)$. However, as it comes out from (3.6)-(3.7), the increase of the amplitude of excursions, turns $\beta(A)$ closer and closer to $0.5$. Then, $P(A)$ gradually gets insensitive to the details of the transition rate statistics. Thus, whenever $A \gg a$, the behavior of the excursions becomes totally insensitive to it. Then the symmetric random walk appears as global attractor for the fractal Brownian walk regardless to whether it is super- or sub-diffusional. Note, however, that this is true only for bounded fractal Brownian walks, not for arbitrary one!

Thus, so far we have established that the state space is a dense transitive set of periodic orbits. In addition, the trajectories in the state space are BIS. Hence the state space exhibits all chaotic properties established so far. Moreover, on the coarse-grained level these properties are universal and independent on the particularities of the transition rate statistics. Moreover, it has an invariant measure that is given by the normal distribution. Now I shall establish more: I shall derive the condition for the asymptotic stability of the invariant measure.

### 3.5. Lyapunov Exponent of the "U-turns"

It is obvious that the issue about the asymptotic stability of the invariant measure (the normal distribution of the amplitude of excursions) is straightforwardly related to the matter of the "U-turns" defined in Chapter 1. Obviously, the invariant measure is asymptotically stable if and only if the system makes "U-turn" at the approach to the thresholds of stability so that not to involve any specific physical process and/or additional matter/energy. Our task is to derive functional relation among the parameters of a BIS that provides its asymptotic stability.

The boundedness as the fundament for the state space chaoticity suggests strong parallel with the low-dimensional deterministic chaos - phenomenon that occurs at the dynamics of simple deterministic systems. It is associated with unpredictability and great sensitivity to the initial conditions introduced by the stretching and folding. However, the deterministic chaos also exhibits boundedness: the folding is provided by the fact that the dynamics of the discussed systems is confined to a finite volume of the state space. Along with it, the stretching happens along the unstable directories and gives rise to the unpredictability.

My attention is particularly focused on the folding because it sustains the evolution of a chaotic system to be permanently confined in a finite attractor. Intuitively it looks like that the folding in a chaotic state space is automatically ensured by the thresholds of stability. However, one may argue that the particularities of the boundary conditions make the folding sensitive to them and hence not universal. The question now becomes to present crucial arguments that the folding is indeed insensitive to the details of the boundary conditions. I consider this problem along with the issue about the folding viewed as a necessary condition for keeping the evolution permanently confined to a bounded attractor. From this point of view, the folding is to be associated with the largest excursions, namely those whose amplitude is of the order of the thresholds of stability.

Below I shall find out that the target folding does exist whenever certain relation among the thresholds of stability, a parameter set on the short-range statistics and the variance of a BIS holds. The derivation of that relation involves characteristics of the excursions established in the previous section.

From the viewpoint of the deterministic chaos the folding is to be associated with a negative value of the Lyapunov exponent. Being measure of unpredictability, the latter is the average measure how fast a trajectory deviates under infinitesimally small perturbation of the initial conditions. On the other hand, from the point of view of a BIS, it is to be associated with the largest excursions, namely those whose amplitude is of the order of the thresholds of stability. Thus, our first task is to define the Lyapunov exponent in terms of the excursions and to show explicitly the dependence of its value and sign on their characteristics.

Evidently, the universality of the folding requires that it does not involve any special physical process in its execution, "U-turns" included. In other words, its implementation must neither involve nor introduce any long-range correlations among the time scales. It has been already established that every coarse-grained BIS has 3 specific parameters: the threshold of stability $A_{tr}$, the variance $\sigma$ and the power $\beta(\Delta)$ in the relation amplitude$\leftrightarrow$duration of the excursions. Next I shall prove that when certain relation among these parameters holds, the target folding exists. Note once again that the power $\beta(\Delta)$ in the relation amplitude$\leftrightarrow$duration is set on the short-range statistics. This justifies our expectation that the target folding neither introduces physical correlations among time scales nor requires additional physical process for the execution of the "U-turns".

From the viewpoint of the deterministic chaos the folding is associated with a negative value of the Lyapunov exponent whose rigorous definition reads:

$$\xi = \lim_{t \to \infty} \frac{1}{t} \ln |U(t)| \tag{3.12a}$$

where

$$U(t) = X(t) - X^*(t) \tag{3.12b}$$

$X^*(t)$ is an unperturbed trajectory and $U(t)$ is the average deviation from it. So $|U(t)|$ is the measure of all available deviations from given point $X^*$.

On the other hand, from the point of view of BIS, the Lyapunov exponent is to be associated with the excursions since they give rise to essential deviations from the expectation value.

Now I am ready to write down explicitly the asymptotic expression for the average deviation from a trajectory that starts at $X^*$. The corresponding $|U(t)|$ set on the terms of the excursions reads:

$$|U(t)| = \int_{A^*}^{A_{tr}} AP(A)dA + \int_{A_{cgr}}^{A^*} AP(A)dA \tag{3.12c}$$

$A_{cgr}$ is the level of coarse-graining, i.e. averaging over all scales smaller than $A_{cgr}$. This "smoothes out" all the excursions whose size is smaller than $A_{cgr}$ and renders their contribution to the Laypunov exponent zero. Thus, by scanning the value of $A_{cgr}$ we can study only the contribution of the excursions whose amplitude exceed $A_{cgr}$.

The separation into two terms each of which represents the deviations from $A^*$ to larger and smaller amplitudes is formal. It is made only to elucidate the idea that starting at any point of the attractor one can reach every other through a sequence of excursions. Hence the Lyapunov exponent $\xi$ reads:

$$\xi = \ln \int_{A_{cgr}}^{A_{tr}} AP(A)dA \tag{3.13}$$

At this point we come to the same result as the Oseledec theorem [3.4], namely: the value of the Lyapunov exponent for the chaotic systems does not depend on the initial point in the state space.

It should be stressed that the properties established in the previous sections are derived under the condition that all scales larger than the blob size $a$ contribute uniformly to the stochastic properties of a BIS. Then the chaotic properties also do not involve any specific scale larger than the blob size. However, it seems that it brings contradiction: how the scale-free process of excursion sequence interferes with the boundary conditions imposed by the presence of the thresholds of stability. The contradiction is solved by the presence of folding that makes the approach to the boundary a "U-turn". Thus, the folding being a necessary condition for keeping the evolution of a BIS permanently confined in a finite attractor ensures that the chaoticity produced by the stretching and folding is a scale-free process.

It is to be expected that the size of an excursion determines its contribution to the stretching or folding of a trajectory. Indeed, since the frequency of the small size excursions is essentially high, figuratively speaking they "hold" every trajectory permanently deviated from the expectation value. So, the small size excursions most probably contribute to the stretching of the trajectories. On the contrary, the largest excursions are rather occasional and the corresponding trajectory spends most of its time as close as possible to the expectation value. So, they contribute rather to the folding. The explicit revealing of the role of small and large excursions is made by the use of coarse-graining: the role of the excursion size is carried out by scanning the ratio $A_{cgr}/\sigma$. The ratio $A_{cgr}/\sigma$ has two extreme cases:

(i) $\dfrac{A_{cgr}}{\sigma} \ll 1$, i.e. the contribution of the small excursions prevails. By the use of the steepest descent method, eq.(3.13) yields:

$$\xi \approx \ln \sigma \qquad (3.14)$$

Eq.(3.14) tells that asymptotically every trajectory visits every point in the attractor so that the mean deviation from the initial point is the same for every trajectory and is bounded by the thresholds of the attractor itself. The positive value of $\xi$ justifies our speculation that the small size excursions contribute predominantly to the stretching. Further, visiting of every point of the attractor starting anywhere in it makes the motion on the attractor ergodic.

(ii) $A_{cgr} \gg \sigma$, i.e. large scale excursions contribution prevails. Eq.(3.13) yields:

$$\xi \approx \left(\dfrac{1}{\beta(A_{cgr})}+1\right) \ln \dfrac{A_{cgr}}{\sigma} + \ln \sigma - \dfrac{A_{cgr}^2}{\sigma^2} \qquad (3.15)$$

While $\xi$ from eq.(3.14) is always positive which provides stretching, eq.(3.15) opens the alternative for $\xi$ being both positive or negative depending on the relation among $\beta(A_{cgr})$, $\sigma$ and $A_{cgr}$. Then, since its natural measure is the negative value of the Lyapunov exponent, the folding that permanently keeps the evolution bounded in a finite attractor is provided if and only if $A_{tr}$, $\beta(A_{tr})$ and $\sigma$ are such that $\xi < 0$:

$$\xi \approx \left(\dfrac{1}{\beta(A_{tr})}+1\right) \ln \dfrac{A_{tr}}{\sigma} + \ln \sigma - \dfrac{A_{tr}^2}{\sigma^2} < 0 \qquad (3.16)$$

I should immediately admit that eq.(3.16) makes the realization of the "U-turns" automatic. This renders the folding insensitive to the details of the thresholds of stability and the way they are approached. That is why I call the condition (3.16) Lyapunov exponent of the "U-turns".

It should be stressed once again that the power $\beta(A)$ in (3.16) is set on the short-range statistics. Then, the "U-turns" neither require nor introduce long-range physical correlations among the time scales. This result agrees with the fundamental assumption about the uniform contribution of all time scales. In the present context it implies lack of any physical process, "U-turns" included, that yields long-range physical correlations among the time scales.

It is worth noting that the folding is broader notion than the tangent approach to the boundary. Both folding and the tangent approach produce the same effect: they contribute to the convergence of a trajectory making it to depart from the threshold. Yet, the tangent approach itself is a property of the random walk that creates the excursions entailed with the appropriate boundary conditions, while the folding is provided by eq.(3.16) without any implication of the particularities of the boundaries.

### 3.6 Numerical Simulation of the Dynamical Systems

The considerations in the previous section are based on the strong parallel between the low dimensional chaos and the boundedness that introduces stretching and folding in the state space. In order to elucidate better the above results it is to the point now to consider the chaos associated with the dynamical systems of the type:

$$\frac{d\vec{x}}{dt} = \vec{\alpha}\hat{A}(\vec{x}) - \vec{\beta}\hat{R}(\vec{x}) \tag{3.17}$$

where $\vec{\alpha}$ and $\vec{\beta}$ are control parameters. Eqs.(3.17) are a system of ordinary differential equations that fulfill the Lipshitz conditions. Then its solution should be smooth line whose course is predetermined by the initial conditions in an arbitrarily long span both back and forward in time. However, it turns out that numerical solution of some dynamical systems behaves differently at certain values of the control parameters $\vec{\alpha}$ and $\vec{\beta}$: it is highly irregular and exhibits strong sensitivity to the initial conditions. It has been established that this behavior arises around unstable solution(s) where the dynamical system has at least one positive Lyapunov exponent. But what drives the solution to behave rather as a solution of stochastic then of deterministic equations? It is to the point to mention that the name deterministic chaos comes namely from the following vicious circle: though eqs.(3.17) are deterministic, i.e. they do not involve any stochastic functions, their solution behaves as a stochastic function that has chaotic properties. It has been established that its power spectrum is a continuous band; its embedding dimension is finite and fractal. However, the great mystery remains: why the solution behaves as an irregular function?

To reveal the mystery let us remind that the numerical simulation involves inevitable round-off at every step. When the solution is stable (and the Lyapunov exponent is negative) this round-off is negligible. However, when the Lyapunov exponent is positive, the round-off is amplified and rapidly becomes compatible to the values of the smooth part. Further, it appears as multi-valued function since the round-off around $0.5$ is executed as a random choice among two selections: the round-off to the lower value and the round-off to the higher value. Then, the amplification of the round-off causes "transformation" of the deterministic equations (3.17) into the following stochastic ones:

$$\frac{d\vec{x}}{dt} = \vec{\alpha}\hat{A}(\vec{x}) - \vec{\beta}\hat{R}(\vec{x}) + \vec{\alpha}\hat{\eta}_{ai}(\vec{x}) + \vec{\beta}\hat{\eta}_{rj}(\vec{x}) \tag{3.18}$$

where $\hat{\eta}_{ai}(\vec{x})$ and $\hat{\eta}_{rj}(\vec{x})$ are the stochastic terms that come from the amplification of the round-off. The only thing necessary for the complete equivalence to the eqs.(2.15) is their boundedness. It is not a problem since it is well established that solution of the chaotic dynamical systems is confined to the so called strange attractor that has finite volume. Thus, indeed, (3.18) is completely equivalent to (2.15).

Yet, there is a difference! Eqs.(2.15) have stochastic terms that persist at every value of the control parameters, while the stochastic terms in (3.18) are introduced by the numerical simulation and are significant only at certain values of the control parameters, i.e. where the solution is unstable. This difference can be traced in the power spectra where continuous band of shape $1/f^{\alpha(f)}$ (chaotic property) appears only at certain values of the control parameters for the simulated dynamical systems while it persists at every value of the control parameters for eqs.(2.15).

### 3.7 Chemical potential

Turning point in theory developed in the present book is the definition of the chemical potential so that to involve the functional relations among the entities and the factors which provide long-term stability. The reason is twofold: on the one hand the interacting systems in the reality are entangled by the processes that proceed among them. Hence, sharp separation between a system and environment is impossible which makes important to define the chemical potential so that to involve the functional relations created in process of the interaction. Furthermore, since the functional relations are not constant in the course of time, their modifications changes the chemical potential and can make it to turn to zero. The latter, however, implies collapse of the system. This calls for explicit relation between the chemical potential and the stability of a system. On the other hand, the traditional definition of the chemical potential does not consider stability under associating/dissociating entities which makes possible associating/dissociating of arbitrary number of entities without any effect on the stability of the system. In order to involve the stability I define the chemical potential through the properties of the state space:

$$\mu_i = -\Omega \frac{\delta L}{\delta n_i} \quad (3.19)$$

where $\Omega$ is the volume of the system; $n_i$ is the number of the entities of the $i-th$ sort; $L$ is the Langrangian. Eq.(3.19) describes a process initiated by small deviations from the general equilibrium condition $\delta S = 0$, where $S$ is the action. The advantage of this definition is that it is not related to the thermodynamics but has deeper origin, namely: since the functional relations in a many-body system are built-in in the state space, eq.(3.19) is an explicit expression for taking into account the intrinsic "relations" among the constituting entities as well as their modification under the interaction of the system with its environment. By means of the new definition I shall demonstrate that a system stays stable if and only if the amount of energy/matter that it exchanges with the environment is bounded.

Since the action and the Lagrangian are explicitly related to the motion in the state space, it is to be expected that the chemical potential is also explicitly related to the properties of the state space and in sequel to the stability.

The further considerations are grounded on the explicit relation between the notion of chemical potential and stability of the system. Indeed, the chemical potential must be measure how strong the stability of the system is "affected" by association or dissociation of an entity. To compare, the thermodynamics leaves the chemical potential intact during the interaction. In turn, this allows a system to associate/dissociate arbitrary number of entities without any affect on its stability. In the previous section it has been found out that the chaoticity of the state space is interrelated with the stability of the system - whenever certain relation among 3 general characteristics of the chaotic motion holds, the system is asymptotically stable (eq.(3.16)). The question now is how the stability is related to the problem about associating and dissociating of entites.

The general condition for breaking the stability of a system is turning its chemical potential to zero. Let us now come back to the structure of the chaotic state space: in general it can be separated to a "bulk" and a "surface" part. The former one is associated with the core of the state space where (3.16) holds in any direction. Therefore, the chemical potential of the "bulk" part is not affected by the number of exchanging entities; it is rather to be associated with that part of the system that remains stable under the interaction. On the contrary, the "surface" part is associated with the current boundary of the chaotic state space; in other words it is to be associated with the interaction with the environment. The corresponding part of the chemical potential is strongly sensitive to the current curvature that in turn strongly depends on the current number of exchanged entities. Now I shall present an example how the total chemical potential turns to zero under the exchange of finite number of entities. Since any exchange of an entity modifies the shape of state space "surface", the natural measure of the surface part of the chemical potential $\mu_s$ is its local curvature:

$$\mu_s = \int_S \alpha k ds \quad (3.20)$$

where $\alpha$ is the density of the surface energy; $S$ is the area of the state space surface. The permanent variations of the number of entities result in permanent modification of the value and sign of the local curvature. So, $\mu_s$ permanently varies and eventually turns the total chemical potential $\mu_{tot}$ to zero. This immediately yields the system falling apart. The next task is to illustrate that this happens at finite values of the state space variables, i.e. when the number of exchanged entities is limited. According to the above considerations the destruction of the system happens whenever:

$$\mu_{tot} = 0 \quad (3.21)$$

Since the bulk part of the chemical potential is insensitive to the variations of the state variables, eq.(3.21) holds whenever the following relation holds:

$$\mu_s = -\mu_b \quad (3.22)$$

where $\mu_b$ is considered *apriori* set constant.

Hereafter my task is to illustrate that the present definition of the chemical potential indeed involves functional relations that appear in any process. Let us consider a chemical reaction that selects two relevant sorts of entities: their numbers are denoted by *x* and *y* correspondingly. The state space is two-dimensional and its "surface" can be parameterized as follows:

$$x = r^{a(\theta)} \cos\theta$$
$$y = r^{b(\theta)} \sin\theta \qquad (3.23)$$

where the powers $a(\theta)$ and $b(\theta)$ comprise the permanent change of the local curvature through the dependence $\theta = \theta(t)$. Then the current local curvature $k$ reads:

$$k = \frac{\left(\dot{x}^2 + \dot{y}^2\right)^{3/2}}{\dot{x}\ddot{y} - \dot{y}\ddot{x}} \qquad (3.24)$$

where the derivation is with respect to time.
Simple algebraic calculations yield:

$$k = \frac{1}{r^{2a-b}} \frac{1 - A - \dfrac{B}{r}}{\left(\sin^2\theta + r^{2(b-a)}\cos^2\theta\right)^{3/2}} \qquad (3.25)$$

where:
$$A = \dot{a}a\dot{b}^2 b(b-1)\sin^2\theta - \dot{b}b\dot{a}^2 a(a-1)\cos^2\theta \qquad (3.26)$$
$$B = \left(\dot{b}^2 + b\ddot{b}\right)\sin^2\theta + \left(\dot{a}^2 + a\ddot{a}\right)\cos^2\theta \qquad (3.27)$$

Since $a$, $b$ and their derivatives permanently vary, due time course eqs.(3.25)-(3.27) select finite $\vec{r}_{cr} = (x_{cr}, y_{cr})$ so that eqs.(3.21)-(3.22) are satisfied. The values of $a$, $b$, their derivatives and $\mu_b$ are set on the particularities of the reaction kinetics. Evidently, the system stays stable until its state variables remain bounded so that:

$$x < x_{cr}$$
$$y < y_{cr} \qquad (3.28)$$

where I called $\vec{r}_{cr} = (x_{cr}, y_{cr})$ kinetic threshold of stability. It should be stressed that (3.28) is to certain extend analogical to eq.(3.16) though not equivalent. The kinetic threshold of stability selects the largest possible size so that the system in question remains stable under the exchange of matter/energy with the environment. Then eq.(3.28) can be viewed as necessary condition for a system to stay permanently stable in the process of an interaction. Besides, since the fluctuations develop also through exchange with the environment, it can be viewed also as necessary condition for limiting their spatial size, amplitude and lifetime. On reaching the critical size, all fluctuations that initially have smaller size are destroyed; those of larger size cannot be created at all. In sequel, we answer the important question about the uniformity of the limit (2.1) - the finite size and finite lifetime of the fluctuations ensures that the limit (2.1) holds for every stable system. In turn, the uniformity of the limit (2.1) makes the notion of intensive state variable plausible for the stable systems. Consecutively, this justifies the setting the evolutionary equations in terms of state variables appropriate for the stable systems.

Besides, now we are able to explain why only small size fluctuations with finite lifetime are developed at the Brownian motion. Let me remind the considerations in &2.3 about the pollen random walks under the "sources" whose origin is the discrete structure of the solvent - these random "sources", viewed as fluctuations, should have finite size and finite lifetime so that to provide coherent "efforts" of the water molecules in order to move randomly and independently from one another the "gigantic" pollens. So, the coherent "efforts" are provided by those fluctuations whose size does not exceed the kinetic threshold of stability; to remind that it is impossible to create larger ones. The finite lifetime appears in the course of the interaction with either the pollens or the other fluctuations: sooner or later the number of exchanged molecules will exceed the kinetic threshold of stability which will result in the collapse of the corresponding fluctuation.

It should be stressed that the spatial boundedness of fluctuations is insensitive to the dimensionality of the system because it is set on dimensionality of the state space which is not proportional to the spatial dimension!

### 3.8 Efficiency of Nanoparticles in Catalysis

The ubiquity of the macroscopic fluctuations is a very dramatic result of our considerations. In particular, from the viewpoint of the technology. Usually, the industry reactors are constructed so that reaction to proceed at steady stable regime. However, the development of fluctuations seriously bewilders the situation. Indeed, fluctuations of large enough size could give rise to undesirable effects such as phase transitions, change of the dynamical regime etc. Hence an indispensable part of the reactor construction must be the development of appropriate feedback so that to prevent those fluctuations. However, this matter is a complicate task that requires a lot of efforts and elaborate mind. That is why the question is whether there is another way out.

Yes, there is another way out and it is straightforwardly related to the interplay of the kinetic threshold of stability and modern nanotechnology. One of the aims of the nanotechnology is deep miniaturization so that to have more effective use at less costs. Applied to the catalysis it implies that the catalyst, an expensive material, is deposited on the support as nanosize clusters whose concentration is so little that they can be considered isolated. It turns out that if the nanoparticles do not agglomerate, such catalysts are very stable. But how about the macroscopic fluctuations? Hereafter I shall outline that to the most surprise, such systems behave steadily and do not exhibit macroscopic fluctuations.

The kinetics of the reaction that proceeds on each nanoparticle is very complicated because of the complex effects of: the boundary, possible fractal dimension etc. However, we certainly know that in general it is described by equations of the type (2.15). Hence, the products that come out from each nanoparticle permanently fluctuate. However, whenever the size of nanoparticles exceeds the kinetic threshold of stability, each of them operates steadily and permanently in the course of time. In addition, the large distance between the nanonparticles, makes them isolated one form another. So, the nanonparticles contribute to the efficiency of the catalyst independently from one another. Yet, in the course of time the sum of huge number of random numbers does not signal out any significant fluctuation. Thus, we come to surprising and highly non-trivial result: the kinetics whose outcome are permanent fluctuations brings about a steady non-fluctuating regime!

Let us now consider the important issue why the efficiency of the nanoparticles as catalysts is so sensitive to their size and to the external constraints. Indeed, it is well known that the nanoparticles are efficient only when their size is confined to certain specific to the system range. The general condition for stable proceeding of a reaction is provided whenever the "kinetic" threshold of stability is smaller than the "spatial" size. However, the value of the kinetic threshold of stability is highly sensitive to the values of the external constraints such as temperature, pressure of the reactants in the gas phase etc. through their explicit involvement in eqs.(2.15). Therefore the ratio between the "spatial" and the "kinetic" size as a function of the external constraints yields the range of efficiency of the nanoparticles. Evidently, this range depends strongly on the values of the external constraints.

NOTE: the requirement that the spatial size of the nanoparticles must exceed the kinetic threshold of stability imposes general constraint to further miniaturization of the nanoparticles!

### 3.9 What Comes Next

The goal of the present chapter is to elucidate how far the fundaments of the statistical mechanics are affected by involving of the boundedness as the most general factor that renders the long-term stability.

An immediate consequence of the boundedness is that the invariant measure of the state space is the normal distribution. The central result is that the boundedness makes the normal distribution ubiquitous. The condition that guarantees the asymptotic stability of the invariant measure (i.e. the normal distribution) reveals interrelation between the boundedness as necessary condition for the state space to have chaotic properties and the presence of stretching and folding for keeping the evolution permanently confined to a finite attractor. From this viewpoint a permanently stable evolution is possible if and only if the folding does not "feel" the boundary conditions imposed by the presence of the thresholds of stability. Otherwise, the evolution strongly depends on the way how the excursions approach the threshold. Hence, the asymptotic stability of the invariant measure is strongly related to the execution of "U-turns" without involving any special physical process. We have demonstrated that when certain relation (eq.(3.16)) among the major characteristics of a BIS holds, a folding with the desired properties exists. As a result, the presence of the thresholds is imperceptible that ensures all scales, thresholds included, uniformly to participate in setting the chaoticity of the state space. In turn, the uniform participation of all scales ensures permanent confinement of the evolution to a finite attractor.

It is to the point to stress that the association of the stretching and folding with the boundedness makes it universal tool for bringing about the chaoticity. Remember, that so far it has been associated with the chaos of the low-dimensional chaotic dynamical systems only. However, here the chaoticity emerges without any reference to the properties of these systems. I have used only the incremental boundedness. Thus, we essentially expand the idea of stretching and folding. Moreover, we strongly associated it with the matter of the long-term stability though eq.(3.16). Consequently, on meeting eq.(3.16) the state space subject to boundedness appears as chaotic global attractor whose invariant measure is the normal distribution. The universality along with the stationarity makes the attractor transitive dense set of periodic orbits insensitive to the particularities of the system. Thus, the structure of the attractor has steady properties insensitive to the development of any specific trajectory and/or the way of approaching the boundaries.

The chaoticity of the state space gives rise to remarkable new property: a system remains stable only under exchange of limited amount of matter/energy with the environment. This result sharply interferes with the traditional statistical mechanics where noting prevents associating/dissociating of arbitrary number of entities. Moreover, being sensitive to the current curvature of the state space boundary, the total chemical potential is not constant during interaction. This result again opposes the traditional statistical mechanics where the chemical potential is supposed dependent only on the chemical identity.

A very important outcome of the properties of the state space is the proof that the chaotic properties provide stability through making the fluctuations limited both in their spatial size and amplitude. Thus, my theory seems to be built on stable grounds. However, in the present considerations I have taken for granted the assumption that the joint action of the static, dynamical boundedness and the spatial coherence makes the systems to exert macroscopic fluctuations permanently. So, the self-consistency requires to derive eqs.(2.15) and to establish independently the boundedness of those fluctuations. This is not a trivial task. We will see in the next Chapter that the premise of only short-range interactions yields divergence of the transition rates of every process. And in Chapter 6 I shall prove that the introduced there coherence mechanism renders boundedness of the amplitude of the macroscopic fluctuations.

# Chapter 4: Selected of Non-Equilibrium Statistical Mechanics. Weak-coupling limit. Divergence of the Scattering Length

### 4.1 Why to Read Chapter 4

In the previous Chapter we studied the fundamental properties of a state space subject to all 3 aspects of the boundedness: static and dynamical ones and the spatial coherence. It has been found out that the coarse-grained structure of every such state space is universal. Furthermore, the boundedness viewed as a necessary condition for long-term stability and the integration of the functional relations among entities into the state space bring about the fundamentally new result that every system (fluctuation) stays stable if and only if the exchanged with the environment number of entities is bounded. This immediately imposes limitation on growing of the system (fluctuation) to arbitrary size under its interaction with the environment. It should be stressed that this result drastically interferes with the traditional thermodynamics where nothing prevents growing of a system (fluctuation) to arbitrary size. The revealing of this dramatic contradiction needs rigorous arguments: to recall that the limitation on the size of a system is obtained on the grounds of the taken for granted validity of eqs.(2.15) as the type of equations that describe the evolution of a system subject to all 3 aspects of the boundedness. This makes my goal to be the presentation of credible arguments in favor of the validity of eqs.(2.15). Since the static and the dynamical boundedness has been already explicitly involved in the construction of the state space, the major step in this direction is to consider in details the mechanism that brings about spatial coherence. The building of a successful approach requires answers to the following top questions:

(I) it is well known that the interactions in the many-body systems are short-range. Then the fluctuations develop under local rules and therefore they remain permanently uncorrelated both in space and time. Then, what "interactions" run long-range coupling among them?

(ii) it is well known that the short-range interactions are highly specific to a system. Then, is the spatial coherence also specific or it proceeds under universal operational protocol? If latter, how the specific short-range interactions contribute to it?

(iii) what is the role of the stochistising interactions in operating the long-range coupling?

It is to the point to mention that the above questions prompt the paramount role of the dynamics in setting the macroscopic behavior of the many-body systems. This is an entirely new property that has no analog in the traditional thermodynamics. Let me recall that the latter assumes the macroscopic behavior of any system insensitive to the details of its dynamics. The first step in building a self-consistent approach to the spatial coherence is to present decisive arguments that it needs entirely new viewpoint on processes that develop in the many-body systems. Prior to that is to make evident that the used so far approaches that account for the interactions in the many-body systems fail in this task. That is why I start with some basic ideas of the traditional non-equilibrium statistical mechanics that set the role of the dynamics in the macroscopic evolution. The relation with our task will become clear later, in &4.3.

### 4.2. Non-Equilibrium Statistical Mechanics. Selected Topics

The non-equilibrium statistical mechanics puts forth the idea that starting from any initial state a system evolves so that eventually reaches equilibrium state that has a global attractor. In the traditional statistical mechanics the only way to make this possible is to presume that the system develops through states that form a stationary Markov process. The latter is defined as stochastic process with the property that for any set of successive times one has:

$$P_{1|n-1}(x_n, t_n | x_1, t_1; \ldots; x_{n-1}, t_{n-1}) = P_{1|1}(x_n, t_n | x_{n-1}, t_{n-1}) \qquad (4.1)$$

Thus the conditional probability density at $t_n$, given the value $x_{n-1}$ at $t_{n-1}$, is uniquely determined and is not affected by any knowledge of the values at earlier times. $P_{1|1}(x_2, t_2 | x_1, t_1)$ is called transition probability.

A Markov process is fully determined by two functions $P_1(x_1,t_1)$ - the probability that at the initial moment $t_1$ the system is in the state $x_1$ and $P_{1|1}(x_2,t_2|x_1,t_1)$; then, the whole hierarchy can be reconstructed from them. Indeed, one has for instance, taking $t_1 < t_2 < t_3$,

$$P_3(x_1,t_1;x_2,t_2;x_3,t_3) = P_2(x_1,t_1;x_2,t_2)P_{1|2}(x_3,t_3|x_1,t_1;x_2,t_2) = \\ = P_1(x_1,t_1)P_{1|1}(x_3,t_3|x_2,t_2)P_{1|1}(x_2,t_2|x_1,t_1) \quad (4.2)$$

Integrating the identity (4.2) over $x_2$ one obtains for $t_1 < t_2 < t_3$

$$P_{1|1}(x_3,t_3|x_1,t_1) = \int P_{1|1}(x_3,t_3|x_2,t_2)P_{1|1}(x_2,t_2|x_1,t_1)dx_2 \quad (4.3)$$

This is called Chapman-Kolmogorov equation. It is an identity, which must be obeyed by the transition probability of any Markov process.

The statistical mechanics presumes that the equilibrium state is described by stationary Markov process. For the latter the transition probability $P_{1|1}$ does not depend on two times but only on the time interval; for this case we introduce a special notation:

$$P_{1|1}(x_2,t_2|x_1,t_1) = T_\tau(x_2,x_1) \text{ with } \tau = t_2 - t_1 \quad (4.4)$$

The Chapman-Kolmogorov equation becomes:

$$T_{\tau+\tau'}(x_2,x_1) = \int T_{\tau'}(x_3|x_2)T_\tau(x_2|x_1)dx_2 \quad (4.5)$$

But why the equilibrium is so interesting for a system out of equilibrium - because of the so called fluctuation-dissipation relation. It assumes that any small perturbation relaxes the same way as an equilibrium fluctuation. Hence the study of the equilibrium fluctuations is the basic implement for finding out the macroscopic behavior of the systems out of equilibrium.

The Chapman-Kolmogorov equation is not easy to handle in actual applications. That is why another equation, called master equation, that is more convenient version of the (4.5), is in use. It is a differential equation obtained by going to the limit of vanishing time difference $\tau'$. For this purpose it is necessary first to ascertain how $T_{\tau'}$ behaves as $\tau'$ tends to zero. It is supposed that:

$$T_{\tau'}(x_2|x_1) = (1 - a_0\tau')\delta(x_2 - x_1) + \tau'W(x_2|x_1) + o(\tau') \quad (4.6)$$

Here $W(x_2|x_1)$ is the transition probability per unit time. The coefficient $1 - a_0\tau'$ in front of the delta function is the probability that the transition takes place during $\tau'$; hence

$$a_0(x_1) = \int W(x_2|x_1) \quad (4.7)$$

Now insert this expression for $T_{\tau'}$ in the Chapman-Kolmogorov equation (4.5) and going to the limit $\tau' \to 0$ one obtains:

$$\frac{\partial}{\partial \tau}T_\tau(x_2|x_1) = \int \{W(x_2|x_1)T_\tau(x_2|x_1) - W(x_2|x_3)T_\tau(x_3|x_2)\}dx_2 \quad (4.8)$$

The differential version of the Chapman-Kolmogorov equation, valid for the transition probability of any stationary Markov process obeying (4.6), is called master equation.

It is useful to cast the equation in a more intuitive form. First note that $T_\tau(x_2|x_1)$ is identical with the distribution function $P_1(x_2)$. Hence we may write:

$$\frac{\partial P(x,t)}{\partial t} = \int \{W(x|x')P(x',t) - W(x'|x)P(x,t)\}dx' \quad (4.9)$$

This is the customary form of the master equation.

Yet, the master equation is not only a form of Chapman-Kolmogorov equation. Actually its wide use is grounded of another interpretation of (4.9), namely it is considered as a linear equation for the evolution of the probability $P(x,t)$. Further, it is supposed that there are two well separated time scales: a short time scale of the relaxation of the fast dynamical variables and the slow time scale associated with the relaxation of the state variables. The linearity of eq.(4.9) indicates also that the two time scales can be considered separately - at the expense of assuming Markovian property. As a result, $W(x|x')$ and $W(x'|x)$ are the probabilities for averaged over the *dynamical* variables rates of transition between any two states $x$ and $x'$. The averaging is completed by means of an appropriate for the system quantum-mechanical approach.

Summarizing, the wide use of the master equation is built on separation of the dynamical variables from the state ones. In turn, the evolution is driven by the dynamics included in the transition rates $W(x|x')$.

The master equation is particularly convenient to account for the impact of the spatio-temporal configurations of the entities on the transition probabilities $W(x|x')$ and $W(x'|x)$.

The general result of the linearity of the master equation is that it always yields a stationary distribution on the condition that detailed balance holds. As a result, the system in question reaches equilibrium. However, detailed balance does not always exist - remember the antithermodynamic behavior of the open systems considered in &2.1! This gives rise to the question whether the dynamics of the closed and the open systems plays different role in constituting their macroscopic behavior. My task is to demonstrate that indeed the dynamics plays different role in constituting the macroscopic behavior of closed and open systems but its appropriate taking into account calls for an entirely new approach.

### 4.3 Weak-Coupling Limit

One of the most fundamental assumptions in the physics is that the bearer of the identity of entities that constitute a system is their Hamiltonian. Furthermore, it is taken for granted that it does not change in any physical process. The explicit expression of this statement is the separation of the Hamiltonian of any interaction into two additive parts:

$$\hat{H} = \hat{H}_0 + \lambda \hat{H}_{int}(t) \tag{4.10}$$

where $\hat{H}_0$ is the Hamiltonian of a single entity: it is supposed that $\hat{H}_0$ upholds its identity; $\hat{H}_{int}(t)$ is the Hamiltonian of the interaction; $\lambda$ is supposed be measure of the intensity of interaction. The fundamental assumption is: being carrier of the identity that remains intact during the interaction, $\hat{H}_0$ is supposed time-independent. Thus, its eigenvalues are time-independent invariants that exhibit only minor changes under weak interactions, i.e. when $\lambda \ll 1$. Therefore, the eigenvalues serve as local invariants. So, it seems that this justifies the basic notions of the statistical mechanics discussed in Chapter 2?! In particular, it is well known that the above considerations justify the use of the Boltzmann distribution for the classical systems, Bose and Fermi-Dirac statistics for bosons and fermions correspondingly. In each of these cases the local invariants are presented trough the eigenvalues of $\hat{H}_0$. Then the Boltzman statistics prescribes that the probability for a system to be in the state of energy $E_n$ reads:

$$P(E_n) = \frac{\exp\left(-\frac{E_n}{kT}\right)}{\sum \exp\left(-\frac{E_n}{kT}\right)} \tag{4.11}$$

where $T$ is the temperature; $k$ is the Boltzmann constant.

The separation of the Hamiltonian according to (4.10) is the fundamental assumption of so called weak-coupling limit.

The assumption that (4.10) holds for every interaction presumes that $\hat{H}_0$ remains the same after the interaction. Hereafter I call this property rigidity of the Hamiltonian. An immediate effect of the "rigidity" is that it ensures the scattering threshold for any interaction to be specific but always non-zero. This is so because under the assumption of the time-independence of $\hat{H}_0$ every interaction is considered perturbation. In turn, the perturbations always leave the distance between the levels finite which provides the scattering threshold to be also finite. In turn, the finite scattering threshold suggests that the weak interactions, i.e. low-energy processes, have finite scattering length $a$.

Further, the properties of a system (transition probabilities in particular) can be calculated as expansions in terms of the dimensionless parameter $ak$, where $k$ is an appropriate wave-length number and $a$ is coupling constant. Generally, the scattering length is supposed comparable in magnitude to the range $r_0$ of the interaction: $a \approx |r_0|$. In this case the interaction is perturbative weak-coupling phenomenon. The scattering length $a$ plays the role of a coupling constant; $k$ takes into account the spatio-temporal configuration of the entities. Note, however, that finite values of $k$ imply that NO long-range correlations are established in the system! However, since $k$ is straightforwardly related to the notion of mean separation, its finite values strongly entangle the perturbative weak-coupling approach with the thermodynamical limit. The very idea of mean separation is related to the notion of concentration as an intensive state variable and to the thermodynamic limit. On the other hand, the use of the mean separation and/or concentration as a small parameter explicitly implies the *apriori* validity of the thermodynamical limit. So, it turns out that the concentration must be well-defined for arbitrary initial conditions and interactions!? Moreover, do not forget that this theory is supposed valid both in equilibrium and out of it! This very important issue and I shall prove in &4.5 that the perturbative weak-coupling approaches fail because they are grounded on the controversial combination of assumptions about the conservative dynamics described by (4.10) and the supposition that the thermodynamical limit holds independently of the dynamics. It should be stressed that this would be possible only if the system is subject of permanent intensive stirring. Some authors suppose that the random motion of the entities between interactions is the factor that produces the required stirring. Yet, this suggestion lacks ubiquity: in dense systems and on interfaces the motion of the entities is very slow to serve as a stirring factor.

So, the above established problems give rise to the question about the role of the stochastising interactions: can they help revealing the controversies and if so, how. The task of the following sections is to present decisive arguments that their role is completely different in closed diluted systems and open dense ones.

### 4.4. Stochastising Interactions and the Weak-Coupling Limit

Stochistising interactions are key notion in our approach. Being interactions whose outcome is multi-valued, it is to be expected that their role is crucial for the explanation and description of wide variety of phenomena. In &2.5 we have already described the fundamental idea how the multi-valuedness appears. Its major effect, the duality stochasticity-determinism, was immediately elucidated. It implies that though we can compute each selection of a multi-valued function, its realization is random choice of one selection among all available. Let us recall that we associate the stochastising interactions with $n-$body interactions $(n \geq 3)$. The particularity of these interactions compared to the two-body ones is that the entities do not enter interaction simultaneously. Let us consider a $3-$body interaction. Most probably the interaction starts with the collision between two entities. Whenever the $2-$body interaction is not completed at the arrival of the third entity, its contribution to the total Hilbert space is given by the projection of the $2-$body Hilbert space onto that subspace whose properties are determined by the moment of arrival of the third entity. The multi-valuedness arises from the independence of the arrival moment of the third entity on the stage of proceeding of the $2-$body interaction. Therefore, the contribution of the $2-$body interaction can be consider as random projection of the entire $2-$body Hilbert space onto one of its subspaces. Let us now consider how this assumption is related to the idea of the "rigidity" of a Hamiltonian.

Up to date, the $n-$body interactions have been handled by the weak-coupling limit. A general consequence of that consideration is that the final characteristics such as the energies and wave-numbers of the scattered entities turn out to be always single-valued. And this is due to the assumption that the interaction contribution is additive and the Hamiltonian $\hat{H}_0$ remains intact during the interaction. Indeed, the Hamiltonian of a $3-$body interaction in the weak-coupling limit reads:

$$\hat{H} = \hat{H}_0 + \lambda_{12}\hat{H}_{12}(t) + \lambda_{13}\hat{H}_{13}(t) + \lambda_{23}\hat{H}_{23}(t) \tag{4.12}$$

Note that the "rigidity" of the Hamiltonian $\hat{H}_0$ ignores the moment of arrival of the third entity! It is to the point to mention that the linearity of the Schrodinger equation with respect to the wave-function and the linear superposition of the interaction terms in (4.12) provides the additive contribution of all terms into the formation of the final characteristics. However, this approach fails to explain the reactions, both chemical and nuclear. Indeed, the reactions are associated with certain modification of the identity of the reacting entities which can be achieved only via certain exchange of matter/energy among interacting entities. Thus, the reaction interactions are to be inevitably associated with some dissipation. However, this means that $\hat{H}_0$ changes in one way or another during reaction!

The idea about compound two-body Hamiltonian that changes non-smoothly under the arrival of other entity is put forth to take into account explicitly the dissipation in the dynamical processes. By virtue of the idea about the compound Hamiltonian I presume that each interaction involves steps of relaxation, excitation and scattering. So, the interaction is never considered as scattering only. Actually, I suppose that the quantum processes are a result of much more complicated formation of temporary compound Hamiltonian that is "flexible" and permanently varies under the exchange of matter/energy among the interacting entities. Formal expression of the "flexibility" is the participation of only a projection of the $2-$body Hamiltonian to the total interaction. To compare, in the weak-coupling limit the entire $2-$body Hamiltonian participates to the total interaction. I adopt the idea about the compound Hamiltonian because this is the only way to avoid the total "rigidity". Actually the major difference between the weak-coupling limit and the compound Hamiltonian is that the latter is considered "flexible" under dissipation. The weak-coupling limit considers the entities absolutely rigid bodies that can only scatter but cannot dissipate. Besides, it considers $\hat{H}_0$ always "recovered" after a perturbation ceases its action! Thus, indeed no dissipation on the dynamical level is presumed in the weak-coupling limit! And no changes in the identity! And indeed the entities are supposed absolutely rigid bodies!

An immediate result of the weak-coupling limit withdrawal and the adoption of the idea about the compound Hamiltonian, is that the arrival time of the third, forth- etc. entity becomes crucial for the appearance of the multi-valuednes of the output characteristics. And this is completely new property of the many-body interactions compared to the always single-valued output given by the weak-coupling limit. The presumption about the "flexibility" of the compound Hamiltonian renders its sensitivity to the moment of arrival of the third entity. In addition, though the concrete range of selections is sensitive to the arrival moment, the multi-valuedness is universal, i.e. insensitive to the details of the interaction.

Now I am ready to elucidate the correspondence between the idea of the identity and the idea of the "flexibility" of the compound interaction Hamiltonian. The very idea of interaction makes the notion of identity to certain extend ambiguous. For example, what characterizes $NaCl$ as a molecule that consists of one atom $Na$ and one atom $Cl$? Nowadays it is established that the valent electrons of an $Na$ atom pass from it to an $Cl$ atom under the interaction. Hence, the Coulomb interaction between ionized atoms "glues" the ions into a molecule of $NaCl$. So, though the valent electrons are part of the identity of an atom, the molecule that emerges from the chemical interaction shares that property. Thus, the twofold role of the valent electrons in the identification of both the molecule and the atoms serves as justification of our idea that dissipation should be involved in the compound interaction Hamiltonian. Indeed, the pass of the valent electrons from one atom to the other is a process of matter dissipation. On the contrary, the additivity in the weak-coupling limit ((4.10)) renders the identity expressed by $\hat{H}_0$ intact whatsoever the interaction is. So, the lack of matter dissipation results in the lack of identity modification. Thus, the indispensable modification of the identity that the entities undergo in a reaction remains not taken into account. It makes the weak-coupling limit inappropriate for description of the reactions and any dynamical process that involves dissipation.

Summarizing, the above considerations elucidate the major contemplation that any interaction violates the identity of the interacting entities in one way or another. The result of the interaction may be two-fold: the first option is that the identity of the entities is restored up to minor exchange of energy only and the entities depart and behave as the same objects: this is scattering. The other option is that the entities exchange matter and form a stable compound like molecule constituted by atoms. Our interest in this subject is that in both cases the output is multi-valued if the interaction proceeds among three or more entities.

It is to the point to stress that the weak-coupling limit is a good approximation for the weak perturbations when the entities can be considered as rigid bodies without a significant mistake. However, the fundamental difference between the idea about the compound Hamiltonian and the weak-coupling limit is that the former always produces multi-valuedness of the final characteristics while the latter yields always single-valued outcome. The crucial test for the relevance of both approaches is how they help to explain the behavior of the many-body systems. We have already mentioned that the successful use of the perturbative weak-coupling approach requires *apriori* validity of the thermodynamical limit. In the next section I shall illustrate that the dynamical interactions based on weak-coupling limit do not help to restore the thermodynamical limit if it is violated by a local perturbation or a fluctuation. Thus, the immediate question is whether the stochistising interactions help to settle the problem. Can they be the driving mechanism for establishing of the thermodynamical limit? If so, at what conditions? What is they role they play in open systems? These are the issues that we will consider in the further two sections.

### 4.5. Divergence of the Scattering Length and the Thermodynamical Limit

My present task is to elucidate why the weak-coupling limit fails in recovering of the thermodynamical limit when the latter is violated by external local perturbations or local fluctuations. Let us consider a system that is a closed vessel of arbitrary size $L$. Let us suppose that a perturbation of size $l$ is developed somewhere in the volume. The perturbation consists of entities whose "density" and energies are very different from that of the rest of the volume. Further, let us suppose only short-range interactions and ballistic motion between successive collisions. Then the relaxation of a perturbation proceeds only through the inelastic collisions. The master equation approach requires uniformity of the thermodynamical limit for the separation of the time scales on which the relaxation of the dynamical and state variables proceeds. Now I pose the question: is the weak-coupling approach able to provide independent relaxation of the dynamical and state variables? The major factor of the dynamical relaxation is the cross-section of the collision the value of which is settled on the scattering length $a$. The state variables set the probability for a collision to happen. And here comes the following vicious circle: the probability for the collision is set on the supposition about the well-defined concentration of the entities everywhere in the system. This is plausible when the perturbations rapidly relax. Now I shall prove that the weak-coupling limit makes the time scale of the relaxation dependent on the size of the perturbation.

The relaxation of a local perturbation (fluctuation) proceeds through inelastic collisions. Since the scattering length in the weak-coupling limit is always finite, the relaxation takes place predominantly at the boundary between them. The rate of the relaxation is proportional to the ratio between the area of the boundary and the volume of the perturbation:

$$R \propto \frac{S}{V} \tag{4.13}$$

where $S$ is the area of the boundary of the perturbation and $V$ is its volume. Evidently, when the perturbation is a compact volume in the space:

$$R \propto l^{-1} \tag{4.14}$$

where $l$ is the size of the perturbation.

The dynamical factor that governs the rate of relaxation is the rate of the energy relaxation. Let us recall that the property of the inelastic collisions is the transformation of certain amount of the kinetic energy into internal one. Thus, the rate of relaxation is proportional to the ratio between the transformed energy to the difference in the energy of the every pair of entities one of which belongs to the perturbation and the other one to the outside volume. Finally we come to the following estimation of the relaxation rate:

$$R \propto l^{-1} \frac{\Delta E}{E_p - E_v} \tag{4.15}$$

where $\Delta E$ is the transformed energy in the collision; $E_p$ is the kinetic energy of the perturbation; $E_v$ is the kinetic energy of the entities outside the perturbation. The estimation (4.15) explicitly shows that $R$ can be made arbitrarily large through concentrating of the energy into smaller volumes. Besides, the size dependence makes the time scale of the relaxation of the dynamical variables ill-defined: it varies up to infinity depending on the size of the perturbation and the energy concentrated in it. In turn this violates the idea about the existence of two separate time scales on which the relaxation to the equilibrium proceeds: the scale on which the dynamical variables relax and a scale on which the state variables relax. Let us recall that the assumption that the local fluctuations relax on the *dynamical* level is crucial for the master equation approach because it brings about the thermodynamical limit on the macroscopic level. And the latter is necessary for the relaxation of the state variables. Actually, the master equation supposes the following succession of steps: (I) relaxation of the dynamical variables so that to validate the thermodynamical limit; (ii) the thermodynamical limit justifies the definition of the intensive state variables; (iii) the relaxation of the state variables.

So, we come to the conclusion that when the interactions are short-ranged there is no single well-defined time scale for the dynamical relaxation because the rate of relaxation of the local fluctuations is size dependent.

Outlining, the master equation does not agree with the effect of local fluctuations because their relaxation violates its fundamental requirement for automatic separation of the time scales so that the relaxation of the dynamical variables is completed prior to that of the state variables. Moreover, the dependence on the size of the fluctuation (perturbation) renders the state variables time and space dependent. Is there any way out? Can the thermodynamical limit survive? If so, what is the mechanism that provides its validation? The estimation (4.15) indicates the way out: it goes through elimination of the dependence on the size of the perturbation. But it seems absurd: the very idea of local fluctuation suggests focusing of energy/matter in a compact volume. Then the relaxation goes via scattering with the rest of the system that proceeds at the boundary of the perturbation. However, this is so if the scattering length is finite. What happens if it diverges? Then the entities from the perturbation are figuratively "shuttled" everywhere in the vessel where they rapidly relax. Another interpretation reads that the divergence of the scattering length implies high sensitivity to the interaction with distant entities. In other words, it is to certain extend equivalent to involvement of long-range interactions. Then, indeed, the divergence of the scattering length seems to provide the target elimination of the dependence on the size of the perturbation. As a result, only dynamical factors would govern the relaxation rate. And our dream would come true: there would be a single time scale at which all the local fluctuations relax! The major property would be that this time scale depends only on dynamical factors and would be insensitive to the spatial size of the fluctuations! But how it is possible to have a divergent scattering length? And if it is possible, can it indeed provide the target separation of the time scales necessary for justifying the master equation.

Usually the divergence of the scattering length is associated with *fine-tuning*. It means that there is some parameter that if tuned to a critical value would give a divergent scattering length $a \to \infty$. I shall start with revealing the role of the stochastising interactions in the activation of the fine-tuning. The multi-valuedness of the stochastising interactions makes their outcome non-uniform even in a dilute homogeneous system because different selections are established even at closest points. In turn, the established variability introduces imediately spatial non-homogeneity. However, the impact of the stochastising interactions is radically different for diluted and dense systems. When the "concentration" of the $n-$body interactions is low, the local non-homogeneities do not overlap and they rapidly relax in the surrounding homogeneous "sea". In sequel, the stochastising interactions cannot activate the fine-tuning in diluted systems. Yet, their major effect is to break the long-range correlations among the positions and velocities of the entities. In turn, this lack of long-range correlations is the key factor that permanently sustains the homogenization of the system. And the latter is vital for the validation of the thermodynamical limit. So, indeed, though the stochastising interactions are not frequent enough for activation of the fine-tuning, their role for establishing the thermodynamical limit in the diluted systems is crucial. Note, however, that this way of restoration of the thermodynamical limit does not suggest any separation of the time scales: instead it operates through breaking of the correlations continuously on the whole range of the time scales!

Now we come to the other extreme, i.e. dense systems. We suppose that the energy and "concentration" of the entities in the dense fluctuations are much larger than outside but still the interactions are short-ranged and the entities move ballisticaly between the collisions. Then the non-homogeneities produced by the stochastising interactions overlap and so they produce strong non-homogeneity at closest points. Moreover, the spatial configuration varies in the time course because of the permanent non-correlated mobility of the entities. As a result, the Hamiltonian of every entity exhibits strong fluctuations in the time due to the perturbations induced by the permanent variations of the spatio-temporal configuration. Hence, the fine-tuning becomes most likely: the permanently varying perturbations sooner or latter certainly "tune" every single-particle Hamiltonian to its critical value. In turn, each entity of the perturbation (fluctuation) effectively starts to "feel" the entire system. An effect that goes along with the fine-tuning is that the concentration of the entities in the fluctuations never exceeds some critical value. Indeed, upon reaching certain concentration the fine-tuning is activated and the scattering length becomes divergent and the fluctuation rapidly relaxes.

To outline, the mechanism of relaxation of dense fluctuations involves the following steps: (I) at certain critical concentration the non-homogeneities produced by stochastising interactions begin to overlap. In turn, this produces spatial non-homogeneity in every point of the fluctuation volume; (ii) the permanent wild variations of the spatio-temporal configuration produce significant permanent perturbations to every single-particle Hamiltonian. In turn, these perturbations inevitably "activate" fine-tuning that is equivalent to involving of long-range interactions. This helps the fluctuation to make an effective contact with the entire system that enormously speeds its relaxation rate. However, though the effective "long-range" interactions make the relaxation rate size-independent, it is true only for dense enough perturbations (fluctuations). On the other hand, we consider isolated in space and time perturbations of arbitrary size and density which makes the process of relaxation discontinuous both in space and in time: at every instant and every point the mechanism of relaxation switches from "dilute" to "dense" one and vice versa.

Summarizing, it turns out that the fine-tuning is not the expected panacea that saves the master equation because it is activated only at large enough concentration. Thus, indeed, we need entirely new approach to the behavior of the many-body systems. Though, it should be stressed that when perturbations are isolated in space and time, i.e. the system in question is closed, the fine-tuning and stochastising interactions indeed ensures fast relaxation of the perturbations. Thus, closed systems reach equilibrium and never deviate significantly from it. Note that, though it seems that we come to the traditional thermodynamics, our result is radically different from its – it is obtained by the use of an entirely new approach to the dynamics of the interactions.

Now it is to the point to focus your attention on another aspect of the triggering effect of the concentration on the fine-tuning. It is consistent with our fundamental idea about the local boundedness. Indeed, fine-tuning-assisted relaxation prevents any further concentrating of matter and/or energy in a given volume. This solves one of the greatest mysteries of the fluctuations behavior: how the fluctuations "feel" the local thresholds of stability. Actually, upon reaching the critical concentration, the fine-tuning triggers the relaxation. This is good dynamical explanation why the fluid fluctuations that drive pollen species at the Brownian motion have always finite size and finite lifetime. In &3.7 I derived the same result on the grounds of the assumption that the macroscopic evolution of any system involves only steps consistent with local and dynamical boundedness. Now this idea gets strong support by the fine-tuning assisted relaxation.

Yet, this not the whole story because the fine-tuning-assisted relaxation is a mechanism that yields fast size-independent relaxation only when the successive perturbations are dense and are independent from one another both in space and in time. Another important type of systems is the open ones: these are systems coupled to a reservoir that is a permanent source of stochasising interactions. My further task is to consider the effect of the fine-tuning in the case of open systems.

### 4.6. Divergence of the Scattering Length in Open Systems

Let us first consider the general construction of an open system so that it can operate steadily arbitrary long time. Obviously the steady operation of a system coupled to a reservoir that permanently accumulates the matter/energy requires their draining in order to prevent the accumulation that goes beyond the thresholds of stability of the system. The question is whether the draining is sufficient implement to meet the dynamical and local boundedness. Let us recall that the dynamical boundedness demands finite rates of all the steps through which the system evolves. The local boundedness imposes constraints on the spatio-temporal configurations through which the system evolves.

Further I shall consider only constructions such that each point of the system is uniformly coupled to the reservoir and the draining device. It is well known that real open systems exhibit a variety of macroscopic behavior. They manifest self-sustained oscillations and highly irregular variations upon varying the intensity of the reservoir and the drain. Such behavior evidently requires coherent response to a perturbation. Otherwise, how the entity in point $A$ "knows" what the entity in distant point $B$ is doing? In order that $A$ "feels" $B$ it is necessary to have an analog of long-range interactions. However, remember that the distant entities "feel" one another only in dense enough systems, i.e. when the density reaches the critical concentration for the fine-tuning. However, the open systems cannot relax: the permanent exposure to the flux from the reservoir sustains the concentration of the stochastising interactions permanently high. In addition, it cannot be compensated by the "draining" since the latter becomes irregular under the permanent wild variations of the spatio-temporal configurations. Thus, it is most likely to expect an inevitable exceeding of the local thresholds of stability and the formation of local defects that will result in fast breakdown of the system. So, the fine-tuning is not sufficient to provide local boundeness.

However, there is something more. Not only is the local boundedness violated but the dynamical as well. The reason is evident: the transition rates of the elementary processes diverge because the enduring fine-tuning keeps the corresponding cross-sections permanently divergent.

Summarizing, in the open systems the fine-tuning cannot automatically provide the relaxation of the wild non-homogeneities produced by the permanent action of the stochistising interactions. Let me recall that the fine-tuning is analog to long-range interactions. So, now it becomes evident the suppression of the arbitrary variations of the spatio-temporal configurations demands coherent response of the entire system. Obviously, the latter immediately would stop destabilization. Yet, does it provide the dynamical boundedness? The answer to this question is a key point in our considerations. In the next two chapters I propose a dynamical mechanism of coherence based on completely novel fundament. It turns out that it exhibits remarkable properties: on the one hand it provides coherent behavior which is consistent with local boundedness and gives rise to the dynamical boundedness in a natural way. On the other hand, the coherence becomes a source of permanent bounded variations of the rates of all elementary process.

### 4.7 What Comes Next

The aim of this chapter is to outline the inconsistency of two major approaches: the master equation and the weak-coupling limit. The importance of revealing in details the controversies that their interplay runs into is their wide use. But there is a more fundamental reason: the goal of the book is to present a coherence mechanism that is supposed to operate in the general case of open systems since they prevail in the reality. On the other hand, the thermodynamics and the statistical mechanics have been created to explain the behavior of the closed systems. That is why it is crucial to find out whether there is any fundamental difference between the behavior of the open and closed systems. The conclusion drawn from our considerations is that indeed their behavior is radically different. It turns out that only diluted closed systems always reach steady state and never deviates from it. Credible arguments in favor of this assertion are based on the role played by the stochastising interactions and by the fine-tuning.

It should be stressed that the weak-coupling fails to provide the thermodynamical limit prior to the relaxation of the state variables. Hence, it is not able to provide a self-consistent description of neither open nor closed systems.

However, our definitions of the stochastising interactions and the fine-tuning are still vague. That is why our further task is to give a rigorous idea of their explication in the construction of the coherence mechanism that operates in the open natural systems. This is the task of the next chapter.

# Chapter 5: Coherence Mechanism I.
## Properties of the "Flexible" Hamiltonian

### 5.1. Why to Read Chapter 5

The major task of the previous chapter was to illustrate the failure of the weak-coupling limit in providing rapid restoration of the thermodynamical limit perturbed by the development of a local perturbation (fluctuation). Besides, I have promised to present credible arguments that the dynamics of the stochistising interactions results in very different macroscopic behavior of open and closed systems. In &4.5 I have already clarified that the stochistising interactions make closed systems to reach equilibrium and after that never to deviate significantly from it. Note that, though it seems that it confirms the zeroth law of the thermodynamics, our result is radically different from it – it is obtained by the use of an entirely new approach to the dynamics of the interactions. Another major dissimilarity with the thermodynamics is coming now: my next task is to illustrate that in open systems the stochistising interactions bring about a destabilization the elimination of which calls for coherence that has no thermodynamical analog.

Since the current density of the entities in an open system is governed by the intensities of the coupling to the reservoir and the drain, upon reaching critical concentration the stochistising interactions are intensified so that the fine-tuning is activated. The fine-tuning is a phenomenon that happens when some interaction parameter tuned to a critical value makes the scattering length divergent. The fine-tuning is straightforwardly related to the permanent spatio-temporal non-homogeneity produced by stochastising interactions in the following way. Since their outcome is multi-valued, different selections are established at closest points. In turn, the spatio-temporal variability of the selections produces non-homogeneity on every scale which sustains fine-tuning because the permanent variations of all interactions parameters make most likely that the required interaction parameter is tuned to the critical value so that the scattering length to become divergent. However, the fine-tuning attacks the dynamical boundedness because the divergent scattering length makes the cross-sections of the elementary processes divergent. In turn, the transition probabilities of all elementary processes diverge as well. However, this divergence corresponds to an infinite rate of exchange of the energy/matter with the environment. Thus, indeed the divergence of the scattering length violates the dynamical boundedness. Furthermore, the permanent exposure to the flux from the reservoir sustains the concentration of the stochistising interactions permanently high. In addition, it cannot be compensated by the "draining" since the latter becomes irregular under the permanent wild variations of the spatio-temporal configurations. Thus, it is most likely to expect an inevitable exceeding of the local thresholds of stability and the formation of local defects that if not suppressed would result in a fast breakdown of the system. However, the fine-tuning alone is not sufficient to eliminate the induced by the stochistising interactions destabilization. My goal is to propose such mechanism and to delineate that both local and dynamical boundedness are some of its generic properties. Yet, as we shall see latter, the substantiation of a mechanism that eliminates the destabilization is available only in the frame of entirely novel viewpoint on the low-energy limit.

The first step in the implementing of the required mechanism is to outline the general construction of an open system so that to provide a steady behavior at a broad range of intensities of the reservoir and the drain. It is obvious that the system needs a physical agent that serves as an implement for the elimination of the spatio-temporal non-homogeneity produced by the stochistising interactions. For this purpose, embedding of the entities into a media is necessary so that a non-local feedback between the entities to be created. The need of the non-local feedback is brought about by the spread of the spatio-temporal non-homogeneity over every scale. Hence, its elimination certainly must involve non-local coupling among the distant entities. The non-local coupling trough media materializes the idea of long-range interactions "created" by the fine-tuning. Indeed, it serves as physical agent that makes the distant entities to "feel" each other. It is to the point to stress that the coupling through media should produce coherent response to further perturbations rather than evening of the current states! The latter does not yield the required non-homogeneity elimination because the short-range interactions make the evening local and non-correlated both in space and time. Besides, let me focus your attention on the major similarity between local $\delta$-like (acting very shortly in the time) and permanent spatially extended perturbations. The former are subject widely considered in the traditional statistical mechanics. In the previous chapter we have established that the stochastising interactions and the fine-tuning damp any local $\delta$-like fluctuation in closed systems. Note that in this case the local boundedness is guaranteed by the automatic "activation" of the fine-tuning upon reaching critical concentration. Further, the fine-tuning "couples" the fluctuation to the entire system because the divergence of the scattering length is equivalent to the involvement of the long-range interactions. Now it is obvious that the system outside the perturbation plays the role of the media coupling to which provides the relaxation. Thus, the necessity a "media" as a provider of non-local feedback between distant entities is general for both closed and open systems. It is to the point to stress, however, that the need of coherence is a result of the drastically different effect of local $\delta$-like perturbations and spatially extended perturbations that act permanently in time. Alone the fine-tuning is enough to provide damping in the former case while in the case of spatially extended permanent perturbations it causes destabilization.

Now we come to the question about the properties that make a media suitable for providing coherence. Our daily experience teaches that we cannot fit everything everywhere. However, now we are interesting in those properties of the media that ensure coherence. An obvious property is that the media should participate through its cooperative excitations, in other word it should be a piece of condensed matter. Why I insist so much on the role of the cooperative excitations? Suppose that the feedback between the entities and the media goes through the following succession of transitions: the colliding entities dissipate through the excitation of the cooperative modes, e.g. local acoustic phonons. In turn, they participate to the interacting Hamiltonian so that to induce a new transition. The latter dissipates through excitation of local acoustic phonons that in turn participate to the Hamiltonian and so on. The feedback ceases its action upon reaching coherent response to further perturbations. The important point is that the target coherence goes through making the response of all interaction Hamiltonians coherent! The advantage of participation of the cooperative excitations is that they "couple" distant Hamiltonians making them to behave coherently. The idea about this feedback is central for the coherence. I shall study its operational properties in the next chapter.

Now I am going on with establishing the general properties necessary for providing a successful feedback. An obvious condition is that the embedding of the entities should keep their identity and the identity of the media. For this purpose they should be either adsorbed on the surface or absorbed in the bulk of the system. Though the adsorption is still an enigmatic phenomenon its modeling is based on some fundamental assumptions strongly supported by the outcomes of its wide study. The adsorption (absorption) implies that the interaction adentities-surface (bulk) goes via formation of independent from one another potential wells so that the relaxation of an adentity proceeds only at one of these potential wells called active site. Further, the adentites manifest hard-core repulsion: no more than one entity can be adsorbed at the same potential well. The most successful modeling of that phenomenon is given in the frame of the so called lattice-gas model. Originally it has been introduced to model the adsorption-reaction phenomena but gradually has been gaining more and more popularity by means of more elaborative models for the hard-core repulsion. For the sake of simplicity now I shall present the lattice-gas model for the surface reactions. The reason is two-fold: on the one hand I have in hand a concrete system; on the other hand the model keeps its universality. I shall prove that the developed approach to the coherence is universal and does not depend on the chosen particularity.

The lattice-gas model is grounded on the following general assumptions:
(I) the adsorption sites are displayed at the vertices of a lattice that represents the interface gas/solid;
(ii) no more than one entity can be adsorbed at a single adsorption site ( hard-core repulsion)
(iii) the reaction proceeds only between already adsorbed entities of required types brought about at the same site by their mobility;
(iv) the reaction products are immediately removed from the system (violation of the detailed balance)

The coupling to a reservoir is brought about by the exposure of the lattice to a steady flux of reactants. Further, the exposure is such that the intensity of bombardment of each adsorption site is uniform both along the interface and in time. My next task is to illustrate that the intensity of the stochistising interactions is always of the order of the concentration of the adsorbed entities. Furthermore, I shall show that the violation of the local and the dynamical boundedness happens at every control parameter choice. The consideration is presented for the adsorption, since it is a step prerequisite of any surface reaction at any control parameter choice. At first, let me illustrate in details how the stochastising interactions appear in the adsorption. They are based on the interplay between: (I) the lack of correlation between moments and points of the gas phase entities trapping at the interface; (ii) the generic property of any adlayer that no more than one entity can be adsorbed at a single active site. That interplay causes fundamental changes of the properties of the overall probability for adsorption (correspondingly the adsorption rate). Given is an entity trapped in a vacant site. Its further relaxation to the ground state can be interrupted by an adentity that arrives at the same site and most probably occupies it. Thus, the adentity violates the further trapped entity relaxation at that site since no more than one entity can be adsorbed at a single site. The trapped entity can complete the adsorption if and only if after migration it finds another vacant site. The impact of the adentity intervention to the trapped entity probability for adsorption is twofold: first, it cannot be considered as a perturbation, since it changes the adsorption potential qualitatively, namely from attractive it becomes repulsive. Second, the lack of coherence between the trapping moment and the moment of adentity arrival makes the probability for adsorption multi-valued function: each selection corresponds to certain level of relaxation at which a stochistising interaction happens. Therefore, this interaction brings about fundamental duality of the probability for adsorption (and of the adsorption rate correspondingly): though each selection can be computed by an appropriate quantum-mechanical approach, the establishing of a given selection is stochastic process since it proceeds as random choice of a single selection among all available. This is how the multi-valuedness of the stochastising interactions appears. Now it is obvious that the concentration of the stochastising interactions is of the order of the concentration of the adsorbed entities. Since the stochastising interactions are local events, the non-correlated mobility of the adentites produces permanent lack of correlation among the established selections at every point. The range of non-homogeneity is of the order of the distance between the adsorbed entities.

Outlining, the non-correlated stochistising interactions always make the adsorption rates that come from different adsorption sites non-identical that immediately produces spatial non-homogeneity. In addition, the induced non-homogeneity is permanently sustained by the lack of coherence between the trapping moments and the adentites mobility. In turn, the adlayer configuration varies in uncontrolled way so that in short time it causes either the reaction termination or system breakdown. Thus, stable long-term evolution is available if and only if there is mechanism that suppresses the induced non-homogeneity.

Sound familiar? This type of stochistising interactions have been already introduced in the Preface under the name diffusion-induced non-perturbative interactions. Now I repeat their presentation in order to elucidate better their role in clarifying the necessity of coherence. The first important conclusion drawn by the above considerations is that the spatio-temporal non-homogeneity is insensitive to the details of the single-site Hamiltonian of adsorption, to the chemical identity of the adentites and the particularities of the interface. In addition, it persists at each value of the control parameters. So, the spatio-temporal instability exists in every point of the state space and is generic for all surface reactions. In sequel, the mechanism that is able to suppress the permanent non-homogeneity should be also insensitive to the details of the single-site Hamiltonian of adsorption, chemical identity of the adentites and the particularities of the interface and should persist at every value of the control parameters.

A successful coherence mechanism needs feedback that acts toward evening of the initially non-identical adsorption rates through making the coupled entities "response" to further perturbations coherent. It should be grounded on strong coupling adlayer-interface, namely: the energy of colliding entities dissipates to local cooperative excitations of the interface. In turn, the impact of these local modes on the colliding entities is supposed large enough to induce new transition that dissipates through the excitation of other local cooperative modes and so on. The feedback ceases its action whenever the colliding entities response becomes coherent.

However, the used so far weak-coupling approach renders any feedback set on the interaction adlayer-interface local and non-correlated both in space and time. It provides only local evening of current states of the colliding entities. Indeed, suppose that the collision energy dissipates through the excitation of local cooperative modes. The weak-coupling approach considers any change of a single-site Hamiltonian of adsorption a perturbation. Consequently, the latter cannot produce a large enough change of the entity state to induce a new transition. As a result, the interaction colliding entity-interface stops. Since only short range interactions are considered, new transition happens only when other entities collide brought together by their mobility. Hence, the weak-coupling limit is not able to provide strong enough coupling adlayer-surface coupling so that to sustain successful feedback.

The formulation of the basic principles of successful feedback is great challenge because it should give rise not only to universality of coherence mechanism but at the same time it should preserve the chemical identity of the entities! I proceed with the presentation of radically novel assumptions that makes such feedback possible.

### 5.2. "Flexible" and "Rigid" Hamiltonian

I have already stressed on the point that the weak-coupling approach fails in providing a successful feedback between the surface and the adlayer because it treats the coupling adlayer-surface as perturbation. So, the feedback is possible only under fundamentally novel assumption about the Hamiltonian response to the coupling adlayer-surface. For this purpose I put forth the idea that every single-site Hamiltonian is decomposable into two parts: "rigid" one and "flexible" one. The "rigid" part is associated with the low-lying states of the spectrum where the interactions with the environment contribute additively, i.e. (4.10) holds. Thus, the "rigid" part of the Hamiltonian preserves the identity of the adsorbed entity. The "flexible" part of the Hamiltonian substantiates the idea of fine-tuning to be "sensitive" to the environment. In other words, it is supposed that the environment influence is long-ranged and thus it is comparable with the "forces" that create the Hamiltonian itself. In sequel, each single-site Hamiltonian $\hat{H}$ is the following direct sum:

$$\hat{H} = \hat{H}_{rigid} \otimes \hat{H}_{flexible} \tag{5.1}$$

where $\hat{H}_{rigid}$ and $\hat{H}_{flexible}$ are the "rigid" and "flexible" parts of the total Hamiltonian. The major property of the "rigid" part $\hat{H}_{rigid}$ is that the interactions are supposed to contribute additively. The latter immediately implies that the identity of an entity is supposed steady in the process of the interaction with the environment. Vice versa, at every instant we are able to separate the entities additively and to determine the interactions of each pair. But why do I need decomposition? The reason is two-fold and now I am going to explain the first one: the relation between the "rigidity" and identity.

In general, the relation between the "rigidity" of a Hamiltonian and the identity gives rise to the following vicious circle. The property of all studied Hamiltonians such as centro-symmetrical potential, Column interaction, harmonic potential etc. is that they are characterised by *infinite* number of quantum numbers. This infinity is an explicit manifestation of the hidden total rigidity "encapsulated" in those Hamiltonians. It means that it takes infinite time to pass level by level an atom, molecule etc. whose properties are governed by such Hamiltonian. In sequel, no reaction, no excitation, no process would be possible! The world would be dead! In order to make the world to "wake up" it is necessary that every real Hamiltonian has only finite number of specific quantum numbers that characterizes it. On the other hand, the spectroscopy has proven in very categorical way the relevance of the above listed Hamiltonians! Is there a way out from this vicious circle? In order to reconcile the controversy I am introducing the idea of separation of single-site Hamiltonian into "rigid" and "flexible" part so that the "rigid" one keeps the identity being subject to (4.10). The second major assumption is that the decomposition goes so that the rigid part of the Hamiltonian always has only finite number of states.

Let us now have a closer look on highly excited states (low-energy limit). Should we expect that the environment influence can be treated as perturbation as well? Obviously, the answer is negative: no perturbation approach is available for highly excited states and resonances where the binding energy is low, comparable with that of the perturbations. Moreover, the crucial question becomes: should we keep the primary role of the linear superposition of the interactions in the low-energy limit? The answer is strongly negative: no, because the linear superposition does not prevent unlimited grow of the interaction energy upon an increase of the interaction entities number! Furthermore, upon operating of the fine-tuning, each entity effectively interacts with all entities in a system. In turn, it would yield infinite amount of energy focused in every point of the system?! Since this apparently breaks the local boundedness, we should face once again the concept about the boundedness. My major assumption about the "flexible" Hamiltonian asserts: whatever the interactions in low-energy limit are, their amplitude remains bounded. To point out once again, the linear superposition does not preserve boundedness: on increase of the interaction intensity and/or the number of interaction entities, the interaction energy can be made arbitrarily large. Since the operation that preserves boundedness is the coarse-graining, I replace the linear superposition of the interactions with the operation of coarse-graining so that to provide permanent boundedness of the interaction energy.

Further, a distinctive property of the "flexible" Hamiltonian is that unlike the traditional notion, it has no steady shape. This immediately rises the question whether the 'flexible" Hamiltonian is a source of scattering only or can be associated with some binding. Further I shall illustrate that the scattering and binding are entangled in a tricky interplay. At first, let me list the fundamental assumptions that delineate the generic properties of a "flexible" Hamiltonian. The first one is that the low binding energies make the de Broglie wavelength much larger than the range of a single-site Hamiltonian. In turn, the particularities of the interaction are lost and the interacting entities should be considered as structureless points. The second general assumption is that all directions in the Hilbert space are equivalent and remain equivalent under the transitions caused by the permanent spatio-temporal variations of the "flexible" Hamiltonian. This assumption immediately implies that the latter does not have any steady discrete or continuous symmetry. Indeed, the permanent variations of the spatio-temporal configurations of the adlayer certainly break any accidentally created symmetry. Furthermore, the lack of permanent symmetry leaves the "flexible" Hamiltonian permanently non-separable. So, the assumption about the equivalence of all directions in the Hilbert space is grounded on both lack of steady symmetry and the permanent non-separability.

Now we are ready to work out the properties of $\hat{H}_{flexible}$ that come out from the three basic assumptions listed above. Moreover, because of the permanent variations of $\hat{H}_{flexible}$, the target properties should be studied only statistically. Nevertheless, I promise to demonstrate that the statistical approach rather helps than interfere with establishing of the important for the feedback properties. Yet, it is an exciting task because the properties of $\hat{H}_{flexible}$ are highly non-trivial and unexpected.

Note that the above assumptions prompt universality of the statistical properties of $\hat{H}_{flexible}$ expressed by insensitivity to the chemical identity of the entities and the particularities of the lattice. In sequel, a step further is to expect that this universality serves as fundament for operational equivalence of the coherence mechanism in systems of different nature. In the next chapter I shall justify that the successful feedback is grounded explicitly on these universal properties of $\hat{H}_{flexible}$. Let us just remind that the carrier of the chemical identity is $\hat{H}_{rigid}$ that remains unaffected by the processes that go on in the low energy limit!

### 5.3 Decoherence

The sensitivity of $\hat{H}_{flexible}$ to the environment renders it dependent on any change of the adlayer spatio-temporal configuration. However, the latter varies widely in every moment and on every spatial scale under the permanently sustained by the gas bombardment high concentration of the stochistising interactions. So, the shape of $\hat{H}_{flexible}$ also varies permanently. Than why at all I call it Hamiltonian: the traditional notion of a Hamiltonian asserts that it is an operator that acts on Hilbert space so that its eigenvalues are invariants of the motion. So, shall we expect that permanent variations of $\hat{H}_{flexible}$ can produce invariants?! In the general case it is certainly impossible. Then, shall we expect scattering only? I shall prove that in the case of $\hat{H}_{flexible}$ binding and scattering are entangled in a tricky interplay.

The major problem that I face is whether all $\hat{H}_{flexible}$ associated with different spatial points share some characteristics: in other words, are their properties statistically equivalent. Evidently, the spotted in different active sites single-site $\hat{H}_{flexible}$ are different even in the academic case of identical "rigid" parts. Yet, I expect that their spectra are statistically equivalent. This implies that they are characterized by the same type of quantum numbers and obey the same distribution. What is the ground for such strong assertion? It comes from the lifted by the lack of any steady discrete and continuous symmetry degeneracies and makes plausible to assume that the spectrum of $\hat{H}_{flexible}$ is characterized by single quantum number. This supposition is strongly supported by the association of $\hat{H}_{flexible}$ with the highly excited states (the low-energy limit) whose structure is "smoothed out" because their de Broglie wavelength is much larger than the range of interaction.

Further, the lack of any discrete and continuous symmetry renders lack of any specific time or space scale associated with any single-site $\hat{H}_{flexible}$. The formal expression of this property is that the $\hat{H}_{flexible}$ is scaling invariant upon coarse-graining. In particular, the properties of the $\hat{H}_{flexible}$ remain invariant under any local spatio-temporal averaging:

$$\hat{H}_f\left(T,\vec{R}\right) \cong \hat{H}_f'\left(\widetilde{T},\widetilde{\vec{R}}\right) \qquad (5.2)$$

where $\hat{H}_f\left(T,\vec{R}\right)$ is $\hat{H}_{flexible}$ before the local averaging and $\hat{H}_f'\left(\widetilde{T},\widetilde{\vec{R}}\right)$ is $\hat{H}_{flexible}$ after that operation.

The requirement that $\hat{H}_{flexible}$ remains permanently bounded renders it transformational invariant as well. To recall, the transformational invariance implies preserving the boundedness upon coarse-graining. Applied to $\hat{H}_{flexible}$, the coarse-graining is to be associated with the modification of the spatio-temporal configurations. Here I replace the idea of linear superposition with the much more general idea of coarse-graining. Note that the linear superposition is an operation that preserves linearity while the coarse-graining preserves the boundedness through involving local non-linearities! However, the invariance of the linearity does not ensure preserving the boundedness! On the contrary, on adding more and more terms every sum can be made arbitrarily large though each of the terms is bounded! The impact of the replacement of the fundamental idea about the linear superposition with the idea about preserving boundedness is milestone in all considerations. It has far going consequences one of which is the obvious violation of (4.10). Let us point out that the coarse-graining is still compatible with the linear superposition if the latter preserves the boundedness in one way or another! So, the idea of a perturbation survives but under radically novel fundamental concept, namely the concept of preserving boundedness.

In sequel the basic assumptions yield that $\hat{H}_{flexible}$ obeys both scaling and transformational invariance. Then, its eigenfunctions must be BIS both before and after a transition. Furthermore, since the operation that preserves boundedness is the coarse-graining, the transitions must be considered as result of operations of coarse-graining. These assumptions are self-consistent since, as established in &1.2, the entanglement of the transformational and scaling invariance makes the set of BIS dense transitive set so that the coarse-graining appears to be an operation that transforms one BIS into another. The lack of any symmetry renders them zero-mean. But do they compose a complete orthonormal basis in the Hilbert space? Next I shall prove it. The completeness is automatically ensured by the transitivity of the set of BIS. Let us now estimate the orthogonality:

$$I = \int_V \psi_i * \psi_j dV \qquad (5.3)$$

Evidently when $i = j$:

$$I = \frac{1}{V}\int_V |\psi_i|^2 dV = \sigma^2 \qquad (5.4)$$

where $\sigma$ is the variance. We already know that every BIS has finite variance. In turn, it ensures the normalisability of BIS as functions that can set an orthonormal basis.

For $i \neq j$ the product $\psi_i * \psi_j$ can be considered as an operation of coarse-graining over either $\psi_i *$ or $\psi_j$:

$$I = \frac{1}{V} \int_V \hat{G} \psi_j dV = \frac{1}{V} \int_V \psi_l dV = o\left(\frac{1}{V}\right) \qquad (5.5a)$$

$$I = \frac{1}{V} \int_V \hat{G} \psi_i * dV = \frac{1}{V} \int_V \psi_k * dV = o\left(\frac{1}{V}\right) \qquad (5.5b)$$

where $V$ is the volume of the system and $\hat{G}$ is the operator of coarse-graining that transforms one BIS into another. The estimation $o\left(\frac{1}{V}\right)$ comes from the dynamical boundedness. Let us remind that the latter imposes finite distance between the zeroes of every BIS. In turn, the average of every BIS is always of the order of $o\left(\frac{1}{V}\right)$. So, each BIS is orthogonal to every other. Preserving of the boundedness requires the eigenfunctions after every modification of the adlayer spatio-temporal configuration to remain BIS. However, this is a very unusual result because it implies that there is no quantum interference! Indeed, the quantum interference means that a given flux is split and it interferes constructively when the path distance is appropriate. Yet, in the present case, the path distance permanently varies both in space and in time. Consequently, in the low-energy limit the quantum objects escape from one of the hallmarks of the quantum mechanics - the quantum interference. The obtained decoherence is the first of the novel properties that are brought about by the leading role of the boundedness.

The decoherence renders to expect radically new properties in the low-energy limit. It should be stressed that the low-energy limit is not the quasi-classical limit! The latter is grounded on taking the limit $\hbar \to 0$ where $\hbar$ is the Planck constant (the eikonal approach). It indicates that the de Broglie wavelength is small comparable to the scale of the problem and varies slowly compared to that specific scale. On the contrary, in the low-energy limit de Broglie wavelength is much larger than any specific scale. Moreover, because of the variations, $\hat{H}_{flexible}$ has no steady specific scale as we will see in the next section. Further, the quasi-classical limit does not produce any decoherence. However, the latter is one of the properties that radically separate classical from quantum objects.

The obtained decoherence is my strongest reason to put forth the idea that the classical and quantum phase space of a "flexible" Hamiltonian are identical. This equivalence will help me to derive the statistical properties of the spectrum. My next task is to make evident the equivalence of the energy distribution derived classically and quantum-mechanically. This demonstration plays the central role in the justification of the entire conjecture.

### 5.4 Classical Phase Space in Low-Energy Limit

Let me point out once again the fundamental idea that every single-site Hamiltonian decomposes into two-parts: the "rigid" one $\hat{H}_{rigid}$ where the interactions with the environment contribute linearly, i.e. (4.10) holds. In turn, the latter ensures preserving of the chemical identity of the adsorbed entities. The "flexible" part of the Hamiltonian $\hat{H}_{flexible}$ is supposed "sensitive" to the environment: the forces that create $\hat{H}_{flexible}$ are comparable with the environmental influence. This renders pivotal role of the boundedness in the formation of the interaction energy. That is why the linear superposition is replaced by coarse-graining, the operation that preserves boundedness. Note that the linear superposition of interactions expressed through the additivity of (4.10) is incompatible with the idea about permanent boundedness. In Chapter 1 we proved that the sustaining of permanent boundedness is met under the operation of coarse-graining. The latter radically differs from the linear superposition in two points: (i) the interactions that come from different pairs participate non-linearly to the total interaction; (ii) the coarse-graining is a non-homogeneous operation, i.e. it can amplify the interaction in one point and damp it in the other. However, when acting on BIS it leaves it BIS. We called this property transformational invariance. Hence, indeed, I introduce a radically novel concept about interactions in the low-energy limit.

Now I am going further and pose the question about the properties of the classical counterpart of a single-site Hamiltonian in the low-energy limit. My working hypothesis is that the boundedness remains the major property of $\hat{H}_{flexible}$ both in the quantum and the classical case. Then the classical counterpart of $\hat{H}_{flexible}$ is subject to permanent variations remaining at the same time permanently bounded. So, in the general case, the classical equations of motion read:

$$\frac{dq_i}{dt} = \frac{\partial \hat{H}_{flexible}}{\partial p_i}$$
$$\frac{dp_i}{dt} = -\frac{\partial \hat{H}_{flexible}}{\partial q_i} \quad (5.6)$$

where $q_i, p_i$ $(i=1,\ldots,g)$ are the canonical variables and their conjugates; $g$ - the degrees of freedom. Any "flexible" Hamiltonian $\hat{H}_{flexible}$ has two crucial properties: (I) each $\frac{\partial \hat{H}_{flexible}}{\partial p_i}$, $\frac{\partial \hat{H}_{flexible}}{\partial q_i}$ ($\forall i, j$) varies irregularly in the space and the time; (ii) both $\frac{\partial \hat{H}_{flexible}}{\partial p_i}$, $\frac{\partial \hat{H}_{flexible}}{\partial q_i}$ are permanently bounded. The boundedness of the $r.h.s.$ of (5.6) guarantees that the solution of (5.6) has the properties of a BIS established in Chapters 1 and 3. Let us remind that the equations (5.6) are of the same type as equations (2.15). Indeed, the boundedness renders that both $\frac{\partial \hat{H}_{flexible}}{\partial p_i}$, $\frac{\partial \hat{H}_{flexible}}{\partial q_i}$ can be presented as:

$$\frac{\partial \hat{H}_{flexible}}{\partial p_i} = \frac{\partial \overline{H}}{\partial p_i} + \frac{\partial H_s(\vec{r},t)}{\partial p_i}$$
$$\frac{\partial \hat{H}_{flexible}}{\partial q_i} = -\frac{\partial \overline{H}}{\partial q_i} - \frac{\partial H_s(\vec{r},t)}{\partial q_i} \quad (5.7)$$

where $\overline{H}$ is the expectation value of $\hat{H}_{flexible}$ and $\hat{H}_s(\vec{r},t)$ is zero-mean permanently varying in space and time term. The assumption that the "flexible" Hamiltonian is to be associated with lack of any steady symmetry suggests that $\overline{H} \equiv 0$. If otherwise, its properties would be dependent on the particularities of $\overline{H}$. Thus, (5.7) becomes:

$$\frac{\partial \hat{H}_{flexible}}{\partial p_i} = \frac{\partial \hat{H}_s(\vec{r},t)}{\partial p_i}$$
$$\frac{\partial \hat{H}_{flexible}}{\partial q_i} = -\frac{\partial \hat{H}_s(\vec{r},t)}{\partial q_i} \quad (5.8)$$

The major property of $\hat{H}_s(\vec{r},t)$ and its derivatives $\frac{\partial \hat{H}_s(\vec{r},t)}{\partial p_i}$ and $\frac{\partial \hat{H}_s(\vec{r},t)}{\partial q_i}$ is their boundedness. The limitation over the variability of the derivatives is to be associated with the dynamical boundedness. Indeed, regardless how irregular the spatio-temporal variations of $\hat{H}_s(\vec{r},t)$ are, the energy involved currently in the exchange with the environment is finite. In turn, this constitutes the boundedness of derivatives $\frac{\partial \hat{H}_s(\vec{r},t)}{\partial p_i}$ and $\frac{\partial \hat{H}_s(\vec{r},t)}{\partial q_i}$. As we have already mentioned above, eqs.(5.8) are of the same type as eqs.(2.15). In Chapter 3 we studied the properties of their solution. There we established that the coarse-grained structure of the solution of (5.8) consists of excursions that have three distinctive characteristics: size $A$, duration $\Delta$ and embedding time interval $T$. I have illustrated these characteristics in the sketch presented in Fig.3.1. Let us recall their meaning: excursion is a trajectory that starts at the expectation value at moment $t$ and returns to it for the first time at the moment $t + \Delta$. The embedding is a property exclusive for the boundedness and implies that each excursion is embedded in time interval larger than its duration so that no other excursion is developed in that time interval. The major property of the embedding is to prevent overlapping of the successive excursions and thus to preserve boundedness under the coarse-graining. The three characteristics $A$, $\Delta$ and $T$ are not independent: on the contrary there exist relations between them. The most peculiar one is that the relation $\Delta \leftrightarrow T$ is multi-valued function. Thus, it acts as an abstract stochastising interaction - it preserves boundedness breaking the formation of long-term correlations among distant excursions!

Yet, the major question is how to interpret the above characteristics when they come from the solution of eqs. (5.8). The formation of well-separated one from another excursions suggests to associate each excursion with confined motion. On the other hand, there is no binding because $\overline{H}$ is zero. Then how does a confined motion appear? The answer comes from the so called chaotic scattering intensively studied recently [5.1]. The chaotic scattering appears when a mass point scatters elastically from non-collinear and non-overlapping discs. A trajectory that enters the space between the discs follows for some time arbitrarily long segments of the periodic orbits till it scatters out. The longer it gets to a trapped trajectory the longer it stays in the interaction region. In our case, the scattering out is governed by the spatio-temporal variations of $\hat{H}_s(\vec{r},t)$. So, there is temporary confinement when the local configuration of $\hat{H}_s(\vec{r},t)$ is appropriate. Then the permanent variations of $\hat{H}_s(\vec{r},t)$ causes scattering out to another appropriate local configuration where new confinement happens. So, the excursion itself is a random walk that starts at zero energy and is executed as random walk from one confinement to another until the entity is scattered either to the gas phase or to $\hat{H}_{rigid}$. The interplay of binding and scattering can be considered as "resonances" of an effective "coarse-grained" time-independent Hamiltonian so that the size of the excursions can be considered as binding energy of the effective Hamiltonian while the duration of the excursion is associated with the width of the resonances. The meaning of the embedding is two-fold:

(I) the relation with the size of the excursion associates the embedding time interval with the nearest level spacing. Thus, the distribution of the nearest level spacing $E$ coincides with that of the amplitude of the excursions $A$. In Chapter 3 we have already established that the distribution of $A$, correspondingly $E$ share the same distribution (3.10):

$$P(E) = cE^{1/\beta(E)} \frac{\exp(-E^2/\sigma^2)}{\sigma} \qquad (5.9)$$

where $c$ is the normalizing factor; $\beta(E)$ is to be associated with the details of the spatio-temporal variations of $\hat{H}_s(\vec{r},t)$; $\sigma$ is the variance of $\hat{H}_s(\vec{r},t)$. The presence of the prefactor $E^{1/\beta(E)}$ implies level repulsion: the most probable nearest distance between levels is non-zero. So, our classical considerations bring about a truly quantum property: discreteness of the spectrum of the effective Hamiltonian!

(ii) The property of the embedding time interval to be always larger than that the duration of the embedded excursion prevents the overlapping of the successive "resonances".

And last but not least, the study of the bounded irregular sequences in Chapter 3 gives us the "size" of $\hat{H}_{flexible}$. This is an important issue since on the one hand the assumption that distant points "feel" one another requires the size of $\hat{H}_{flexible}$ to be infinite. On the other hand, the obtained result about temporary binding implies temporary localization and calls for finite size of $\hat{H}_{flexible}$. The revealing of the controversy is easy and is as follows: following the definition of the chemical potential, the size that turns its to zero demarcates the largest size of $\hat{H}_{flexible}$. The illustration that the kinetic threshold of stability is always finite, assures us that this size is always finite. Then, though we suppose that in the low-energy limit the entities "feel" distant variations of $\hat{H}_s(\vec{r},t)$, the impact of this "sensitivity" is two-fold: the variations inside the volume of $\hat{H}_s(\vec{r},t)$ yield temporary confinement whilst outside they always destabilize it.

### 5.5. Quantum Phase Space in Low-Energy Limit

Let me recall that the present goal is to prove the equivalence of the energy distribution derived classically and quantum-mechanically. The idea about that equivalence comes out from the fundamental conjecture that in the low-energy limit the motion in the classical and quantum phase space is identical. This idea is strongly supported by the proven in &5.3 decoherence. In the frame of this conjecture the considerations in the previous section show that the nearest level spacing is distributed according to (5.9). So, the task is to derive the same distribution from quantum-mechanical viewpoint.

I start with reminding that difference between the low-energy limit and the quasi-classical limit. It should be stressed that the latter is associated with the eikonal approximation which does not imply necessarily low energies. The eikonal approach is set on the condition about smooth variation of the de Broglie wavelength compared to the specific for the problem scale. This condition is certainly violated in the considered low-energy limit because of the lack of specific scale of the variations of $\hat{H}_{flexible}$. Then, how to describe the low-energy limit? Obviously it is not an eigenvalue problem since the Hamiltonian $\hat{H}_{flexible}$ involves the time implicitly. Moreover, the Hamiltonian $\hat{H}_{flexible}$ has no ground state because it has not stationary part: remember that it forms a zero-mean irregular stochastic sequence in the time course - (5.7). So, its variations cause transitions until the entity either is scattered back to the gas phase or passes to $\hat{H}_{rigid}$. Then the task is to find out the spectrum of the transition energies. The chaotic scattering considered in the classical case suggests that its quantum counterpart is temporary binding so that the transitions happen between bind states. The temporary binding can be described in the following way. The boundedness and the lack of any steady symmetry of the variations of $\hat{H}_{flexible}$ make its eigenfunctions to be always BIS. This assumption is consistent not only with the boundedness but also with the general conjecture of the quantum mechanics that after a transition, the eigenfunction of the final state is superposition of the eigenfunctions at the initial moment. So, we have:

$$\Psi_i = c_{ij}\psi_j \tag{5.10}$$

where $\Psi_i$ are the final eigenfunction; $\psi_j$ are the initial eigenfunctions. Since $\hat{H}_{flexible}$ involves the time implicitly, its evolution is subject to the time-dependent Schrodinger equation:

$$\frac{\partial \psi_j}{\partial t} = \frac{\hbar^2}{2m}\Delta\psi_j + \left(\hat{H}_{flexible} - E\right)\psi_j \tag{5.11}$$

The transition probabilities $c_{ij}$ can be determined through multiplying (5.11) by integrating over the entire volume:

$$c_{ij}\int_V \psi_i * \frac{\partial \psi_j}{\partial t}dV = c_{ij}\int_V \psi_i *\Delta\psi_j dV + c_{ij}\int_V \psi_i *\hat{H}_{flexible}\psi_j dV - c_{ij}E\int_V \psi_i *\psi_j dV \tag{5.12}$$

Next I shall evaluate each term in (5.12) starting with:

$$I_1 = \int_V \psi_i * \frac{\partial \psi_j}{\partial t}dV \tag{5.13}$$

The understanding of the spatial and time derivatives in (5.11) is new: since the operation that preserves boundedness is the coarse-graining, the spatial and time derivatives must be considered as operations of coarse-graining. Let us start with the time derivation $\frac{\partial}{\partial t}$ regarded as operation of coarse-graining. Applied to BIS it commutes with the integration over the volume. In turn this yields:

$$I_1 = \int_V \psi_i * \frac{\partial \psi_j}{\partial t} dV = \frac{\partial}{\partial t} \int_V \psi_i * \psi_j dV = \frac{\partial}{\partial t} a\delta_{ij} = 0 \tag{5.14}$$

where $a$ is the expectation of the BIS that serve as initial eigenfunctions. Since all eigenfunctions are zero-mean BIS, $a = 0$. It should be stressed that $\frac{\partial}{\partial t}$ and integration over the volume considered as operations of coarse-graining commute only when applied to BIS!

Similarly:

$$I_2 = \int_V \psi_i * \Delta \psi_j dV = -\int_V \nabla \psi_i * \nabla \psi_j dV = -\delta_{ij} a' \tag{5.15}$$

where $a' = 0$ is the expectation value of the gradient of the initial eigenfunctions.

The evaluation of the term $\int_V \psi_i * \hat{H}_{flexible} \psi_j dV$ requires additional supposition about the size of the Hamiltonian $\hat{H}_{flexible}$. Evidently, the matrix elements $G_{ij}$:

$$G_{ij} = \int_V \psi_i * \hat{H}_{flexible} \psi_j dV \tag{5.16}$$

are non-zero only if the size of $\hat{H}_{flexible}$ is finite. A finite size of $\hat{H}_{flexible}$ is rendered by the suggested parity between the classical and quantum phase space. To remind that in &5.4 we derived that the classical counterpart of $\hat{H}_{flexible}$ has finite size. On the other hand, the boundedness along with the lack of any symmetry ultimately yields the matrix elements $G_{ij}$ non-correlated and bounded. Then, the equation for the transition probabilities $c_{ij}$ reads:

$$c_{ij}(G_{ij} - E\delta_{ij}) = 0 \tag{5.17}$$

Equation (5.17) needs additional attention because it has two-fold meaning. On the one hand, it is equation for evaluating the eigenvalues $E$. On the other hand, it gives the transition probabilities. Though the eigenvalues can be evaluated for any given $\hat{G}$, the permanent variations of the matrix elements make this task meaningless. That is why I shall focus the attention on the statistical properties of the eigenvalues. The next step is to find their distribution. They are the roots of the characteristic equation of (5.17) that reads:

$$P(E) = \sum_k s_k E^k = 0 \tag{5.18}$$

where $s_k$ is the coefficients of the polynomial. The most important property of $P(E)$ is that it is random polynomial. This is an immediate result of the variability of the matrix elements $G_{ij}$. So, it is to be expected that at every instant all coefficients $s_k$ are non-zero. It has been proven [5.2] that the roots of a random polynomial with non-zero coefficients tend to cluster near unit circle and their angles are uniformly distributed. This clustering immediately renders boundedness of the distance between the successive roots. Since the distance between roots equals the transitions energies, the latter are bounded as well. Thus, the boundedness of the transition energies gives rise to dynamical boundedness!

The "binding" is associated with the resonances of an effective time-independent Hamiltonian whose "binding energies" are set on (5.18). The permanent variations of $\hat{H}_{flexible}$ result in permanent rearrangement of the resonances that in turn produces permanent transitions between them. This makes the motion in the phase space to be executed as fractal Brownian walk constituted by transitions between resonances. Note that the successive steps of the "walk" are given by the transition energies. It is to the point to stress that the binding is provided by the commuting of the integration and the time derivative in (5.14). Note that this is a particular property of the coarse-graining applied to BIS and it is not hold for arbitrary eigenfunctions!

Thus, from quantum-mechanical viewpoint, the understanding of the interplay of binding and scattering is the same as in the classical phase space: on the one hand the excursions are to be associated with the resonances of an effective time-independent "coarse-grained" Hamiltonian so that the distribution of their amplitude to coincide with the nearest level spacing distribution of the "coarse-grained" resonances:

$$P(E) = cE^{1/\beta(E)} \frac{\exp(-E^2/\sigma^2)}{\sigma} \qquad (5.19)$$

where $E$ is the nearest level spacing. On the other hand, the relation between the embedding time interval and the duration of the excursion can be considered as level repulsion, i.e. avoiding the overlapping of successive resonances. Indeed, the formal expression of the level repulsion is the presence of the power prefactor $E^{1/\beta(E)}$.

Note that there is only one "quantum number" $E$ that characterizes the states of the effective Hamiltonian. In turn, it substantiates the expected universality of the low-energy limit – the distribution (5.19) is invariant under the chemical identity of the entities. It implies that the entities appear as structureless objects characterized by a single quantum number. Yet, note that the lost of the chemical identity is temporary: it gets until the entity enters the "rigid" part of the Hamiltonian or scattered to the gas phase where their chemical identity is restored.

An important sequel of the level repulsion is that the number of states in the low-energy limit is always finite. Since the number of states of $\hat{H}_{rigid}$ is also finite, the level repulsion guarantees finite total number of states. In turn, the latter ensures that each entity is characterized by finite number of characteristics. So, the world is alive indeed! Let us recall that limitation over the number of states is the requirement necessary for providing that any excitation and relaxation is to be completed in finite time interval and with the involvement of finite amount of energy/matter.

Summarizing, an entity moves from one binding region to another under the permanent variations of $\hat{H}_{flexible}$. Though the binding energy varies in space and in time, it manifests some common properties. Indeed, eq.(5.19) can be interpret as level spacing distribution of ensemble of Hamiltonians that share the property to have random but bounded matrix elements. The randomness and boundedness of the matrix elements come from the interplay of the fundamental assumptions about the lack of any steady symmetry and about the boundedness of the energy/matter involved in any process. It should be pointed out that the idea about ensemble of random Hamiltonians has been introduced already in the 1930`es by Wigner. He put forth them to study the complex behavior of the many-body interactions in the highly excited states of heavy nuclei. More precisely, he introduced the following concept: the behavior of a complex many-body system is subject to the properties of ensemble of different Hamiltonians that share the same symmetry properties. In addition, he supposed that the distribution of the random elements should be independent of the basis in the Hilbert space. It turns out that this is possible only with the Gaussian (normal) distribution. These are the basic postulates of the famous Random Matrix Theory (RMT). In the beginning it has been intensively studied for the purposes of the nuclear physics only. Nowadays its popularity is growing up and it has been already applied to the problem of localization in the condensed matter physics, mesoscopic physics, supersymmetry theories. An extensive review on these topics can be found in [5.3]. Yet, I would like to point out some fundamental differences between our and RMT basic assumptions.

The first fundamental difference is that our Hamiltonian $\hat{H}_{flexible}$ is time-dependent. On the contrary, the RMT deals with time-independent Hamiltonians whose matrix elements are *apriori* statistically defined. The second fundamental difference is that I assume lack of any steady symmetry while the RMT *apriori* sets the symmetry.

The major difference in the outcomes of our and the RMT approach is that the nearest level spacing distribution (5.19) comes out straightforwardly from (5.18) while such relation cannot be derived in the frame of the RMT. Actually, the distribution (5.19), with $\beta(E)=1$, is well known under the name Wigner distribution. It very well fits the level distribution of heavy nuclei but its explicit relation to the RMT still remains mystery. The great advantage of our approach is that it straightforwardly relates the randomness of the matrix elements in eq.(5.17) and the distribution of the level spacing (5.19). Moreover, the imposed boundedness naturally gives rise to the normal distribution of the matrix elements. Note that in RMT the normal distribution of the matrix elements is necessary for bringing about equivalence of all basis in the Hilbert space. On the contrary, in our approach, the normal distribution appears naturally due to the boundedness. At the same time, the equivalence of all basis in the Hilbert space is brought about by the interplay of the lack of any steady symmetry and the boundedness that result in setting BIS as eigenfunctions of $\hat{H}_{flexible}$.

The key consequence of all considerations is the remarkable natural emerging of the dynamical boundedness (the boundedness of the transition energies). In turn, we can conclude that the boundedness is not only an assumption: it turns out to be a property! Indeed, we suppose boundedness of the matrix elements and our approach yields the boundedness of the transition energies.

### 5.6 Fine-Tuning and the Low-Energy Limit

Our considerations started with the idea about decomposition of a single-site Hamiltonian into "rigid" and "flexible" part. We supposed that the "flexible" part $\hat{H}_{flexible}$ is sensitive to the distant disturbances and this sensitivity is strong to be considered as perturbation. The concept about the high sensitivity of $\hat{H}_{flexible}$ comes to substantiate the idea about the fine-tuning. Let us recall that the fine-tuning happens when some interaction parameter tuned to a critical value makes the scattering length divergent. The latter corresponds effectively to involvement of long-range interactions. Vice versa, the emerging of long-range interactions is brought about by the high sensitivity of $\hat{H}_{flexible}$ to distant interactions.

However, in the course of our considerations we have been facing temporary binding in a finite spatio-temporal region. Have we run into a conflict: on the one hand, we impose high sensitivity to distant interactions and on the other hand, the approach yields binding in a finite region?! The answer is that not only there is no conflict but on the contrary, we manage to formulate mathematically the vague idea about the divergence of the scattering length. This is due to the properties of BIS as eigenfunctions. Let us recall that in the weak-coupling limit the scattering length of an entity trapped in a bounded region is of the order of the size of that region. However, this is true because the probability for finding the entity outside the interacting region tends to zero. In the low-energy limit we encounter a very peculiar situation. On the one hand, the size of binding region is always finite as established in &5.4. Then, it seems that the scattering length is to be finite. However, the scattering length is proportional the probability for finding the entity anywhere in the system after a transition:

$$p = \frac{1}{V} \int_V |\psi_i|^2 dV \tag{5.20}$$

where the eigenfunction are BIS subject to dynamical boundedness. The latter makes the distance between successive zeroes of eigenfunctions (that are BIS) finite and independent on the length of the BIS (eigenfunction). In turn the probability for finding the entity anywhere in space and time is finite: its value is roughly equal to constant and varies very slowly. Moreover any local averaging smoothes more and more the variations. So, on increase of the volume where we look for the entity, the probability becomes more and more uniform.

### 5.7 What Comes Next

The first task of the present chapter was to make evident that the multi-valuedness of stochastising interactions is a source for destabilization in open systems at any value of the control parameters. Thus the pivotal task is to find out mechanism that eliminates the induced destabilization. It turns out that it should be grounded on non-local feedback that couples distant entities. The need of non-local feedback is brought about by the spread of the induced spatio-temporal non-homogeneity over every scale. So its elimination certainly must involve non-local coupling among distant entities. The non-local coupling trough media provides an effective physical agent that renders each entity to "feel" distant ones, i.e. it is analog to long-range interactions. Yet, the coupling should produce coherent response to further perturbations rather than evening of current states! The latter does not yield required non-homogeneity elimination because the "coupling" to reservoir is permanent "source" of stochastising interactions that correspondingly produce permanent spatial non-homogeneity.

Further we have pointed out that the weak-coupling limit is not able to activate required feedback grounded on the coupling adlayer-interface. That is why I put forth the idea that any single-site Hamiltonian is decomposed into two parts: the "rigid" one associated with the low-lying states and the "flexible" one associated with the highly excited states. The "rigid" part $\hat{H}_{rigid}$ is specific to the system and the interactions that come from the environment are considered as perturbations. On the contrary, the "flexible" part $\hat{H}_{flexible}$ is associated with replacement of the idea about linear superposition with the conjecture that the interactions behave so that their impact at every point remains bounded. Let us recall that the linear superposition that is the ground of any perturbation approach, does not guarantee boundedness. Furthermore, I proved that the motion in the classical phase space and in its quantum counterpart is equivalent. The obtained decoherence gives additional support to that result making the entire frame self-consistent. Our evaluations delineate explicitly the properties of the temporal behavior of an entity in the low-energy limit. It turns out that it can be considered as temporary binding altered with scattering in other binding region. The corresponding level spacing statistics is presented by (5.19). This behavior is temporary since the entity eventually is either scattered back to the gas phase or passes to $\hat{H}_{rigid}$.

A great advantage of our approach is the emergence of the dynamical boundedness viewed as boundedness of the transition energies. Note that it appears naturally as a result of the evaluations: it has not been involved in the list of the fundamental assumptions of the new concept for the low-energy limit.

Nevertheless, we swamp into a problem: on the one hand the dynamical boundedness appears on the quantum-mechanical level. On the other hand, the fine-tuning causes divergence of the scattering length which makes cross-sections divergent. In turn, the rates of all elementary processes become divergent as well. As a result, the divergence of the rates violates the dynamical boundedness on macrolevel. Can we get over the problem? This issue is one of the key points in the next chapter where I shall present the operational protocol of coherence. One of its advantages is that it ensures the boundedness of the cross-sections and hence assists the dynamical boundedness on macrolevel.

## Chapter 6: Coherence Mechanism II. Operational Protocol

### 6.1 Why to Read Chapter 6

In the previous chapter we revealed the necessity of coherence for open systems. It is supposed to be the implement for elimination of the induced spatial non-homogeneity that arises from the multi-valuedness of the stochastising interactions - different occupation is established independently at closest points of a system so that the induced non-homogeneity is spread over every scale of the system and is sustained by permanent bombardment that comes from the gas phase. I put forth the idea that the coherence operates as mechanism that yields "self-induced stabilization" through coupling among distant entities. Since the physical agent of the coupling is the cooperative excitations of a media, the presence of media appears as fundamental ingredient of a successful coherence mechanism. An obvious general requirement is that it should preserve the chemical identity. The only processes that meet this demand are adsorption and/or absorption. Hence they acquire enormous importance that goes beyond any concrete model. Successful modeling of these processes is given by the lattice-gas approach presented in &5.1.

A successful mechanism for elimination of the destabilization driven by the induced non-homogeneity should be grounded on a strong coupling adlayer-interface, namely: the energy of colliding entities dissipates to local cooperative excitations of the media. In turn, the impact of these local modes on the colliding entities is supposed large enough to induce a new transition that dissipates through the excitation of other local cooperative modes and so on. The feedback ceases its action whenever the colliding entities response becomes coherent.

However, the used so far weak-coupling approach renders any feedback set on coupling adlayer-interface local and non-correlated both in space and time because it provides only local evening of current states of the colliding entities. Indeed, suppose that the collision energy dissipates through excitation of local cooperative modes. The weak-coupling approach considers any change of adsorption Hamiltonian due to these excitations a perturbation. Consequently, the latter cannot produce large enough change of the entity state to induce a new transition. As a result, the interaction colliding entity-interface stops. Since only short range interactions are considered, a new transition happens only when other entities, brought together by their mobility, collide. In order to open the door for strong coupling adlayer-surface I put forth the assumption that the adsorption Hamiltonian is decomposed into two parts: the "rigid" one where the environmental influence is regarded as perturbation and the "flexible" one where the linear superposition of the interactions is replaced with permanent boundedness. The radical novelty of these assumptions becomes clearer under the consideration that the operation that preserves the boundedness is the coarse-graining, an essentially non-linear operation. An important outcome is that unlike the linear superposition, the coarse-graining does not preserve the chemical identity. In the previous chapter I made evident that the chemical identity in the low-energy limit is got lost and all effective "flexible" Hamiltonians share the same level spacing statistics (5.19). This gives rise to the expectation that the feedback operates namely in the low-energy limit where it is insensitive to the particularities of the system.

Note that the suppression of the chemical identity makes the differentiation between entity and media spurious. Indeed, the non-linear response due to application of the coarse-graining as operation that preserves boundedness strongly suggests inseparability between the flexible Hamiltonian and the media. This consideration supports the idea that the transitions in the low-energy limit are non-radiative. Summarizing, I suppose that the feedback involves the following succession of steps: the collision energy of the entities in the low-energy limit dissipates through excitation of local cooperative excitations of the surface; in turn they participate to the flexible Hamiltonian so that to induce a new transition. The latter dissipates non-radiatively again trough excitation of other local modes and so on. The process stops on making all the flexible Hamiltonians response coherent.

My further goal is to demonstrate that the feedback is a scale-free process: i.e. it does not contribute with any specific temporal or spatial characteristics to the rates of the elementary processes. It implies that though the coherence is an essentially cooperative process, the feedback acts towards imposing individual characteristics throughout the system. On my way of establishing the above property I shall rigorously prove that it is naturally related with the issue about the self-stabilization of a system.

### 6.2 General Premises of the Coherence

The major goal of this chapter is to illustrate explicitly how the feedback operates. The present task is to discuss two key premises for its successful execution. The feedback is driven by the collisions between entities in the low-energy limit. The uniform probability for finding an entity anywhere in the system obtained in &5.6 suggests that each entity collides effectively with every other, i.e. it substantiates the idea of fine-tuning as an analog of long-range interactions. So, it is natural to suppose high collision rate at every point of the system. The small energies associated with the low-energy limit suggest considering the collisions as perturbations. Then the target perturbation is regarded as sudden inclusion [6.1] because a collision is active in a very short time interval:

$$P = \frac{|\hat{V}_{ij}|^2}{\hbar^2 \omega_{ij}^2} \tag{6.1}$$

where $\langle j|\hat{V}|i\rangle$ is the matrix element of the transition from the initial state $i$ to the final state $j$ under the perturbation $\hat{V}$; $\hbar\omega_{ij} = E_j - E_i$, $E_i, E_j$ are the energies of the initial and final states correspondingly.

In &5.5 we have established that the nearest level spacing distribution obeys (5.19). Now I can make a step further: the feedback sets the value of $\beta(E)$ equal to unity. This happens because the succession of non-radiative induced transitions acts always towards making the response to further perturbations coherent. So, the nearest level spacing distribution in the low-energy limit coincides with the famous Wigner distribution:

$$P(E) \propto \frac{E}{\sigma}\exp(-E^2/\sigma^2) \qquad (6.2)$$

The same nearest level spacing distribution (6.2) shared by the effective "flexible" Hamiltonians that comes from different adsorption sites renders the transition energies $\hbar\omega_{ij}$ equiprobable in the range $[0,\sigma]$. The range $[0,\sigma]$ determines the types of available cooperative excitations through which the feedback operates. Since the low-energy limit is associated with the weakest bound states, it is to be expected that $\sigma$ is very small. Therefore the acoustic phonons which are gapless modes certainly are an available type of cooperative excitations.

Now the question about size of the area of the feedback operation comes. Since the local excitations are collective modes, it is expected that any excitation is spread throughout the entire system. However, the effect of any of them is pronounced only up to a wave-length distance because the excitations at different sites are not coherent. Evidently, the dissipation through cooperative excitations makes the feedback non-local: its "response" covers wave-length distance while the transition happens at a given site, i.e. at a point. Thus, the coupling among neighbor "flexible" Hamiltonians is carried out by the non-local feedback "response".

Very important for the further considerations is the particular case when the local excitations that participate to the feedback are acoustic phonons. Their dispersion relation:

$$\lambda = \frac{c}{\omega_{ij}} \qquad (6.3)$$

provides the specific property of the coupling, namely it covers larger areas on decrease of $\omega_{ij}$; $c$ is the sound velocity. Hence, the smoothing out of the energy difference between colliding entities results in spreading of the coupling over larger distances and eventually yields global coherence. Indeed, the driver of the feedback is the collisions in the low-energy limit. According to eqs.(6.2) and (6.3), the feedback based on excitation of acoustic phonons ensures expansion of the coupling size $\lambda$ on the decrease of the colliding energy $\omega_{ij}$. An obvious outcome is that the local configuration of the smallest $\omega_{ij}$ "couples" the largest number of "flexible" Hamiltonians. Since the coupled Hamiltonians behave coherently, they can be united in a single temporary Hamiltonian whose range of interaction is proportional to the number of coupled Hamiltonians. Outlining, the smallest change of $\omega_{ij}$ drives the most distantly spread coupling. So, more and more entities "act" coherently with the entity that initially is in the most favorable configuration. The process is completed when all "flexible" Hamiltonians becomes identical which in turn renders their further response coherent.

The above consideration is heuristic one and needs further specification in order to serve as grounds for rigorous mathematical derivation of the operational protocol. The major missing points are two: the specification of measure of coherence and explicit relation between it and self-induced stabilization.

### 6.3 Adsorption Rate

Now I am going to present decisive arguments that the adsorption (reaction) rate is the desired measure for coherence straightforwardly related to the self-stabilization. The most widely spread approach to the notion of a rate of an elementary process is the so called transition state theory. It has been introduced by Eyring and Wigner in order to explain the Arrhenius type thermal activation in the chemical reactions. Up to now this approach cherishes great popularity. A lot of modifications are proposed for its best adaptation to challenges such as liquids, solids, interfaces etc. The theory is based on the concept of a transition state intermediate between the reactants and products. The reactants are assumed to be in thermal equilibrium with the transition state. The resulting theory, transition state theory, takes into account only the statistical properties of the reactive system, not the microscopic details of the molecular collisions. Transition state theory has the advantage that the entire potential surface is not needed. Only the potential energy and the properties of the reactants as they proceed along the reaction pathway of the multidimensional potential energy surface are needed. Summarizing, the fundamental assumptions of the transition state theory assert:

(I) transition state entities that originate as reactants are in local equilibrium with the reactants;

(ii) any system passing through the transition state does so only once before the next collision or before it is stabilized or sermonized as reactant or product (no recrossing)

However, the transition state theory appears as totally inconsistent with the proposed by us approach to the low-energy limit. Indeed, since the reaction pathway always goes through low-energy limit where the chemical identity is lost, sharp demarcation between reactant and products is missing. Furthermore, it is impossible to ignore the microscopic details of the collisions because the multi-valuedness of the stochastising interactions violates the equilibrium between the reactants and the intermediates. Let me recall that the induced non-homogeneity yields not only destabilization. It makes the intensive state variables such as concentration ill-defined. Moreover, in the absence of mechanism that suppresses induced non-homogeneity, the lack of thermodynamical equilibrium is permanent. In turn, the determination of the reaction rates requires radically new approach, different from the transition state theory.

Our considerations show strong entanglement between the microscopic dynamics in the low-energy limit and the statistical properties of a system. Indeed, as we already mentioned, the lack of thermodynamical equilibrium is due to the destabilization induced by the multi-valuedness of the stochastising interactions. Measure of this destabilization is the difference in the occupation. In general, the latter is proportional to the adsorption rate the major property of which is the multi-valuedness. Yet, a closer look on the definition of stochasising interaction (&2.5 and &5.2) shows that the multi-valuedness has been introduced as a property of the cross-section. I assume parity between the probability for adsorption and the cross-section. Consequently, the adsorption rate is determined as the ratio between the cross-section and the time interval in which the adsorption is completed. An immediate outcome of this definition is that the local adsorption rate varies from one active site to another in the absence of coherence. So, the measure of successful coherence is the evening of the local adsorption rates.

Yet, these considerations give rise to number of questions the most important of which are: how does coherent response of the "flexible" Hamiltonians stabilize the system? Since the coherent response of the "flexible" Hamiltonians is supposed accompanied by evening of the local adsorption rates, is the coherence a scale-free process?

### 6.4 Stabilization through Coherence

Now I am ready to discuss the central issue of the present chapter, namely how the coherence provides stabilization. Let us recall that the implement of the coherence, the non-local feedback, has the specific property to couple more and more "flexible" Hamiltonians on decrease of the colliding energy. Since the already coupled "flexible" Hamiltonians respond coherently, they can be united in a single temporary Hamiltonian the size of which is set upon the condition for turning its chemical potential zero. Therefore, according to the considerations in &3.7, the temporary Hamiltonian decomposes into two parts: the stable "bulk" one which is insensitive to the interaction with the environment and the unstable "surface" one which is subject to changes under the interactions with the environment. Note that in configurational space the size of the "bulk" temporary Hamiltonian is proportional to the number of the coupled "flexible" Hamiltonians and correspondingly to the coupled area; the "surface" part is located in a strip around the boundary. Due to the interactions with the neighboring temporary Hamiltonians, the total chemical potential eventually turns to zero and the coherence is destroyed. This consideration gives rise to key dilemma: does the interaction among temporary Hamiltonians yields formation of a new temporary Hamiltonian or some averaging out occurs? The major implement of revealing that dilemma goes through study of the stability of the temporary Hamiltonians. Now we are ready to work out the details of disclosing the dilemma.

The interaction between any two adjacent temporary Hamiltonians is constituted by the collisions at their boundaries. The intensity of each collision is determined by (6.1) where $\hbar\omega_{ij} = E_j - E_i$, $E_i$ ($E_j$) is the energy of an entity of the temporary Hamiltonian $i$ ($j$). Since the chemical identity of entities in the low-energy limit is lost, the collisions neither introduce nor involve any specific scale associated with the details of their dynamics. Formally this is carried out by associating the collision rate with the power $X^{\nu(X)}$, where $X$ is the size of a given Hamiltonian. $\nu(X)$ is non-constant exponent determined so that $X^{\nu(X)}$ gives the collision strip "weight" expressed trough the current size of the given temporary Hamiltonian. The advantage of that presentation is that it makes the collision rate scale-free. Since the collision strip has finite width, the gradual increase of the size of the temporary Hamiltonian is accompanied by gradual decrease of $\nu(X)$. The justification of the range of $\nu(X)$ is discussed further.

The premise that initially the entities are randomly distributed over the low-energy limit renders all $\omega_{ij}$ equiprobable. Then the initial iterations of the coupling of $N$ ($N$ arbitrary) temporary Hamiltonians are governed by the following system of ordinary differential equations:

$$\frac{dX_i}{dt} = \sum_{j=1}^{z_i} \frac{|V_{ij}|^2}{\hbar^2 \omega_{ij}^2} X_i^{\nu(X_i)} \operatorname{sgn}(i,j) \tag{6.4a}$$

$$\frac{dX_j}{dt} = \sum_{k=1}^{z_j} \frac{|V_{jk}|^2}{\hbar^2 \omega_{jk}^2} X_j^{\nu(X_j)} \operatorname{sgn}(j,k) \tag{6.4b}$$

...........................................

$$\frac{dX_N}{dt} = \sum_{l=1}^{z_N} \frac{|V_{Nl}|^2}{\hbar^2 \omega_{Nl}^2} X_N^{\nu(X_{Nj})} \operatorname{sgn}(l,N) \tag{6.5c}$$

where $z_i$ ($z_j$) is the number of the nearest neighbors of the $i$ ($j$) temporary Hamiltonian.
The initial conditions are:
$$X_i = X_j = \ldots = X_N = M \tag{6.6}$$
There is an additional condition, namely:
$$\sum_{i=1}^{N} X_i(t) = \sum_{i=1}^{N} X_i(0) = NM \tag{6.7}$$

In order to eliminate the dimension factor, I rescale all $X_i$ by the factor $NM$. So, $X_i$ becomes $\frac{X_i}{NM}$. Thus the range of rescaled variables is always $(0,1]$. The time is also rescaled with respect to the overall time of the coupling and also varies in the range $[0,1]$.

Further all $V_{ij}$ ($i,j = 1,\ldots,N$) are supposed equal because they come from the same type of interaction and so they are of the same order. Since $\omega_{ij}$ is the same for the $i-th$ and the $j-th$ temporary Hamiltonian, additional considerations are necessary to determine the sign of each term that involve these temporary Hamiltonians. Let us consider each of the equations in (6.4) separately from one another, i.e. to consider all the terms in their *r.h.s.* positive. The temporary Hamiltonian that grows fastest is that of the smallest $\sum_{j=1}^{z_l} \omega_{ij}$. For the sake of simplicity let it be the $i-th$ temporary Hamiltonian. Figuratively, it implies that the $i-th$ temporary Hamiltonian, being in most favorable configuration compared to its neighbors, remains the only one stable during the interaction and expands with accelerating rate at the expense of the other ones that are destabilized by the interaction. Then the contribution of the interaction between the $i-th$ and the $j-th$ temporary Hamiltonian is positive for the $i-th$ temporary Hamiltonian and is negative for the $j-th$ temporary Hamiltonian. Thus, in this particular case, $\text{sgn}(i,j)$ in (6.4a) is positive while $\text{sgn}(j,i)$ in (6.4b) is negative.

Eqs.(6.4) has a single stable solution that is:

$$X_i = 1, \qquad (6.8a)$$
$$X_j = 0 \quad j \neq i \quad j = 1,\ldots,N \qquad (6.8b)$$

where $X_i$ is that temporary Hamiltonian which is initially in the most favorable configuration. Thus, the temporary Hamiltonian that initially is in the most favorable configuration is the only one that remains stable during the interaction with its neighbors. The others are destabilized by the interaction and figuratively speaking, are "swallowed up" by the stable one.

An important property of eqs.(6.4) is that its solution does not depend on $N$. This fact allows considering the coherence of an infinite system in the following way. Initially a partition of the system into areas of finite size is made. The above procedure is applied to each area. The next iteration involves interactions between new (larger) areas at which the previous areas appear as temporary Hamiltonians.

The initial condition implies that the entities in the low-energy limit are homogeneously distributed throughout the surface, i.e. they form perfect mixture. This makes plausible to assert that initially the "bulk and the"surface" part of every temporary Hamiltonian are of the same order. This immediately sets the lower limit of $\nu(X)$ to unity. It was shown above that the enlargement of the temporary Hamiltonian is accompanied by gradual decrease of $\nu(X)$. The monotonic decrease of $\nu(X)$ in eqs.(6.4) provides permanently accelerated growing of the "most favorable" temporary Hamiltonian. Its asymptotic reads:

$$X(t) \propto t^{\frac{1}{1-\nu(t)}} \qquad (6.9)$$

The integration of a power function with non-constant exponent is presented in the Appendix to Chapter 1. The properties of $\nu(t)$ are determined by the diffeomorfism between $\nu(t)$ and $\nu(X)$. Consequently, it provides the scale-free behavior of the solution of eqs.(6.4). So, the monotonic decrease of $\nu(t)$ ensures an accelerated grow rate of the "most favorable" temporary Hamiltonian (see (6.9)). The speeding up of the grow rate of the "most favorable" temporary Hamiltonian makes possible its expansion beyond the percolation threshold in finite time interval. Thus the formation of the infinite cluster sets the large-size limit of $\nu(X)$, namely $\nu = 0$.

The accelerating grow rate of the "most favorable" temporary Hamiltonian ensures its stability to the temporary Hamiltonians created in later moments. Indeed, according to eq.(6.9), even more advantageous than the "most favorable" initial colliding energy $\hbar\omega_{ij}$ cannot overcome the difference in the size due to the time lag. So, the accelerating grow rate of the most favorable temporary Hamiltonian provides the required self-stabilization of the coherence.

The fact that the interaction between adjacent temporary Hamiltonians selects only one of them to remain stable implies that its specific adsorption rate is imposed over the joined new areas. Moreover, the lack of any scale in eq.(6.9) is warrant that the coherence does not impose specific scale(s) onto the established adsorption rate. So, the latter is determined only by the individual properties of the "most favorable" local one.

Now we are ready to answer the question whether the above mechanism of coherence provides dynamical boundedness on macrolevel. Let us recall that the divergence of the scattering length due to fine-tuning makes the cross-section of the elementary processes divergent. In turn, that divergence violates the dynamical boundedness of the adsorption (reaction) rates. Does the coherence resolve the problem with the divergence of the scattering length? The coherent response sustains the boundedness of the scattering length since it is a process that prevents fine-tuning. Indeed, the coherence response is rather equivalent to a specific kind of localization – it stops the non-correlated migration of the entities that is crucial for occurring of the stochastising interactions. In addition, the transitions involved in the feedback, being non-radiative, do not contribute to the cross-sections. In turn, this keeps the cross-sections finite. Therefore, the coherent response is the warrant of the cross-section boundedness. So, the dynamical boundedness is provided on the macrolevel. Note, that the localization is temporary – it lasts until the coherence is going on; after the entities enter the rigid part of the Hamiltonian, the non-correlated mobility is restored.

### 6.5 Size-Independent Fluctuations

The leading outcome of the previous section is that the global adsorption rate equals the individual one that initially is in the energetically most favorable configuration. The imposing of a local adsorption rate throughout the entire system renders the global adsorption rate independent from the particularities of the spatial configuration of the adentities in the low-energy limit. Yet, the major factor that drives the most favorable configuration is the stochastising interactions. Indeed, let us suppose that a stochastising interaction happens at given active site. As a result, it "shuttles" the corresponding entity from the 'rigid' part of the Hamiltonian to the low-energy limit. In result, the local energetic configuration can become the most favorable one. Therefore, the "initial condition" that determines the signs in eqs.(6.4) strongly depends on the probability for undergoing a stochastising interaction. Hence, the global adsorption rate acquires the most distinctive characteristic of the local adsorption rate, namely it becomes multi-valued function.

A foremost property of the global adsorption rate is the size-independence of the fluctuations amplitude. The later is limited by saturation threshold rather than the size of the surface as it is in the thermodynamics. The saturation threshold is a property particular for the adsorption and is straightforwardly set on the thresholds of stability of the system. This immediately renders the margins of the fluctuations independent of the size of the surface. In sequel, the size-independence of the margins renders the boundedness of the fluctuation amplitudes: locally it is ensured by the hard-core repulsion - the property that only one entity can be adsorbed at a single site; globally it is ensured by the presence of saturation threshold. The above considerations support once more the idea that the adsorption is a process that successfully substantiates the idea about coherence since their interplay not only provides self-induced stabilization but it naturally makes the boundedness available both on micro- and macro-level as well.

In addition, I should stress on the point that neither system subject to coherence can reach thermodynamical equilibrium because the coherence makes the states associated with different selections equiprobable. Therefore, there is not single equilibrium state that appears as global attractor; on the contrary, there is permanent motion in the state space among equiprobable states. The basic constraint to that motion is the boundedness, not the attraction to equilibrium state! This result substantiates our considerations about eq.(2.15). Now we come again to the same equations and note - without any reference to the thermodynamics. Moreover, the derivation of eqs.(2.15) is based on the introduced by us radically novel concepts about the multi-valuedness of the stochastising interactions and the leading role of the boundedenss in the low-energy limit.

However, we have not yet finished our story. The coherence changes the temporal behavior of the process of adsorption - instead of continuous in time it turns out to be discrete. Indeed, the adsorption starts with a coherence session - imposing the most favorable selection throughout the entire system; after the coherence session is completed, the entities enter the "rigid" part of the spectrum where the relaxation to the ground state is completed; after that starts a new session of coherence where another selection becomes dominant. However, the change of the temporal behavior from continuous to discrete one gives rise to the question about the validity of eqs.(2.15) – does the "switch" from continuous to discrete time causes only their modification or we should expect some new properties? I shall consider this problem in &7.3.

### 6.6 Operational Universality

The obtained lost of chemical identity in the low-energy limit actually provides the greatest advantage of the proposed coherence mechanism, namely it makes its action universal and operationally equivalent in systems of different nature. Indeed, the universality of the feedback low-energy limit $\Leftrightarrow$ acoustic phonons is set on the insensitivity of the low-energy spectrum to the chemical identity of the reacting entities (the level spacing distribution is characterized by a single parameter, see (6.2)) and the insensitivity of the acoustic phonons to the details of the adentites configuration and the lattice itself (its dispersion relation involves a single parameter - sound velocity, see (6.3)). So, the feedback operates similarly in all natural systems.

In turn, one can recognize the coherence as driving mechanism for permanent macroscopic fluctuations. Further, the universality of the coherence mechanism makes these fluctuations indispensable part of the temporal behavior of all open natural systems at every control parameters choice. The outstanding property of these fluctuations is their boundedness. It is to the point to recall that the dynamical boundedness appears both on micro- and macrolevel – on microlevel it is asocoated with the boundedness of the transition energies, while on macrolevel it is associated with the boundedness of the global adsorption rate.

Yet, our curiosity faces the question whether there is an operational equivalence between the coherence mechanisms that operates in natural and social systems. In the Preface to this book I presented some general properties of the coherence that acts in the social systems. Now I am recalling the operational protocol derived from those considerations:

(I) the coherence operates in those many-body systems that are subjects to both local and global destabilization. The understanding of the local one, derived from its "social" context, implies that the interaction of each entity with its neighborhood executes multi-valued response so that one selection, arbitrarily chosen among all available, is realized at given instant. At the next instant, the entity responses via other selection, again arbitrarily chosen. As a result, the spatio-temporal configuration of the local fluctuations permanently varies which gives rise to the global destabilization.

(ii) the coherence is achieved through very fast process that eliminates the destabilization though making all entities to share the properties of a single one. As a result, all the entities share the characteristics of the same selection.

(ii) Since the succession of the selections is arbitrary, it is to be anticipated that the global characteristics exhibits permanent irregular variations in time.

So, indeed there is total equivalence of the operational protocol proceeding in social and natural systems. Moreover, the destabilization of the personal opinion that exhibits a disoriented response to the signals that come from its immediate neighborhood is very similar to the lost of the identity in the low-energy limit. In other words, the public discussion of the issues that cause local and global destabilization is similar to a transition to the low-energy limit where the identity is lost. Further, the imposing of a single selection throughout both social and natural systems is brought about by the property of the coherence mechanism to be scale-free process. This is amazing property: though the coherence operates under the condition of lost identity, it imposes individual characteristics without blurring them!

### 6.7 What Comes Next

The operational protocol of the coherence developed in the present chapter reveals non-trivial and unexpected properties. The most prominent among them is the execution of size-independent macroscopic fluctuations whose hallmark is their boundedness both on micro- and macrolevel. In turn, since the global rates participate to the evolutionary equations (2.15), their boundedness provides the incremental boundedness introduced in Chapter 3. Bearing in mind that in Chapter 3 we developed the properties of the motion in the state space under the fundamental constraint of the incremental boundedness, note that now it emerges naturally from the operational protocol of the coherence. So, the entanglement between the coherence and the boundedness appears as two-fold comprehension of the functional relation between execution fluctuations and long-term stability. Indeed, we have found out the following interrelation: the coherence provides boundedness and the boundedness provides long-term stability. Yet, the fundamental concept that makes coherence possible is that of the explicit boundedness of the interactions in the low-energy limit. Thus, the circle closes in self-consistent theory.

A property of the coherence that deserves special attention is that it makes impossible reaching of equilibrium due to the execution of macroscopic fluctuations. Unlike the thermodynamical equilibrium that is single state, the coherence makes a range of nearest states equiprobable. This renders the motion in the state space to be executed as multi-valued function: each step is random choice among all available. The advantage of the multi-valuedness is in breaking the long-term correlations among the time scales. Now we can explain naturally why the fluctuations are irregular and scale-free: their irregularity is a result of the multi-valuedness of the motion in the state space. Once again our theory manages to explain successfully a key property of the fluctuations.

However, it seems that there is a serious flaw is our considerations: on the one hand, the use of the rates of the elementary processes as measure for coherence makes the description of the evolution to be in terms of balance equations for all elementary processes, i.e it justifies the use of eqs.(2.15). On the other hand, the change of the temporal behavior from continuous in time to discrete one gives rise to the question about their appropriate mathematical form. I the next Chapter I will consider this issue along with other amazing new properties that the discretization of the time imposes.

### References

6.1. Landau L.D. and Lifschitz E.M.; Quantum Mechanics. Non-Relativistic Theory, Nauka: Moscow; 4 Edition, 1989, p.183 (in Russian)

## Chapter 7: Discretization of Time

### 7.1 Why to Read Chapter 7

So far we have developed the theory of the many-body systems in view of the interplay of the stochastising interactions and the boundedeness. The central result of the study is that the macroscopic behavior separates into two completely different classes: the thermodynamic-like and the antithermodynamic-like one. The first one is associated with the diluted closed systems where the stochastising interactions yields establishing of thermodynamical-like behavior, i.e. once the equilibrium is reached, the system remains in it and never exerts significant deviations. It turns out that this behavior is a result of breaking any long-range correlations among velocities and positions of the entities both forward and backward in time due to the action of the stochastising interactions. Note that the specific role of the stochastizing interactions is brought about by the lack of dynamical time-reversal invariance, assumption that fundamentally disagrees with the thermodynamics. Remember that the latter is grounded on the postulate that every interaction is invariant under the reversal of time. However, this invariance gives rise to one of the greatest paradoxes of the thermodynamics: that of the inconsistency between the monotonic approach to the equilibrium and the possibility for significant departure from it on turning back time. Indeed, the zeroth law of the thermodynamics asserts that the equilibrium is a single state that has global attractor – whatever the initial conditions are the system in question monotonically approaches the equilibrium and once reaching it stays there forever. However, the time reversal invariance of the interactions opens the door for significant departures from the equilibrium on turning back time. On the other hand, the specific role of the stochastising interactions is to break the long-range correlations among the velocities and positions of the entities both forward and backward in time. Therefore, once the equilibrium is reached, the system never exerts significant departures from it.

The second class, the antithermodynamic-like one, is associated with dense open systems where the stochastising interactions play new role: they trigger destabilization of the system that, if not suppressed, would produce breakdown in short time. The elimination of the destabilization is accomplished by means of the coherence. In the previous Chapter we have established that the latter generates permanent macroscopic fluctuations of the rates of the elementary processes. In turn, they bring about the "antithermodynamic-like" behavior of the system. Indeed, permanent executing of fluctuations makes possible transitions between states of different entropy. Furthermore, the realization of the fluctuations as excursions of finite durations renders the transitions in the thermodynamical direction (the direction from the state of lower to the higher entropy) and the antithermodynamical one (the direction from the state of higher to the lower entropy) not only equiprobable but also renders them to happen in finite time intervals. Thus, the antithermodynamical behavior is not an exotic phenomenon but is rather ubiquitous and frequent for the open systems where the coherence operates.

Besides, the coherence makes a local quantum phenomenon (local adsorption rate) to acquire macroscopic impact (the global adsorption rate equals that local one which is in the most favorable energetic configuration). This result is surprisingly away from the thermodynamics. To remind that the latter supposes that the macroscopic behavior of the many-body systems is governed by relations among few variables such as pressure, concentration, temperature that have no dynamical analog. Therefore, the dynamics appears subordinated to the relations established by these variables. On the contrary, in our approach the macroscopic characteristics such as the global rates of the elementary processes come out straightforwardly from the dynamical interactions on quantum level. So, shall we expect that other quantum properties that originate from the coherence can be traced in the macroscopic behavior of a system and, if so, how? Further in this chapter I shall discuss the manifestation of one such quantum effect – the discretiztion of time whose commence is the alternation of coherence sessions and sessions of relaxation through the rigid parts of the single-site Hamiltonians. It is to be expected that the discretiztion of time brings about radically novel and non-trivial properties. At the end of the previous Chapter I have already mention one of them: that about the necessity of modification of eqs.(2.15). The dilemma is how serious is the modification: is it a specific mathematical problem or one should anticipate non-trivial properties. I shall discuss this topic in &7.3.

Along with looking for its macroscopic manifestations, an important task is to find out how the discretization of the time is running on the quantum level, i.e. how it is incorporated in the quantum spectra. My interest in the topic is provoked by the fundamental for the quantum mechanics assumption that all quantum processes are continuous in time. Besides, since the discretization of time is a property specific for the coherence, it is to be expected that its manifestations give rise not only to clear differentiation with the continuous processes but also delineate properties specific for discriminating the coherence.

## 7.2. Continuous Band in Quantum Spectra

I choose to start with elucidating the issue how the discreteness of the time comes into view in the quantum spectra. One of the most widely used methods for studying the properties both of individual atoms/molecules and of many-body systems is their interaction with radiation. Thus, certainly it is worth studying how the coherence and discretization of time are involved in it.

Being a quantum phenomenon, it is to be expected that the coherence is manifested on the quantum level. Next, the issue how the coherence is related to the quantum spectra such as IR, EXAFS etc. is considered. At first sight the issue is controversial since one of the major assumptions advanced in the book is that the transitions in the low-energy limit are non-radiative and their energy dissipates through excitation of appropriate cooperative modes. Then, it seems that the transitions in the low-energy limit do not contribute to the corresponding quantum spectra. Yet, the problem about the influence of the radiation on the coherence stands. My task is to elucidate that it gives rise to a continuous band with atypical for the traditional quantum mechanics properties. The first step is to verify that the coherence is "transparent" to the radiation regardless to its frequency.

The photons do not interact with the entities in the low-energy limit because they have neither charge nor magnetic moment nor polarization: recall that in the low-energy limit they lose their chemical identity and are characterized by a single parameter, energy. So, the radiation elapses unchanged through the "flexible" part of the spectrum where the coherence takes place. Therefore, the emitted radiation comes from the interaction between the incident photons and excited entities at the rigid part of the spectrum. However, the temporal behavior of the photon emission is fundamentally influenced by the strong alternation in the time course of coherence sessions and sessions of relaxation trough the rigid part of the spectrum. Indeed, since the radiation elapses through coherence sessions, the photon emission is available only when the entities are at the rigid part of the spectrum. Then the photon emission is not continuous in time process but exhibits discrete behavior: the subsequent emission sessions are separated by "silent" intervals, i.e. time intervals in which the emitted radiation elapses through coherence sessions. Inasmuch as the length of the silent intervals is much smaller than the length of the emission sessions it is plausible to assume that the successively emitted wave trains interfere. Evidently, that interference is incorporated in the quantum spectra. Next, it is proven that it gives rise to a continuous band of certain shape superimposed on the discrete one that comes from the interaction in the rigid part of the spectrum. It should be stressed that the continuous band does *not* correspond to any real radiation but it is a result of the discrete nature of the photon emission. It is to the point to stress that the discrete nature of the photon emission commences from the different characterization of the entities in the coherence sessions and the sessions of relaxation through rigid part of the spectrum. Indeed, the lost of the chemical identity in the coherence sessions makes the entities not to interact with the radiation. On the other hand, the interaction entities ⇔ radiation is strongly specific to the system in the rigid part of the spectrum where the chemical identity is fully restored. Therefore, the alternation of sessions with different behavior naturally gives rise to a discreteness of the emission.

I start with the consideration how the interference becomes involved in the autocorrelation function of the photon emission. Hereafter $\tau$ denotes the length of an emitted wave train and $r_i$ is the length of the $i-th$ silent interval. The length of the silent intervals exhibits temporal variations because the coherence sessions are not identical. The silent intervals are uniformly distributed over bounded range the margins of which are rendered by the shortest and largest duration of the coherence sessions ($r_{min}$ and $r_{max}$ correspondingly). For the sake of simplicity let us start with the case when only nearest wave trains interfere. Then the autocorrelation function $G(T)$ of the interference reads:

$$G(T) = \sum_{j=1}^{K} \sum_{n=1}^{N} |g(\omega_n)|^2 \exp(i\omega_n r_j) \qquad (7.1)$$

where $T$ is the duration of the measurement, $K$ is the number of the silent periods in the time series; $g(\omega_n)$ is the wave function amplitude of an emitted photon of energy $\hbar\omega_n$. Since the emitted photons comes from the rigid part of the spectrum where the nearest level spacing is always non-zero, the cross-section of any interaction (scattering) between a photon and an excited entity is always finite regardless to the particularity of the interaction (scattering). Boundedness of the cross section renders $g(\omega_n)$ finite. In turn this renders $G(T)$ to be bounded irregular function. Note that the irregularity arises because the length of the silent interval varies in the time course. Thus, being BIS, the autocorrelation function shares their universal properties established in Chapter 1. One of them is that the power spectrum of a time series of arbitrary but finite duration $T$ uniformly fits the shape:

$$S(f) \propto \frac{1}{f^{\alpha(f)}} \qquad (7.2)$$

where $\alpha(f) = 1 + \kappa\left(f - \frac{1}{T}\right)$. A distinctive property of the spectrum, proven in &1.4, is that it has an artificial infrared edge $f_{min}$ associated with the length of the time series $T$ through the relation $f_{min} = \frac{1}{T}$. It has been proven that this edge is related to the boundedness of the time series itself and is insensitive to the particularity of the fluctuation succession.

The presence of silent intervals restricts the number of subsequent wave trains that can interfere at an instant and hence provides the boundedness of the autocorrelation function $G(T)$. That number varies in the range $\left[1, \frac{\tau}{r_{min}}\right]$. On the contrary, any random emission violates boundedness since the lack of silent intervals allows the number of interfering wave trains to become arbitrarily large. Thus, the boundedness of the autocorrelation function is justified only for a discrete emission.

Summarizing, the discrete nature of the photon emission is manifested through persistent appearance in the quantum spectra of continuous band of shape $1/f^{\alpha(f)}$. The lack of chemical identity during coherence sessions makes the presence and the shape of the band insensitive to the particular properties of both incident radiation and details of the studied system. On the other hand, the continuous band is superimposed to a discrete band that is highly specific because it comes out from the interaction between the system and the radiation that takes place in the rigid part of the spectrum where the chemical identity is restored. Once again it is worth noting that the continuous band does *not* correspond to any real radiation.

It should be stressed that the $1/f^{\alpha(f)}$ behavior of the quantum spectra predicted by us is fundamentally different from that mechanism of quantum $1/f$ noise proposed by Handel [7.1-7.2]. He has supposed that the $1/f$ noise comes from interference between elastically and inelastically scattered waves, which emerge when a beam of particles is scattered under the influence of a potential. The model predicts $1/f$ noise in any system whenever the cross section for scattering of particles exhibits an infrared divergence of low-frequency excitations. Later this mechanism has been strongly criticized [7.3 and references therein] and it has been rigorously proven [7.4] that it does not bring about $1/f$ behavior. Our considerations about the "transparency" of the coherence to the radiation are in accordance with that critic. Indeed, in the present model the low-frequency transitions that exhibit infrared divergence are non-radiative and thus they do not contribute to the quantum spectra.

Next I shall present a simple discrimination criterion between Handel and our model. The major property of the $1/f^{\alpha(f)}$ power spectrum is the persistent presence of an infrared edge $f_{min}$ that is inverse proportional to the length of the time series $T$, namely $f_{min} = \frac{1}{T}$. On the contrary, Handel has predicted that the theoretical power spectrum is $1/f$ for all the frequencies in the range $[0, \infty)$. In turn, this yields that the power spectrum coming from a measurement of duration $T$ does not signal out cut-off associated with $T$. Indeed, let the predicted power spectrum is $g(f), f \in [0, \infty)$. Then, the power spectrum that comes from a measurement of duration $T$ is:

$$g(f,T) = \frac{1}{T} \int_0^T S(t) \exp(ift) dt \tag{7.3}$$

where $S(t) = \int_{-\infty}^{\infty} g(f') \exp(-if't) df'$ (7.4)

Then:

$$g(f,T) = \frac{1}{T} \int_0^T \exp(ift) \int_{-\infty}^{\infty} g(f') \exp(-if't) df' dt =$$
$$= \int_{-\infty}^{\infty} \frac{\sin((f-f')T)}{(f-f')T} g(f') df' \tag{7.5}$$

It is obvious that $g(f,T)$ becomes closer and closer to $g(f)$ on the increase of $T$; in other words, the following limit holds:
$$\lim_{T \to \infty} g(f,T) = g(f) \tag{7.6}$$
Note that $g(f,T)$ comprises all frequencies in the range $(0, \infty)$! Thus, the presence of non-zero band to the left of $f_{min} = \frac{1}{T}$ supports Handel approach while the presence of a sharp cut-off at the frequency $f_{min} = \frac{1}{T}$ confirms our theory.

Note that the presence of the $1/f^{\alpha(f)}$ - band in the quantum spectra is associated with the coherence only. So, it can serve as criterion both for the presence of coherence and exhibiting of macroscopic fluctuations while its absence should be associated with thermodynamical-like behavior.

## 7.3 Evolution: Discrete Mappings and Differential Equations

The alternation of coherence sessions and sessions of relaxing trough $\hat{H}_{rigid}$ is fundamental property of the temporal behavior of the open many-body systems subject to coherence. In the previous section we established that it is mapped into all quantum spectra as continuous band of shape $1/f^{\alpha(f)}$. However, my curiosity goes further: can the discretization of time be traced up to macrolevel? This seems puzzling because eqs.(2.15) describe the macroscopic evolution of a continuous in time process. On the other hand, we have just established discretization of time on the quantum level. In result, the evolution on the quantum scale is governed by the following discrete mapping:

$$\frac{\Delta \vec{x}}{\Delta t_i} = \hat{A}_{av}(\vec{x}) + \hat{\eta}_{ai}(\vec{x}) - \hat{R}_{av}(\vec{x}) - \hat{\eta}_{rj}(\vec{x}) \qquad (7.7)$$

where $\vec{x}$ is the vector of the macroscopic variables, e.g. concentration; $\hat{A}_{av}(\vec{x})$ and $\hat{R}_{av}(\vec{x})$ are the rate averages; $\hat{\eta}_{ai}(\vec{x})$ and $\hat{\eta}_{rj}(\vec{x})$ are zero-mean BIS; the index $i$ and $j$ are put to stress the stochastic nature of that terms; $\Delta t_i$ is the length of the $i-th$ session. The alternation of the two types of sessions is taken into account through discretization of time: $\vec{x}$ changes only during the sessions of relaxing through $\hat{H}_{rigid}$; it remains constant during coherence sessions. However, the discretization of time gives rise to the dilemma: is there any specific time scale so that (7.7) holds for the time scales smaller than it and (2.15) holds for larger time scales. Yet, the existence of such time scale would introduce specific correlations among certain time scales whose origin should be traced in specific entanglement of the dynamical and state variables. Note that this entanglement is equivalent to physical process that correlates those time scales. However, any presence of specific correlations among time scales strongly interferes with the advanced in Chapter 1 scaling invariance of the fluctuation succession. Bearing in mind that it implies lack of any physical process that correlates time scales larger than the fundamental one, one should adopt scaling invariance of the evolutionary equations as the only alternative consistent with the approach developed so far. The scaling invariance immediately validates discrete mappings (7.7) to be the only appropriate evolutionary equations that hold on every time scale larger than the fundamental one. Thus, we face the question how far the properties of the discrete mappings deviate from the properties of (2.15) established in &3.2. The solutions of the discrete mappings (7.7) and their differential counterparts (2.15) are entangled in very tricky interplay of similarities and differences. I shall start with the major similarity.

Along with eq.(7.7) let us consider its "deterministic" part:

$$\frac{\Delta \vec{x}_{\det}}{\Delta t_i} = \hat{A}_{av}(\vec{x}_{\det}) - \hat{R}_{av}(\vec{x}_{\det}) \qquad (7.8)$$

I assert that the power spectrum of the solution of (7.7) comprises additively two parts: the discrete one that is brought about by the solution of (7.8) and the continuous one that comes out from the stochastic part that form BIS. The additivity can be derived in completely the same way as we did for its differential counterpart (2.15) in &3.2. Now I am going to prove something more: though the solution of (7.7) is a discontinuous function, the power spectrum of its stochastic part fits the shape $1/f^{\alpha(f)}$ as its differential counterpart does. The first step of the proof is to approximate the discontinuities by a smooth function. This point immediately gives rise to the question how the interpolation interferes with the genuine properties of the original function. The answer comes easily: since the approximation of discrete BIS by a smooth BIS implies that the interpolation acts as an operation of coarse-graining, the power spectrum of the interpolated BIS fits the shape $1/f^{\alpha(f)}$ regardless to the details of the statistics of the original BIS. In result we come to an amazing effect: the shape of the continuous band in the power spectrum is robust to the details of the interpolation! In turn, it makes the process of obtaining the specific information encoded in the discrete band unambiguous and reproducible. In our approach the "noise" is related to both natural processes such as coherence and artificial intervention such as interpolation and recording. Since the latter are always encapsulated in the continuous ("noise") band whose shape $1/f^{\alpha(f)}$ is robust to the details of their statistics, the extraction of the discrete band from a spectrum proceeds with accuracy that is independent of the noise statistics. Let us now suppose for a minute that the shape of the continuous band depends on the statistics of time series. It would immediately give rise to inevitable and inseparable entanglement of the human intervention and the genuine properties of the system. Moreover, if this were the case, no reproducibility of the results would be possible. Note that in our approach the "noise" part of time series is not reproducible but its power spectrum is. It should be stressed once again that the insensitivity of the noise band shape to the time series statistics is property of the BIS only. On the contrary, the lack of boundedness yields strong dependence of the power spectrum shape on the statistics of time series.

Let us now face the major dissimilarity between the discrete mappings (7.7) and their differential counterpart (2.15). The major difference comes out from their deterministic parts, namely eq.(7.8) exhibits broader spectrum of solutions than its differential counterpart that reads:

$$\frac{d\vec{x}_{det}}{dt} = \hat{A}_{av}(\vec{x}_{det}) - \hat{R}_{av}(\vec{x}_{det}) \tag{7.9}$$

One already established new type of solution that (7.8) has while (7.9) has not, is the so called Feigenbaum cascade. The latter appears as solution of the discrete analog of the logistic equation:

$$\frac{\Delta x}{\Delta t} = kx(1-x) \tag{7.10}$$

It turns out that at certain values of $k$ (7.10) comprises atypical discrete band in its power spectrum. Moreover, the frequencies tend to double on changing $k$ so that the power spectrum becomes continuous on $k$ reaching certain value. This intriguing behavior has no analog in its differential counterpart. That is why it has been proclaimed as one of the general routes to deterministic chaos. However, the open problem of the entire theory of the discrete mappings has been how the discreteness of time appears. Here we arrive just to the point: the discreteness of time is an immediate outcome of the approach developed by us. Moreover, it is to be expected that it is presented not only on the quantum level, but on macrolevel as well. This is an effect of the scaling invariance of eqs.(7.7). Remember the pivotal assumption introduced already in Chapter 1: all time scales contribute uniformly! Applied to the present considerations it implies that the evolutionary equations are invariant under rescaling of time. Then, the same eqs.(7.7) describe the evolution starting on quantum level up to arbitrary large time scale. This beautiful result makes our approach self-consistent: we assume that there are no physical correlations among distant time scales. In result, we obtain that the evolutionary equations for a given process are invariant under rescaling of time. So, indeed no physical correlations among time scales appear during a stable evolution.

Now I must admit that the evolutionary equations are rather discrete mappings of the type (7.7) than the differential equations of the type (2.15). The broader spectrum of solutions of the discrete mappings occurs both on quantum spectra and on any other output record on macrolevel. The study of these novel properties is challenging task because the classification of the mappings is still incomplete. Yet, the replacement of (2.15) with (7.7) does not interfere with the obtained so far results because the only difference between them comes out from their "deterministic" parts that give rise to the specific properties. On the other hand, all our efforts so far have been focused on working out general and universal properties introduced by the boundedness.

The major advantage of our approach is that it substantiates natural fundament for the use of discrete mappings as mathematical implement for describing the evolution. The discrete behavior radically differs from the widely accepted and taken for granted continuity of the processes in the Nature. Furthermore, the acquired natural discretization of time is related to every process that involves coherence. Hence, the continuity and discreteness have clear demarcation: the former is available for the systems that do not need coherence for sustaining their long-term stability while the discreteness appears for the systems that require coherence for sustaining their long-term stability.

Note that the obtained discretization of time emerges from the interplay of the coherence and the boundedness, not from a space-time quantization as in the quantum mechanics. That is why it appears both on quantum and classical level.

And last but not least let us consider the relation between the discrete mapping (7.7) and the motion in the state space. Let us begin with the recalling that the multi-valuedness of the rates makes the evolutionary equations of a system subject to coherence to be discrete mappings of the type (7.7). In turn, the permanent presence of bounded stochastic terms in (7.7) renders the motion in the state space to be subject to incremental boundedness. Therefore, the motion in the state space possesses all the properties established in Chapter 3. Besides, the state space is separated into basins of attractions whose range is determined by the deterministic part of the discrete mappings (7.8). Being a non-linear mapping, (7.8) gives rise to different dynamical regimes depending on the control parameters choice. Therefore, there is no single equilibrium state that is global attractor. This result covers the behavior of open systems that still puzzles the modern science. Up-to-date the efforts of the scientific community have been focused on its understanding of being a complicated effect of the non-linearity incorporated in the statistical mechanics. Note that we came to the same result without any reference to the thermodynamics and without involving the notion of the entropy. The latter is available only when local invariants exist so that the entropy is additive with respect to them and insensitive to the way of partitioning the system. On the other hand, the real systems do have specific time and space scales which make the entropy dependent on the partitioning.

### 7.4. Embedding Dimension: Unambiguous Determination

One of the most significant properties of BIS is the existence of embedding dimension. In &1.6 this characteristic has been already considered from the viewpoint of the boundeness. The embedding dimension is the dimension of the attractor of the motion in the phase space. The presence of thresholds of stability makes the size of the attractor finite while the dynamical boundedness sets finite value for its dimension. Let us recall that the dynamical boundedness sets relation between the size and the duration of any stable fluctuation. This relation is the physical foundation for the study of dynamics of BIS by the use of the time delay embedding method. The latter implies to choose a small delay $\tau$ so that the vectors $\vec{R}_n(t) = (X(t), \ldots X(t - i\tau), \ldots X(t - n\tau))$ are embedded in a $n$ dimensional Euclidean space. $X(t)$ are successive points that come from a given time series. Further, the phase space is divided into small size cells and the vectors whose ends are inside each cell are counted. The population in lg-lg scale is plotted vs. cell size. Applied to a BIS, the time delay embedding is noting more than a particular way of coarse-graining. So, the ends of the vectors $\vec{R}_n(t)$ also construct BIS. Then we can associate the value $\overline{X}(t - i\tau)$ of the coarse-grained BIS with the $i-th$ axis, where $i \in [1, n]$. Note that the association of the delay $i\tau$ with the topological dimension is equivalent to the parameterization of the size of the fluctuation through the phase space angle. Yet the size of a fluctuation can also be parameterized trough its relation with the duration. The evident parity of both parameterizations selects a topological dimension so that a fluctuation forms a closed continuous curve by a single revolt. The topological dimension that renders the largest fluctuation to appear as a loop is called embedding dimension. The hallmark of this dimension is that all the smaller fluctuations are closed continuous loops so that each loop has its own embedding dimension smaller than the embedding dimension of the entire attractor. The boundedness of both the size and duration makes the embedding dimension always finite.

A closer look on the above considerations shows, however, that the exact value of the embedding dimension remains ambiguous because it implicitly depends on the delay $\tau$ which can be arbitrary. Note that the dependence on the delay $\tau$ comes out through the dependence of the Euclidean space topological dimension on the value of $\tau$. The discretization of time immediately points the way out. Indeed, setting the value of $\tau$ equal to the duration of a session selects non-ambiguously the value of the embedding dimension.

### 7.5 Choice of Units

The fact that each process has its natural units is far away from the widely accepted so far concept that the processes in the Nature must be independent of the unit choice. This idea has been taken for granted and has been considered as automatically fulfilled. A very few attention has been paid to the problem and even it has not been recognized as a problem at all. However, it is a serious problem. Moreover, it turns out to be a fundamental problem of modern mathematics that involves nonlinearities. To resolve the puzzle let us first elucidate explicitly why this problem has not been recognized for more than a century. Let us consider a system of linear ordinary differential equations:

$$\frac{d\vec{x}}{dt} = \hat{A} \bullet \vec{x} \qquad (7.11)$$

where $\hat{A}$ is the matrix of the coefficients that determines the eigenvalues of the solution. Obviously, the rescaling of the variables $x_1, x_2, \ldots, x_N$ by any vector of parameters $(c_1, c_2, \ldots, c_N)$ leaves (7.11) invariant. Note, however, that the invariance of (7.11) with respect to the choice of units is property of the linear differential equations only! Yet, because of the dominant role of the linear mathematics in all fields of science for more than a century, the problem with the units has not been recognized at all. Only very recently starts the intensive study of the nonlinearities and their adequate incorporation in our knowledge about the Nature. It turns out that unlike their linear counterparts, the non-linear differential equations are not invariant under the rescaling of the units. Furthermore, an already established general result is that the type and the properties of the solution of the non-linear ordinary or partial differential equations strongly depend on the concrete values of their parameters. To illustrate the point let us consider the following example:

$$\frac{dx}{dt} = ax - b(x-k)^3 \qquad (7.12)$$

where $a, b$ and $k$ are parameters. Obviously, the solution of (7.12) depends strongly on whether $\left((ax - b(x-k)^3)\right)$ is positive or negative. If positive, the solution is unstable and it exponentially departs from the steady one determined by $ax - b(x-k)^3 = 0$. On the contrary, in the range of $x$ where $ax - b(x-k)^3 < 0$, the solution is stable and monotonically approaches the steady one. Hence, the demarcation between the stable and unstable solution is determined by:

$$ax - b(x-k)^3 = 0. \qquad (7.13)$$

Let us now rescale the variable $x$ according to $x' = 2x$. Then (7.12) becomes:

$$\frac{d2x}{dt} = 2ax - b(2x-k)^3 \qquad (7.14)$$

Simple calculations show that the equation for demarcation between the stable and the unstable solution becomes:

$$ax - \frac{b}{2}(2x-k)^3 = 0. \qquad (7.15)$$

Obviously (7.15) deviates from (7.13) in its non-linear part. Therefore, indeed, the non-linearities render the differential equations non-invariant under the choice of units. In turn, this gives rise to the fundamental question about the notion of unit itself and its interrelation with such basic concepts such as continuity in space and in time. We have already demonstrated that the discretization of time helps to establish the units in natural way. A step further in this direction is the replacement of the differential equations with discrete mappings as general mathematical implement for description of the evolution. Note that the differential equations describe continuous in time processes while the discrete mappings discrete ones. The natural discretization of time provides huge advantage of the discrete mappings: despite the particularities of the non-linearities involved, the discretization sets the units non-ambiguously. Besides, the discretization of the time sets the units regardless to whether the mapping (7.7) is linear or non-linear. So, it is the fundament that integrates the unit-independent linear processes and the unit-dependent non-linear ones in a single united frame.

### 7.6. What Comes Next

The developed throughout the book approach to fluctuations is not only based on radically novel assumptions but it also yields non-trivial and unexpected properties. Certainly, one of the most fascinating properties is the discretization of time. Note that the discreteness of time is property of a process that evolves in an open system that sustains its long-term stability by means of coherence. Let me now delineate its differentiation from the space-time quantization: (I) the quantization is a static property of an entity confined in a given potential: the quantized levels are those ones at which the entity stays arbitrarily long time if not perturbed. So, the quantization must be associated with closed systems. On the contrary, the discretization of time is related to the evolution of a given process that proceeds in an open system through alternation of coherence sessions and sessions of relaxation through $\hat{H}_{rigid}$. Yet, to certain extend paradoxically, both quantization and discreteness of the time provides the stability of the evolution - the quantization selects stable orbits in closed systems while the coherence eliminates the spatio-temporal destabilization due to the stochastising interactions in the open systems.

The interplay of the quantization and the discreteness of time is best revealed in the properties of the quantum spectra of the open many-body systems. Our considerations give rise to coexistence of discrete and continuous bands: the discrete band comes from the rigid part of the spectrum while the continuous one emerges form the interference of the successive trains of emission. The emission is separated into trains since the coherence sessions are silent for the interaction between the entities in the low-energy limit and the radiation. The obtained discretization of time is mapped in the quantum spectra as persistent continuous band of shape $1/f^{\alpha(f)}$ superimposed to the discrete band that comes from the genuine interaction that proceeds in the rigid part of the spectrum. The hallmark of the continuous band is that neither its presence nor its shape depends on the type of intervention and the properties of the system - its presence is set only on whether the coherence takes place or not. Hence, the presence of a continuous band of shape $1/f^{\alpha(f)}$ serves as discrimination criterion for the reality of the coherence.

Another fascinating property of the discretization of time is that it sets unambiguously the units regardless to whether the mappings (7.7) are linear or non-linear. So, it unravels the enigma why the linear processes are unit-independent while the non-linear ones are not.

The non-trivial effects obtained in the present chapter provoke my interest to look further for other unexpected effects brought about by the boundedness in all its aspects. This will be the goal of the next chapter.

# Chapter 8: Boundedness in the Reality.
# Transmission of Information through Media

## 8.1 Why to Read Chapter 8

So far our attention has been focused on the development of self-consistent theory that accounts for the long-term stability of the systems that exert fluctuations. Despite the plausibility of our approach, it would be bare words if not confirmed by the reality. That is why the aim of the present chapter is to outline how the developed theory can be traced in it. Since the effects of the discretization of the have been extensively studied in the previous chapter, now I aim to consider other effects produced by the boundedness and coherence.

The approach is built on the grounds of advanced by us concept about boundedness which has 3 major aspects. The first one asserts that every system has its thresholds of stability that if exceeded makes a system to undergo breakdown. This assertion imposes limitation on the amplitude of fluctuations: in order to sustain long-term stability, the fluctuations are exerted so that never to exceed the thresholds of stability. However, at this point our intuition runs into problem: what happens when a fluctuation reaches them - do "U-turns" involve specific physical processes? My answer is no, no physical process is needed for executing the "U-turns when an intriguing interplay of the short-ranged dynamics and the chaoticity of the state space takes place. It is explicitly expressed by the relation (3.16) that sets the largest amplitude of fluctuations so that the corresponding "U-turns" occurs without involving any additional physical process. Though the relation (3.16) has been rigorously established, its key role in the entire approach poses the question whether this rather abstract result can be verified in the reality. To the most surprise the answer is positive: it indeed can be verified. Let us recall that the assumption about the automatic undergoing of "U-turns" renders scale invariance of the time scales correlations, i.e. all time scales contribute uniformly to the evolution. I shall elucidate this issue in the next section where the effects of the entanglement between the scaling invariance and the boundedness are considered. Moreover, I shall present a broad spectrum of experiments that support them.

The second specification of the basic concept is the dynamical boundedness. It originates from the fundamental assumption that every process in the Nature develops so that the rate of exchanged with the environment energy and/or matter is always bounded. The formal explication of that assumption implies boundedness of the rate of fluctuation development. Evidently, it sets relation between the size of a fluctuation and its duration. Furthermore, as proven in &1.6, the parity between the parameterization through the phase space angle and the duration of the fluctuation renders every fluctuation to have its own embedding dimension. Hence, the embedding dimension of a time series varies with its length because its current value is set on the size of the largest in the time series fluctuation. On exerting the first "U-turn" the embedding dimension reaches its largest value and stops varying on further increase of the length of time series. In &8.3 I shall consider concrete experiment that manifests variable embedding dimension.

The third specification of our fundamental assumption is that of the spatial coherence. It implies necessity of long-range correlations among local fluctuations so that to prevent development of local defects due to otherwise non-restrained succession of their spatio-temporal configurations. Indeed, when only short range interactions and local rules are involved in the elementary processes, the dynamics in the extended systems is non-correlated both in space and in time. This immediately renders developing of local defects such as strain, overheating, sintering etc. At the next instant the spatial configuration changes and the local defects move. If not suppressed, due time course their interaction would produce local reconstruction, would create mechanical, thermal and/or other defects. Eventually the process would yield the system breakdown. So, the long-term stability apparently calls for persistent coupling of the local fluctuations. A wide spectrum of works aims to explore an effect of correlations of fluctuations in extended systems as interplay among noise correlations, non-linearity and spatial coupling. Yet, the stochastic variables and noise sources in all developed so far approaches are modeled by the use of Wiener process whose increments are independent and unbounded. Thus, though cooperation of the fluctuations is available, the sequence of spatio-temporal configurations through which it arrives to global coupling varies in uncontrolled way that is irreconcilable with the idea of boundedness. These circumstances forced me to suggest an entirely new viewpoint on the interactions that yields unexpected properties. In Chapter 7 we considered one such property: the incorporation of the time discretization in the quantum spectra. The peculiarity of its manifestation is two-fold: (I) it is proved through persistent presence of continuous band of the shape $1/f^{\alpha(f)}$. This band subsists at the entire range of radiation starting at far IR and ending at UV region, i.e. it is insensitive neither to the incident radiation nor to the particularities of the system; (ii) the band does not correspond to any real emission.

An important outcome of the joint action of all 3 aspects of the boundedness is that the macroscopic evolution is described by stochastic discrete mappings of the type (7.7). For example, the suggestion about the lack of physical correlations among time scales (scale invariance) validates their application on every time scale larger than the basic for the process one. A distinctive property of the solution of (7.7) that its power spectrum comprises additively discrete band that comes out from the deterministic part (7.8) and continuous band of shape $1/f^{\alpha(f)}$ that originates from the stochastic terms. It should be stressed that while the band brought about by (7.8) is always discrete and emerges only for certain choices of the control parameters, the continuous band persists at every admissible choice of the control parameters. On the contrary, neither system of ordinary or partial differential/difference equations gives rise to coexisting of a discrete and a continuous band in the power spectra of their solutions at any control parameter choice. Hence, the coexisting of a discrete band and continuous band of shape $1/f^{\alpha(f)}$ is a crucial test for the validity of the entire theory. I shall come back to this topic in &8.4.

### 8.2 Boundedness and Scale Invariance

Before presenting any experiment let us recall how the boundedness of the fluctuation size is interrelated with the scaling invariance. I start with the major characteristics of the correlations among time scales, the autocorrelation function:

$$G(\eta) = \frac{1}{T} \int_0^T X(t+\eta)X(t)dt \tag{8.1}$$

The autocorrelation function $G(\eta)$ is measure for the average correlation among any two points in a time series separated by time interval $\eta$. Yet, more popular is the power spectrum, i.e. the Fourier transform of $G(\eta)$:

$$S(f) = \lim_{T\to\infty} \frac{1}{T} \int_0^T G(\eta) \exp(i\eta f) d\eta \tag{8.2}$$

The power spectrum is due its popularity to its easy and non-ambiguous reading. Indeed, any long-range correlation between two time scales appears as single line whose amplitude is proportional to their correlation. Therefore, it is to be expected that the scale invariance does not signal out any perceptible line in the power spectrum. In Chapter 1 we found out that the power spectrum of a time series of length $T$ that is subject both to boundedness and scale invariance is continuous band of shape $1/f^{\alpha(f)}$ where $\alpha(f) = 1 + \kappa\left(f - \frac{1}{T}\right)$, $f \in \left[\frac{1}{T}, \infty\right)$. Note that the non-zero correlations among time scales are introduced by the boundedness and are not result of any physical process! Besides, if the time series were unbounded, the scale invariance would result in zero correlations among the time scales, i.e. all components of the power spectrum would be zero! Therefore, the interplay of the boundedness and the scale invariance gives rise to continuous band of the shape $1/f^{\alpha(f)}$. So, the persistence and smoothness of that band serve as criterion for the validity of the assumption about the joint action of the boundedness and the scale invariance. Now we are ready to interpret the experiments. Let me start with a particular one. The purpose is to support as much as possible aspects of our theory by examination of the same experiment. It is expected that since the listed in &8.1 aspects of the basic concept are inseparably entangled, they are included in one way or another in the same experiment.

We studied the reaction of oxidation of $HCOOH$ (formic acid) over two modifications of supported $Pd$ catalyst at steady flow of the reactants [8.1]. The control parameters were the partial pressure of the reactants and the temperature of the feedstock at the reactor inlet. The temporal behavior was studied in the temperature interval $110-130\ °C$; the $O_2$ feed concentration was varied from $0.5-12\%$ while that of $HCOOH$ - from $0.5\%$ to $8\%$.

The difference $\Delta T$ between the catalyst bed temperature and the feedstock at the reactor inlet was measured, digitized and continuously recorded. The sampling rate was 2 points per second. 80 time series of that difference have been recorded scanning the values of the feed concentrations and temperature of the feedstock at two charges of the catalyst.

At all 80 time series the difference $\Delta T$ exhibits irregular variations the amplitude of which does not exceed $10\%$ of the average "shift" of the catalyst bed temperature from the feedstock one. However, there are occasional large variations amplitude of which is about $50\%$ of the average "shift".

Our study show that all 80 power spectra comprise continuous band of the shape $1/f^{\alpha(f)}$, where $\alpha(f)\to 1$ at $f\to 1/T$ ($T$ is the length of the time series) and $\alpha(f)$ linearly increases on $f$ increasing. The shape is robust to the catalyst charge, feedstock temperature and the feed concentrations. Let me show you two of those time series along with their power spectra. The time series are presented in digital units and the power spectra in lg-lg scale in relative units: the value of the successive components is divided to the first one. This is made in order to make obvious that the power spectrum is power function of shape $1/f^{\alpha(f)}$ with $\alpha(f)\to 1$ at $f\to 1/T$ ($T$ is the length of the time series) and $\alpha(f)$ linearly increasing on the increase of the frequency.

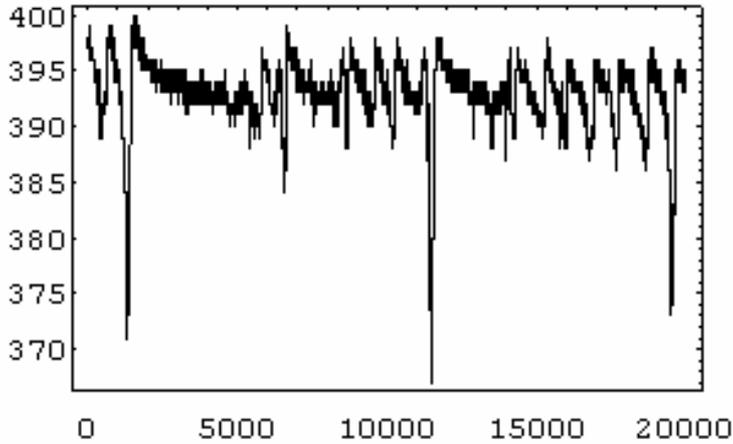

Fig.8.1a First time series

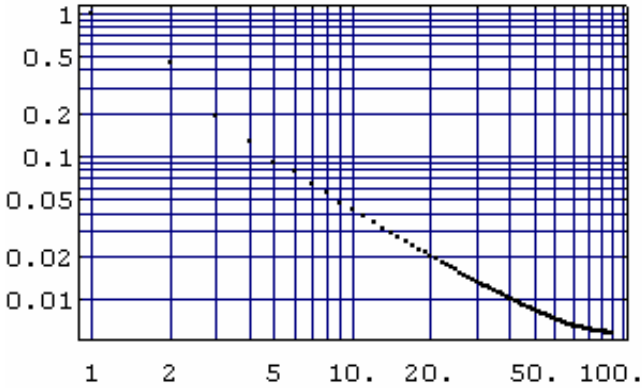

Fig.8.1b The power spectrum of the time series presented in Fig.8.1a

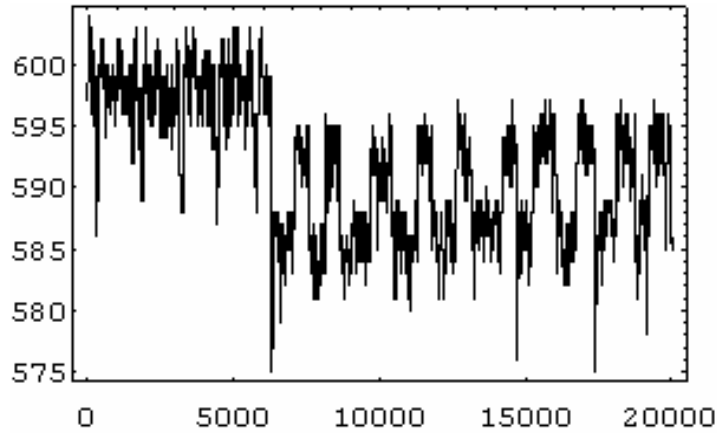

Fig.8.2a Second time series

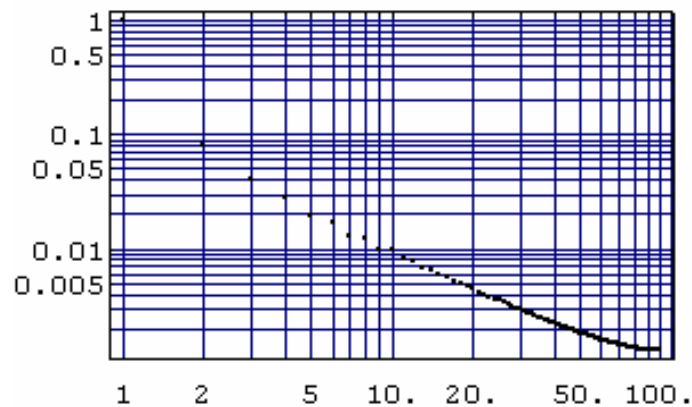

Fig.8.2b The power spectrum of the time series in Fig.8.2b

The power spectra in Fig. 8.1b and Fig.8.2b show that they indeed fit the shape $1/f^{\alpha(f)}$ with $\alpha(f) \to 1$ for the infrared edge (the first component) and $\alpha(f)$ linearly increase on the increase of the frequency. It should be stressed that the use of relative units for presentation of the power spectrum makes revealing of the exponent $\alpha(f)$ non-ambiguous because it makes the shape $1/f^{\alpha(f)}$ insensitive to the choice of the units. Be aware that the shape $1/f^{\alpha(f)}$ is not unit-invariant! The latter implies that on any change of the units $f' = cf$, the shape $1/f^{\alpha(f)}$ is multiplied by factor $c^{\alpha(f)}$ which, however, is different for different frequencies! Hence, $\alpha(f')$ deviates from $\alpha(f)$! The only way to make $\alpha(f)$ unit-invariant is to present the power spectrum in relative units.

Though the above experiment is very convincing, it poses the question whether it is only a particular case. Remember that we have introduced the boundedness as a concept advantageous for a broad spectrum of systems. Correspondingly, credible support must be brought about by a variety of systems of different nature. Such support does exist and comes one of the most ubiquitous phenomena in the world: $1/f$ noise. Its major characteristic is that the infrared edge of the power spectra uniformly fits the shape $1/f$. The fit does not depend on: (I) the incremental statistics, i.e. the details of the irregularity succession in the time series; (ii) the length of the time series; (iii) the nature of the system - the phenomenon is observed in large variety of systems: quasar pulsations, meteorology, financial time series, music and speech etc. Though it has been thoroughly studied for more that a century, it still remains enigma. In &1.4 we have already discussed the relation between the shape $1/f$ and $1/f^{\alpha(f)}$. The purpose to come again to that topic is two-fold: (I) since our conjecture it is able to clarify reliably the above listed properties, the proximity of the shapes $1/f$ and $1/f^{\alpha(f)}$ is a strong argument in its favor. (ii) the entanglement of the coherence, scaling invariance and boundedness makes possible to explain another important aspect of the mysterious $1/f$ noise : the non-stationarity. This is the task of &8.4.

The advantage of the boundedness conjecture is best revealed in the reconciliation of the greatest mystery of the $1/f$ noise: on the one hand, the shape $1/f$ should be associated with instability and breakdown because it renders the variance of the time series, calculated on its grounds, infinite. In turn, the infinity of the variance implies that fluctuations large enough to carry the system beyond the thresholds of stability become most probable. Therefore, any system would rapidly blow up or get extinct. However, it does not happen. Moreover, it is well established that the $1/f$ behavior is spanned over several dozens of orders in the time course. Hence, it should be rather associated with long-term stability. In turn, the latter makes the concept of boundedness advantageous for the $1/f$ noise. The boundedness always sets finite variance of the fluctuations that opposes the infinite variance set on the shape $1/f$ of the power spectra. The considerations in &1.4-&1.5 undoubtedly show that the boundedness along with the scaling invariance gives rise to the shape $1/f^{\alpha(f)}$ not to $1/f$. The merit of the shape $1/f^{\alpha(f)}$ is that it brings about finite variance of the fluctuations as proven in &1.5. On the contrary, the shape $1/f$ yields infinite variance! Besides, the shape $1/f^{\alpha(f)}$ does not depend neither on the statistics of the time series nor its length. Note, that the withdrawal of the boundedness concept yields strong dependence of the power spectrum shape on the statistics of the time series. So, no universal shape would be possible! Hence, our theory takes away the major enigmas of the phenomenon $1/f$ noise. That is why the establishing of non-constant exponent $\alpha(f)$ makes our arguments most persuasive. Our experiment on oxidation of $HCOOH$ is the first step in this direction. Let us just point out that the lack of data on linearity of $\alpha(f)$ is easily understandable: recall Table 1 where some values of $\kappa$ are calculated. Since $\kappa$ is vanishingly small, it is very difficult to be established if not looked for it purposely. Let us point out that we are the first one who have introduced and have been studying boundedness.

Summarizing, we assert that our concept successfully takes away the major mysteries of the $1/f$ noise. Then, the rich variety of systems that exhibit $1/f$ noise behavior strongly supports the advantage and ubiquity of the concept about boundedness.

### 8.3 Variations of the Embedding Dimmension

The second aspect of the boundedness conjecture is the dynamical boundedness. The assertion is that the rate of development of fluctuations is bounded as a result of suggestion that the exchange of energy/matter with environment is kept permanently finite for every process in the Nature. Furthermore, in &1.6 we develop explicit relation between the dynamical boundedness and the embedding dimension: it varies with the length of the time series because the current value of the embedding dimension is set on the size of the largest in the time series fluctuation. On exerting the first "U-turn" the embedding dimension reaches its largest value and terminates its further variations on the increase of time series length. The examination of the experiment on the oxidation of $HCOOH$ shows that the current value of the embedding dimension manifests strong sensitivity to the fluctuation size for all 80 time series. The establishing of that persistent sensitvity is made by means of cutting the time series into pieces and monitoring the corresponding embedding dimension. It turns out that the latter permanently varies. In turn, this straightforwardly support our finding that the current embedding dimension is related to the largest in the time series fluctuation size.

Though the persistent presence of a band that fits the shape $1/f^{\alpha(f)}$ and the variations of the embedding dimension with the length of the time series apparently support our fundamental concept about the boundedness, there are still not enough for sharp discrimination from alternative concept that involves the idea about non-stationarity. The latter emerges from the widely observed sudden changes in the temporal behavior of the systems that exhibit $1/f$ behavior. Therefore, an alternative explanation of the embedding dimension variations might be related to development of non-stationary process(es). So, crucial test for our theory is the plausible explanation of the non-stationarity. Still, the foremost question is whether there are comprehensible manifestations of the non-stationarity that definitely confirm our theory. The considerations how the non-stationarity is related to the concept about the coherence and boundedness are presented in the next section along with experimental results supporting them.

## 8.4 Coherence and Macroscopic Fluctuations. Non-Stationarity

The third aspect of the boundedness conjecture that of the local boundedness, requires mechanism that acts towards correlation of the local fluctuations making their further response coherent. One of the most fascinating outcomes of our theory is that the coherence drives the systems to exert permanently macroscopic fluctuations whose major property is the boundedness of their amplitude. As a result, the macroscopic evolution of the system is described by discrete mappings of type (7.7). The inevitable presence of the stochastic terms renders permanent deviations from the dynamical regime prescribed by the deterministic part (7.8). This is so because any difference $\left(\vec{x}(t) - \vec{x}_{det}\right)$ can effectively be presented as a solution of the mapping (7.8) at "shifted" control parameters. Note that physically the values of the control parameters are kept permanently fixed. The "shifting" causes immediate change either of the characteristics of the original dynamical regime or it even induces a bifurcation. It is obvious that at given parameter choice, an induced bifurcation needs development of a fluctuation of appropriate size. The induced bifurcation will contribute to the power spectrum as a discrete band. Note, however, that the induced bifurcation has temporary effect – it lasts until the fluctuation size is significant. Yet, though the induced by fluctuations temporary effects scores in favor of our efforts to explain the non-stationarities, there is need of decisive arguments that emerge directly from eq.(7.7) and are active along with the temporary effects. For this purpose let us take a closer look on the sequence constituted by $\left(\vec{x}(t) - \vec{x}_{det}\right)$. It is a zero-mean BIS that always contributes to the power spectrum by continuous band of the shape $1/f^{\alpha(f)}$. This is very important property that gives rise to the target discrimination criterion for our theory: according to the considerations in &3.2 and &7.3, the power spectrum of the solution of (2.15) and (7.7) comprises additively continuous band of shape $1/f^{\alpha(f)}$ and discrete band. The latter comes from (7.8) at appropriate values of the control parameters. Note, that neither system of ordinary or partial differential/difference equations can give rise to coexisting of a discrete and a continuous band in the power spectrum of its solution! Then, that coexisting along with the temporary effect of the induced bifurcation serves as criterion that discriminates our theory from any other whose mathematical description is given by systems of ordinary or partial differential equations and/or discrete mappings.

Let me now present an induced bifurcation that lasts until the fluctuation is significant. We took one particular time series recorded at the oxidation of *HCOOH* and cut it into three successive parts of equal length. At Fig.8.3a, b, c are presented those three parts in order of their succession. At Fig.8.4a, b, c are presented the corresponding power spectra.

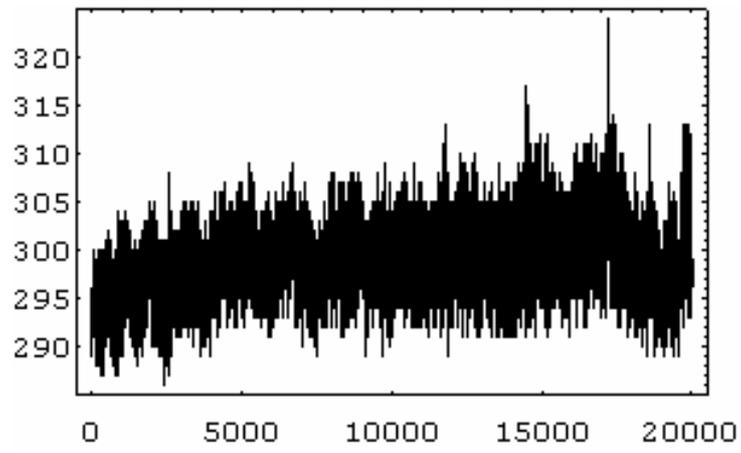

Fig8.3a

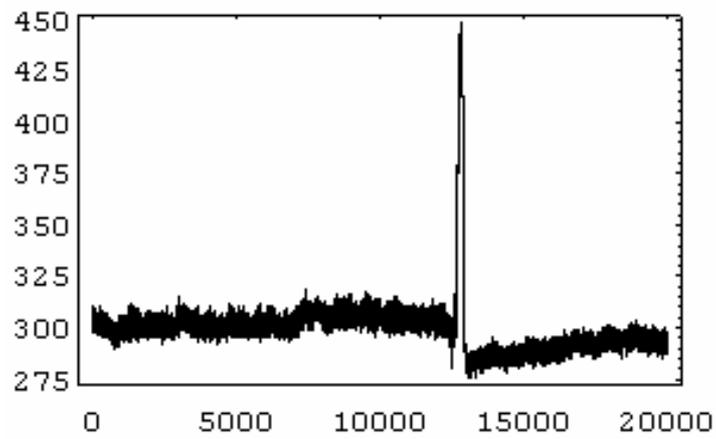

Fig.8.3b

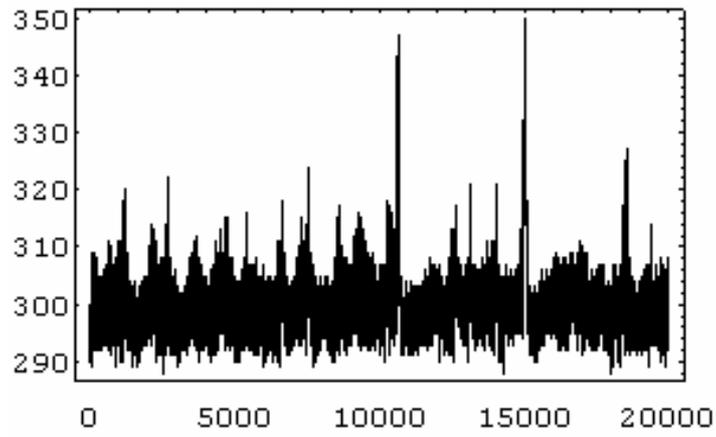

Fig.8.3c

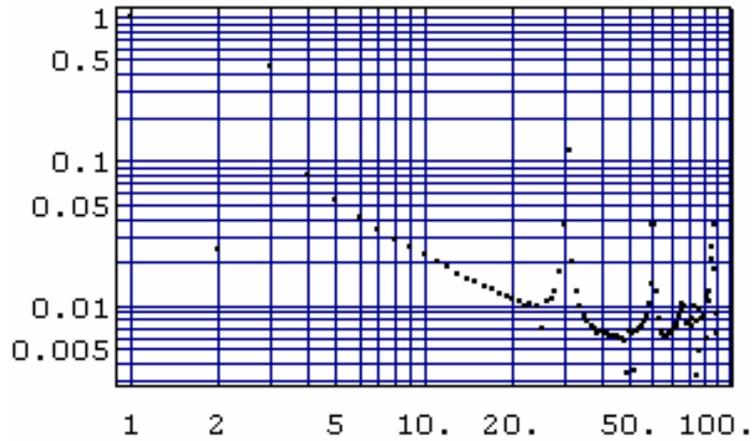
Fig. 8.4a

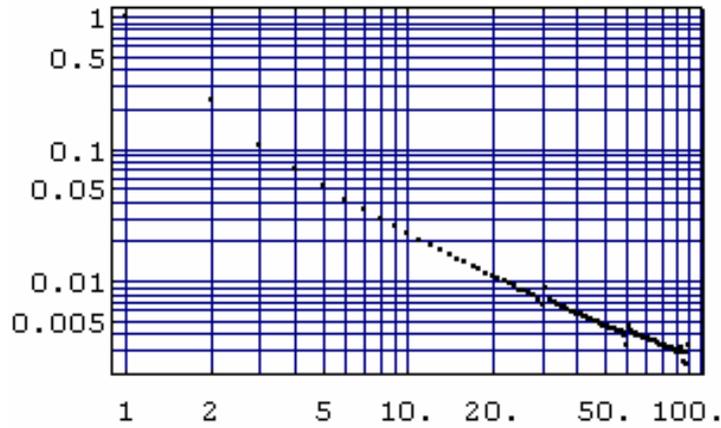
Fig.8.4b

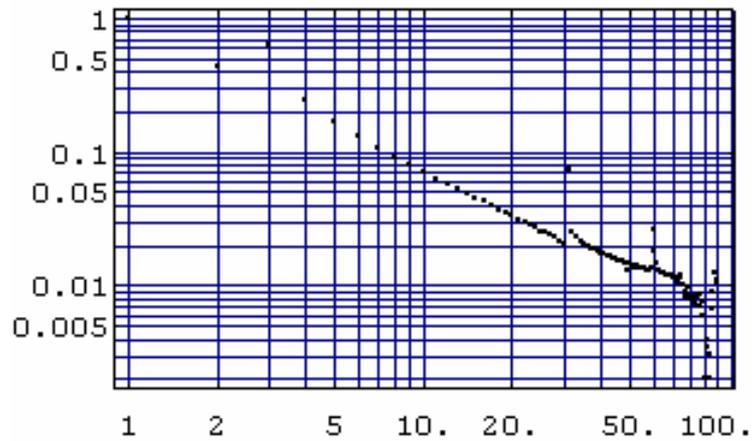
Fig.8.4c

The examination of the Fig.8.3 and Fig.8.4 apparently shows that the power spectrum of all three parts comprises both discrete and continuous band of the shape $1/f^{\alpha(f)}$ where $\alpha(f) \to 1$ at the low frequencies and grows linearly with the frequency. It is obvious that the presence of a large fluctuation (Fig.8.3b) strongly manipulates the amplitude of the discrete band in the corresponding power spectrum (Fig8.4b): it is about 10 times smaller than that of the discrete bands in Fig.8.4a and 8.4c. On the other hand, the period remains the same at all 3 power spectra. The great sensitivity of the amplitude of oscillations to the distance to the bifurcation point along with robustness of the period is a property genuine for a limit cycle. Further, the impact of that large fluctuation is temporary - it lasts as long as the fluctuation is essentially large. Indeed, as we already mentioned, the amplitude of the limit cycle at Fig.8.4a and Fig.8.4c is 10 times greater than that of the Fig. 8.4b. Still, a closer look on Fig.8.4a and Fig.8.4c shows that there is difference in the amplitudes of the discrete bands though not as pronounced as the one in Fig.8.b. Therefore, we come to the conclusion that the effective shift of the control parameters is governed by the current largest fluctuation.

Summarizing, the above presented induced bifurcation and its properties are crucial and convincing evidence for confirmation of our theory. Moreover, it not only substantiates our conjecture but gives successful explanation of the ubiquitous non-stationarity that occurs along with the $1/f$ noise.

The obtained effective shift of the control parameters implies complicated interaction with the environment: though the control parameters are permanently kept fixed, the response of the system varies in the course of the time. Therefore, there is no sharp separation between system, environment and the interaction between them - the weak-coupling limit turns out to be inappropriate for this case. This gives strong support to the arguments advanced in Chapters 4, 5 and 6 for the necessity of a new, more general, viewpoint on the interactions that must replace the weak-coupling limit.

### 8.5 Natural Bits. Transmitting Information Trough Media

Save and accurate transmitting of information trough media is a longstanding and still puzzling problem. We know how an artificial computer works because we have built it. But how the information is transmitted in the Nature? Now we are ready to answer this question: we are able to point out the natural binary code and its operational protocol. Indeed, by means of the idea of induced bifurcation, we can construct two natural bits: one is the state whose power spectrum comprises continuous band only, the other one comprises both discrete and continuous band. Transition from one state to the other is achieved by shifting the control parameters. Note, that apart from those two, no other options for the power spectra are possible. The greatest advantage of the use of the natural binary code is the robustness of the continuous band: it always fits the shape $1/f^{\alpha(f)}$ whatever the internal noise and reading process is. This provides a save and reproducible way of transmitting information and renders constant accuracy of its reading.

However, the natural binary code cannot provide transmitting of arbitrarily long series of bits. The reason is natural and unavoidable: the exerting of induced bifurcations that come out from the fluctuations because they interrupt the sequence of digits and introduce "noise". Furthermore, since the probability for a fluctuation increases on the increase of the length of the time series, mistakes in reading the information increase on increasing the length of the sequence. Therefore, the fluctuations bound the amount of information that can be transmitted without significant distortion. So, we again encounter the boundedness!

I hope that by this book I manage to persuade you that the boundedness not only introduces a novel fascinating viewpoint on both natural and artificial systems but it also opens the door to a creative new prospective for designing our World.